\documentstyle[11pt]{article}

\newcommand{\noi}{\vspace{12pt}\noindent}
\newcommand{\beq}{\begin{equation}}
\newcommand{\eeq}{\end{equation}}
\newcommand{\bea}{\begin{eqnarray}}
\newcommand{\eea}{\end{eqnarray}}

\newcommand{\e}[1]{{(\ref{#1})}}
\newcommand{\eq}[1]{{eq.\ (\ref{#1})}}
\newcommand{\es}[2]{{(\ref{#1}) and (\ref{#2})}}
\newcommand{\eqs}[2]{{eqs.\ (\ref{#1}) and (\ref{#2})}}
\newcommand{\Ref}[1]{{Ref.~\cite{#1}}}
\newcommand{\mb}[1]{{\mbox{${#1}$}}}

\newcommand{\ie}{{${ i.e.\ }$}}
\newcommand{\eg}{{${ e.g.\ }$}}
\newcommand{\ea}{{${ et~al.\ }$}}
\newcommand{\wrt}{{with respect to }}

\newcommand{\rhs}{{right hand side }}

\newcommand{\odd}{{\rm odd}}
\newcommand{\even}{{\rm even}}
\newcommand{\ad}{{\rm ad}}
\newcommand{\str}{{\rm str}}
\newcommand{\sdet}{{\rm sdet}}
\newcommand{\rank}{{\rm rank}}
\newcommand{\pl}{{\sf Pl}}
\newcommand{\ext}{{\rm ext}}

\newcommand{\Hf}{{1 \over 2}}

\newcommand{\Ih}{{i \over \hbar}}
\newcommand{\Hi}{{\hbar \over i}}
\newcommand{\phu}[1]{\tilde{#1}}
\newcommand{\phd}[1]{\tilde{#1}}
\newcommand{\ph}[1]{{#1}_{\sim}}
\newcommand{\gh}{c}
\newcommand{\bgh}{\bar{c}}
\newcommand{\bo}{(0)}
\newcommand{\ho}{[0]}
\newcommand{\bl}{(1)}
\newcommand{\hl}{[1]}
\newcommand{\bz}{(2)}
\newcommand{\hz}{[2]}

\newcommand{\htr}{[3]}
\newcommand{\bnz}{(n-2)}
\newcommand{\hnz}{[n-2]}
\newcommand{\bnl}{(n-1)}
\newcommand{\hnl}{[n-1]}
\newcommand{\bn}{(n)}
\newcommand{\hn}{[n]}
\newcommand{\nth}{\mb{n^{\prime}}{\em th}}
\newcommand{\nlth}{\mb{(n\!-\!1)^{\prime}}{\em th}}
\newcommand{\nzth}{\mb{(n\!-\!2)^{\prime}}{\em th}}
\newcommand{\ND}{N_{D}}
\newcommand{\matrisse}[1]{#1}
\newcommand{\onebyone}[1]{\left[{#1}\right]}
\newcommand{\twobyone}[2]{\left[\begin{array}{c}{#1} \cr
                                {#2} \end{array} \right]}
\newcommand{\onebytwo}[2]{\left[\begin{array}{ccc}{#1}&{#2}
                                        \end{array} \right]}
\newcommand{\twobytwo}[4]{\left[\begin{array}{ccc}{#1}&{#2} \cr
                                  {#3} & {#4} \end{array} \right]}
\newcommand{\stacktwo}[2]{\begin{array}{c}\scriptstyle{#1} \cr 
                                   \scriptstyle{#2} \end{array}}

\newcommand{\lpart}{\raise.3ex\hbox{$\stackrel{\leftarrow}{\partial}$}}
\newcommand{\rpart}{\raise.3ex\hbox{$\stackrel{\rightarrow}{\partial}$}}
\newcommand{\ldr}{\raise.3ex\hbox{$\stackrel{\leftarrow}{\delta^r}$}}

\newcommand{\deder}[1]{{ 
 {\stackrel{\raise.1ex\hbox{$\leftarrow$}}{\delta^r}   } 
\over {   \delta {#1}}  }}
\newcommand{\dedel}[1]{{ 
 {\stackrel{\lower.3ex \hbox{$\rightarrow$}}{\delta^l}   }
 \over {   \delta {#1}}  }}
\newcommand{\papar}[1]{{ 
 {\stackrel{\raise.1ex\hbox{$\leftarrow$}}{\partial^r}   } 
\over {   \partial {#1}}  }}
\newcommand{\papal}[1]{{ 
 {\stackrel{\lower.3ex \hbox{$\rightarrow$}}{\partial^l}   }
 \over {   \partial {#1}}  }}
\newcommand{\ddr}[1]{{ 
 {\stackrel{\raise.1ex\hbox{$\leftarrow$}}{\delta^r}   } 
\over {   \delta {#1}}  }}
\newcommand{\ddl}[1]{{ 
 {\stackrel{\lower.3ex \hbox{$\rightarrow$}}{\delta^l}   }
 \over {   \delta {#1}}  }}

\newcommand{\proofbox}{\begin{flushright}
${\,\lower0.9pt\vbox{\hrule \hbox{\vrule
height 0.2 cm \hskip 0.2 cm \vrule height 0.2 cm}\hrule}\,}$
\end{flushright}}

\newtheorem{theorem}{Theorem}[section]

\newtheorem{corollary}[theorem]{Corollary}
\newtheorem{definition}[theorem]{Definition}

\newtheorem{lemma}[theorem]{Lemma}
\newtheorem{principle}[theorem]{Principle}
\newtheorem{proposition}[theorem]{Proposition}

\begin{document}
\thispagestyle{empty}
\vspace{3cm}
\title{\Large{\bf On Generalized Gauge-Fixing\\ 
in the Field-Antifield Formalism}}
\vspace{2cm}
\author{{\sc I.A.~Batalin}$^1$\\I.E.~Tamm Theory Division\\
P.N.~Lebedev Physics Institute\\Russian Academy of Sciences\\
53 Leninisky Prospect\\Moscow 119991\\Russia\\~\\
{\sc K.~Bering}$^2$\\Institute for Theoretical Physics \& Astrophysics\\
Masaryk University\\Kotl\'a\v{r}sk\'a 2\\CZ-611 37 Brno\\Czech Republic\\~\\
and\\~\\{\sc P.H.~Damgaard}$^3$\\The Niels Bohr Institute\\Blegdamsvej 17\\
DK-2100 Copenhagen\\Denmark}
\maketitle
\begin{abstract}
We consider the problem of covariant gauge-fixing in the most general setting
of the field-antifield formalism, where the action $W$ and the gauge-fixing
part $X$ enter symmetrically and both satisfy the Quantum Master Equation.
Analogous to the gauge-generating algebra of the action $W$, we analyze the
possibility of having a reducible gauge-fixing algebra of $X$. We treat a
reducible gauge-fixing algebra of the so-called first-stage in full detail
and generalize to arbitrary stages. The associated ``square root'' measure
contributions are worked out from first principles, with or without the
presence of antisymplectic second-class constraints. Finally, we consider an
$W$-$X$ alternating multi-level generalization.
\end{abstract}

\vspace{10mm}

\vfill
\begin{quote}
PACS number(s): 02.40.-k, 02.40.Hw, 04.60.Gw, 11.10.-z, 11.10.Ef, 11.15.-q. \\
Keywords: BV Field-Antifield Formalism, Odd Laplacian, Antisymplectic
Geometry, Second-Class Constraints, Reducible Gauge Algebra, Gauge-Fixing. \\ 
\hrule width 5.cm \vskip 2.mm \noindent 
$^{1}${\small E-mail:~{\tt batalin@lpi.ru}} \hspace{5mm}
$^{2}${\small E-mail:~{\tt bering@physics.muni.cz}} \hspace{5mm}
$^{3}${\small E-mail:~{\tt phdamg@nbi.dk}} \\ 
\end{quote}

\vfill
\newpage

\setcounter{equation}{0}
\section{Introduction}

\noi
The field-antifield quantization formalism \cite{BV81,BV83,BV84} has been
given a substantial reformulation, which shows how it fits into a much more
general scheme \cite{BT1,BT,BMS,BBD}. The essential ingredient is a
Grassmann-odd and nilpotent differential operator \mb{\Delta} that is 
symmetric,
\beq
\Delta ~=~ \Delta^{T}~, 
\label{deltasym}
\eeq
\wrt a transposition defined by
\beq
\int \! d\mu~ F (\Delta G)~=~(-1)^{\epsilon_{F}} \int \! d\mu~(\Delta^{T}F)G~,
\eeq
where \mb{d\mu} is a functional measure, whose explicit form we will 
return to in great detail below. The partition function is then given by
\beq
{\cal Z}^{X} ~=~ \int \! d\mu~ e^{\Ih (W + X)} ~,
\label{partition}
\eeq
where both $W$ and $X$ satisfy the Quantum Master Equations
\beq
\Delta e^{\Ih W} ~=~ 0 ~~~~~~~~~~~{\rm and}~~~~~~~~~~~~~ 
\Delta e^{\Ih X} ~=~ 0 ~,
\label{themastereq}
\eeq
respectively.
The important observation is that $W$, which through the boundary conditions
incorporates the classical action, and $X$, which does the required fixing
of gauge symmetries, enter symmetrically.
Viewing $\Delta$ as a generalized odd ``Laplacian'' which may potentially 
have quantum corrections that consist of higher order differential operators,
we see that the integrand in (\ref{partition}) is required to be a product 
of two superharmonic functions.  
One may argue on general grounds that an arbitrary infinitesimal variation 
of the gauge-fixing part \mb{e^{\Ih X}} has the form
\beq
 \delta e^{\Ih X}~=~ [ \stackrel{\rightarrow}{\Delta},\delta \Psi] e^{\Ih X}~,
\label{expmaxvar}
\eeq
or equivalently using \e{themastereq}, the variation \mb{\delta X} is 
BRST-exact,
\beq
\delta X~=~\Hi e^{-\Ih X} \Delta( e^{\Ih X} \delta \Psi)
~\equiv~\sigma_{X}(\delta \Psi )~,
\label{maxvar}
\eeq
where \mb{\sigma_{X}} is a quantum BRST-operator. Surprisingly, the 
independence of gauge-fixing $X$ for the partition function \mb{{\cal Z}^{X}}
can formally be demonstrated \cite{BBD} by just using the above ingredients 
\e{deltasym}, \es{themastereq}{expmaxvar}, without reference to the detailed
form of $\Delta$:
\beq
{\cal Z}^{X+\delta X}-{\cal Z}^{X}~=~ \int \! d\mu~ e^{\Ih W} 
[ \stackrel{\rightarrow}{\Delta},\delta \Psi] e^{\Ih X} 
~=~\int \! d\mu\left[ e^{\Ih W} \delta \Psi ~(\Delta e^{\Ih X})
+(\Delta e^{\Ih W})~\delta \Psi~e^{\Ih X}  ~\right] 
~=~0~.
\eeq 
The analogous statement (with $X$ replaced by $W$) to show independence 
of the choice of gauge generating functional $W$ was first studied by
Tyutin and Voronov \cite{TV}. 

\noi
The fact that the most general description of gauge-fixing $X$ puts it
on equal footing with the construction of the action $W$ means that
there are situations in which gauge-fixing must require special attention.
This happens when, be it for reasons of for example locality or unitarity,
the gauge-fixing function $X$ itself contains gauge degrees of freedom.
It is of interest to clarify what happens in such a situation. What,
in this formalism, fixes gauge symmetries of $X$? Remarkably, it turns out
that the machinery is ready to tackle this more general situation,
and provide the solution to the gauge-fixing problem. How this is achieved
will be described in detail in this paper; the principle is simply that
gauge symmetries of $X$ are fixed by what used to play the r\^{o}le
of only ``action'', $W$. We thus introduce the notion 
of a {\em gauge-fixing} algebra in the $X$-part of the 
field-antifield formalism, 
very similar to the usual {\em gauge-generating} algebra inside the $W$-part.
In order to give a very specific example where the more general quantization
problem needs to be faced, we consider in Section~\ref{secreduc} in detail
the case where the gauge-fixing $X$ is described in terms of a set of 
{\em reducible} gauge-fixing conditions \mb{G_{\alpha}}. The process
may not stop there, since also the new gauge-fixings in $W$ in turn
may contain additional symmetries that need to be fixed as well. 
Then the formalism allows $X$ to take over the gauge-fixing again, and so 
forth, for any finite number of steps in an alternating manner. In this way
a multilevel construction is induced naturally. This is the subject of 
Section~\ref{sechigherlevels}.
We take the viewpoint that the existence of these new classes
of theories must be taken seriously, and that their formal properties
\wrt the quantization program therefore must be established.

\noi
The present paper is devoted to an exposition of this more general situation.
However it is necessary first to establish a systematic and condensed 
formalism before approaching these new and interesting possibilities. 
We therefore begin in Section~\ref{secbvgeometry} with a discussion of some
of the geometrical aspects of the field-antifield formalism from a covariant
perspective. In particular we describe the properties of the measure density
\mb{\rho}, and the use of anticanonical transformations \ie the antisymplectic
analogue of canonical transformations. As an immediate application we extend
the semi-density theory of Khudaverdian \ea \cite{K02} to the degenerate case.
We next show how deformations of solutions to the Quantum Master Equation can
be understood in terms of anticanonical transformations and associated
measure function changes. This establishes a compact formula 
for the changes of gauge in the action, which is essential 
for all subsequent developments in the $W$-$X$ multi-level formalism.
While we shall give the necessary definitions below, we here just briefly 
remind the reader that the multi-level formalism must be introduced
if one wishes to secure the most general and covariant construction 
that in particularly simple gauges reduce to the well-known field-antifield
prescription that was presented in the original papers \cite{BV81,BV83}. 
In Section~\ref{secirr} we then turn to the situation where
gauge-fixings in $X$ are irreducible, exploring gauge-fixing at the 
first-level, and providing a new compact derivation of the form of $X$ 
in that case.  In Section~\ref{secsecondclass} we
return to the possibility of having antisymplectic second-class
constraints in the path integral, a situation quite analogous to the
more conventional case of symplectic second-class constraints \wrt the 
Poisson bracket in the Hamiltonian formalism. In particular, filling out a 
gap in the existing literature, we first establish a reduction theorem which
explicitly demonstrates that the final gauge fixed path integral can be 
expressed, on a physical subspace of antisymplectic coordinates, in precisely
the same form as the partition function \e{partition}. Secondly, we show that
the second-class construction is manifestly invariant under reparametrizations
of the second-class constraints, a vital investigation that taps into the
very foundation of the antisymplectic Dirac construction.
In Section~\ref{secreduc} we consider a reducible gauge-fixing algebra and 
work out a general first-stage reducible theory in detail, 
and determining the associated path integral measure by solving the 
Master Equation. We perform several consistency checks by reduction
techniques, linking reducible and irreducible descriptions of the 
gauge-fixing constraints, and comparing minimal and non-minimal approaches.
Section~\ref{sechigherlevels} discusses the generalization 
of the first-level formalism to the above mentioned multi-level formalism.
{}Finally, Section~\ref{secconc} contains our conclusions.

\setcounter{equation}{0}
\section{Antisymplectic Geometry Revisited}
\label{secbvgeometry}

\noi
Let us start with a covariant, odd \mb{\Delta}-operator of second order,
\bea
\Delta&=&\Delta_{\rho}+V
~,~~~~~~~~~~~~~~~~~~~~~~~~\epsilon(\Delta)~=~1~,\label{deltav} \\
\Delta_{\rho}&\equiv&\frac{(-1)^{\epsilon_{A}}}{2\rho}
\papal{\Gamma^{A}}\rho E^{AB}\papal{\Gamma^{B}}~,\label{deltarho} \\
V&=&V^{A}\papal{\Gamma^{A}}~,
\eea
where \mb{\Gamma^{A}} denotes local coordinates with Grassmann parity
\mb{\epsilon_{A}\equiv\epsilon(\Gamma^{A})}.
The \mb{\Delta}-operator is built with the help of covariant structure 
functions \mb{E^{AB}\!=\!E^{AB}(\Gamma)}, \mb{V^{A}\!=\!V^{A}(\Gamma)},
and \mb{\rho\!=\!\rho(\Gamma)} that transform under 
general coordinate transformations as a 
bi-vector, a vector and a density, respectively. 
We shall assume that \mb{E^{AB}} has a Grassmann-graded skewsymmetry 
\beq
E^{BA}=-(-1)^{(\epsilon_{A}+1)(\epsilon_{B}+1)}E^{AB}~.\label{skewsym}
\eeq
Locally, \eq{deltav} describes the most general second-order odd 
\mb{\Delta}-operator such that
\beq
 \Delta(1)~=~0~.\label{deltaone}
\eeq
The condition \e{deltaone} is not vital for the construction below, but 
since currently there are no applications that would require 
\mb{\Delta(1)\neq 0}, we shall not pursue such a possibility here.  
The antibracket of two functions \mb{F=F(\Gamma)} and \mb{G=G(\Gamma)} is 
defined via a double commutator\footnote{Here, and throughout the paper,
\mb{[A,B]} denotes the graded commutator 
\mb{[A,B]=AB-(-1)^{\epsilon_{A}\epsilon_{B}}BA}.}
with the \mb{\Delta}-operator, acting on the constant function $1$,
\beq
(F,G)~\equiv~-[[F,\stackrel{\rightarrow}{\Delta}],G]1
~=~(F\papar{\Gamma^{A}}) E^{AB}(\papal{\Gamma^{B}}G)~,
\label{antibracket}
\eeq
where use was made of \eq{skewsym}. The square 
\mb{\Delta^{2}=\Hf[\Delta,\Delta]} is generally a third-order operator
with no zero-order term \mb{\Delta^{2}(1)=0}. It becomes of second order 
if and only if a Grassmann-graded Jacobi identity
\beq
\sum_{F,G,H~{\rm cycl.}}(-1)^{(\epsilon_{F}+1)(\epsilon_{H}+1)}
(F,(G,H))~=~0 \label{jacid}
\eeq
for the antibracket holds. We shall assume this from now on. 
A bi-vector  \mb{E^{AB}} that satisfy skewsymmetry \e{skewsym} 
and the Jacobi identity \e{jacid} is called a possibly degenerate 
{\em antisymplectic} bi-vector.
There is an antisymplectic analogue of Darboux's Theorem that states that
locally, if the rank of \mb{E^{AB}} is constant, there exist Darboux 
coordinates 
\mb{\Gamma^{A}\!=\!\left\{\phi^{\alpha};\phi^{*}_{\alpha};\Theta^{a}\right\}},
such that the only non-vanishing antibrackets between the coordinates are
\mb{(\phi^{\alpha},\phi^{*}_{\beta})=\delta^{\alpha}_{\beta}
=-(\phi^{*}_{\beta},\phi^{\alpha})}. In other words, the Jacobi identity
is the integrability condition for the Darboux coordinates.
The variables \mb{\phi^{\alpha}}, \mb{\phi^{*}_{\alpha}} and \mb{\Theta^{a}}
are called {\em fields}, {\em antifields} and {\em Casimirs}, respectively.
Granted the Jacobi identity \e{jacid}, the square of the $\Delta$-operator
is a first order differential operator,
\beq
\Delta^{2}(F G) ~=~\Delta^{2}(F)~G+F~\Delta^{2}(G)~,
\label{delta2leibnitz}
\eeq
if and only if there is a Leibniz rule for the interplay 
of \mb{\Delta} and the antibracket 
\beq
\Delta(F,G)~=~(\Delta(F),G)
-(-1)^{\epsilon_{F}}(F,\Delta(G))~.
\label{deltarholeibnitz}
\eeq 
We shall also assume this to be the case.
It is interesting to note that the Leibniz rule \e{deltarholeibnitz} holds 
automatically for a conventional odd Laplacian \mb{\Delta_{\rho}} 
(still assuming the Jacobi identity \e{jacid}), so the Leibniz rule 
\e{deltarholeibnitz} actually reduces to
\beq
   V(F,G)~=~(V(F),G)-(-1)^{\epsilon_{F}}(F,V(G))~.
\label{vee}
\eeq
We see that $V$ is a generating vector field for an anticanonical 
transformation. 

\noi
In the degenerate case, one would usually proceed by investigating an
antisymplectic leaf/orbit where the values of the Casimirs \mb{\Theta^{a}} are
kept fixed. Then seen from within such leaf the antisymplectic structure
will appear non-degenerate. An example of this is the case of antisymplectic
second-class constraints, which will be the subject of 
Section~\ref{secsecondclass}. On the other hand, if \mb{E^{AB}} is 
non-degenerate, the $V$ in \eq{vee} becomes locally an Hamiltonian vector
field, \ie there exists a bosonic Hamiltonian $H$ such that 
\mb{V\!=\!(H,\cdot)}. It follows that one can locally absorb the $V$-term
into a rescaling of the measure density 
\mb{\rho \to \rho^{\prime}=e^{2H}\rho}. Since \mb{\rho} and $H$ are 
intimately related through this mechanism one may regard the Leibniz rule
\e{deltarholeibnitz} as an integrability condition for the local existence of
\mb{\rho}. In any case, we shall from now on only consider the conventional
odd Laplacian \mb{\Delta_{\rho}} without the $V$-term.

\subsection{Compatible Structures}
\label{compatiblestructures}

\noi
A measure density \mb{\rho} and a possibly degenerate antisymplectic 
\mb{E^{AB}} are called {\em compatible} if and only if the odd Laplacian 
\mb{\Delta_{\rho}} is nilpotent,
\beq
 \Delta_{\rho}^{2}~=~0~.\label{deltanilp}
\eeq
Constant \mb{\rho} and constant \mb{E^{AB}} are the most important example 
of compatible structures. It is interesting to classify the compatible 
structures within the set of all pairs \mb{(\rho,E)}. To this end, consider 
two \mb{\Delta}-operators sharing the same antisymplectic structure \mb{E}, 
and with two different measure densities \mb{\rho} and \mb{\rho^{\prime}}, 
respectively, that are not necessarily compatible with \mb{E}.
They differ by a Hamiltonian vector field, 
\beq
\Delta_{\rho^{\prime}}-\Delta_{\rho}
~=~(\ln\sqrt{{\rho^{\prime} \over \rho }},~\cdot~)~.
\label{reldelta}
\eeq
Also the difference in their squares is a Hamiltonian vector
field \cite{schwarz93,KN93}:
\beq
\Delta_{\rho^{\prime}}^{2}-\Delta_{\rho}^{2} ~=~
\left(\nu(\rho^{\prime};\rho,E),~\cdot~\right)~.
\label{relnil}
\eeq
Here we have introduce a Grassmann-odd function 
\beq
\nu(\rho^{\prime};\rho,E) ~\equiv~\sqrt{{\rho \over \rho^{\prime}}}
(\Delta_{\rho}\sqrt{{\rho^{\prime} \over \rho}})~=~
\frac{1}{\sqrt{\rho^{\prime}}}(\Delta_{1}\sqrt{\rho^{\prime}})-
\frac{1}{\sqrt{\rho}}(\Delta_{1}\sqrt{\rho})
\label{relmass}
\eeq
of a measure density \mb{\rho^{\prime}} \wrt a reference system 
\mb{(\rho,E)}. The quantity \mb{\nu} acts as a scalar under general 
coordinate transformations, and satisfies the following $2$-cocycle 
condition \cite{K02}:
\beq
\nu(\rho_{1};\rho_{2},E)+\nu(\rho_{2};\rho_{3},E)
+\nu(\rho_{3};\rho_{1},E)~=~0~,
\label{cocyclecondition}
\eeq
and, as trivial consequences thereof,
\bea
\nu(\rho_{1};\rho_{1},E)&=&0~,\cr
\nu(\rho_{1};\rho_{2},E)+\nu(\rho_{2};\rho_{1},E)&=&0~.
\eea
In fact, \mb{\nu(\rho^{\prime};\rho,E)} can be written globally as a 
difference of a scalar function \mb{\nu(\rho;E)},
\beq
\nu(\rho^{\prime};\rho,E) ~=~ \nu(\rho^{\prime};E)-\nu(\rho;E)~.
\label{coboundarycondition}
\eeq
To see \eq{coboundarycondition}, go to a coordinate system where \mb{E^{AB}} 
becomes equal to a constant antisymplectic reference matrix \mb{E^{AB}_{0}}, 
which we for simplicity take to be the Darboux matrix.
It follows from the antisymplectic analogue of Darboux's Theorem that
one may cover the manifold with such coordinate charts, except for singular
points where the rank of \mb{E} jumps.
In this Section we shall denote a \mb{\Delta}-operator corresponding to 
constant \mb{\rho\!=\!1} and \mb{E^{AB}\!=\!E^{AB}_{0}} as \mb{\Delta_{0}}. 
Now define 
\beq
\nu(\rho;E_{0})~\equiv~\nu(\rho;1,E_{0}) 
~=~ \frac{1}{\sqrt{\rho}}(\Delta_{0}\sqrt{\rho})~,
\label{absmass}
\eeq
where the arguments \mb{\rho}, $1$ and \mb{E_{0}} all refer to the above 
Darboux coordinate system. {}For the definition \e{absmass} to be
well-defined, one should justify that two different choices of Darboux
coordinates lead to same value of \mb{\nu}. By definition, any two Darboux
coordinate systems are connected by an anticanonical transformation.
According to Lemma~\ref{thebvlemma} below the Jacobian
\beq  
J_{fi}~\equiv~
\sdet\left(\frac{\partial \Gamma_{f}^{A}}{\partial \Gamma_{i}^{B}}\right)
\eeq
associated to an anticanonical transformation 
\mb{\Gamma^{A}_{i} \to \Gamma^{A}_{f}} has a vanishing $\nu$: 
\beq 
\nu(J_{fi};E_{0})~\equiv~\nu(J_{fi};1,E_{0})
~=~\frac{1}{\sqrt{J_{fi}}}(\Delta_{0}\sqrt{J_{fi}})~=~0~.
\label{bvlemma0}
\eeq
Hence it follows from the $2$-cocycle condition \e{cocyclecondition} that the 
$\nu$-definition \e{absmass} does not depend on the particular choice of 
Darboux coordinate system: 
\beq
\nu(\rho_{f};1,E_{0})
~=~\nu(\frac{\rho_{i}}{J_{fi}};1,E_{0})
~=~\nu(\rho_{i};J_{fi},E_{0})~=~\nu(\rho_{i};1,E_{0})-\nu(J_{fi};1,E_{0})
~=~\nu(\rho_{i};1,E_{0})~.
\eeq
In this way one achieves a well-defined function \mb{\nu(\rho;E_{0})} on the
set of all Darboux coordinate charts. It is assumed that the definition can
be extended uniquely to singular points by continuity. One generalizes the
definition of \mb{\nu(\rho;E)} to an arbitrary coordinate system 
\mb{\Gamma^{A}} by requiring that \mb{\nu(\rho;E)} is a scalar under general
coordinate transformations, \ie
\beq
\nu(\rho;E) ~\equiv~\nu(\frac{\rho}{J};E_{0})~, 
\eeq
where
\mb{J\equiv\sdet\left(\frac{\partial \Gamma_{0}}{\partial \Gamma}\right)} 
denotes the Jacobian of a transformation \mb{\Gamma^{A} \to \Gamma^{A}_{0}}
into some Darboux coordinate system \mb{\Gamma^{A}_{0}}. One may easily check
that this definition fulfills \eq{coboundarycondition}. Moreover, the 
definition is independent of the constant reference matrix \mb{E^{AB}_{0}}. 
We shall from now on use a shorthand notation
\mb{\nu_{\rho}\!\equiv\!\nu(\rho;E)}. Next define an operator \mb{\Delta_{E}}
that takes semi-densities to semi-densities \cite{K02}
\beq
 \Delta_{E}(\sqrt{\rho})~\equiv~\sqrt{\rho}~\nu_{\rho}~,
\eeq
\ie for Darboux coordinates it is simply 
\beq
    \Delta_{E}(\sqrt{\rho})~\equiv~\Delta_{0}(\sqrt{\rho})~.
\eeq
We emphasize that the constructions of \mb{\nu_{\rho}} and \mb{\Delta_{E}}
rely heavily on  Lemma~\ref{thebvlemma}. The operator \mb{\Delta_{E}} is 
nilpotent,
\beq
 \Delta_{E}^2~=~0~.\label{deltaenilp}
\eeq
Eq.~\e{deltaenilp} encodes precisely the antisymplectic data 
\es{skewsym}{jacid} without information about any particular \mb{\rho}.
On the other hand, the odd Laplacian \mb{\Delta_{\rho}}, which takes 
scalars to scalars, consists of both the $E$-structure in \mb{\Delta_{E}}
and a measure density \mb{\rho},
\beq
 \Delta_{\rho}(F)~=~
\frac{1}{\sqrt{\rho}}[ \stackrel{\rightarrow}{\Delta}_{E},F]\sqrt{\rho}
~=~(\nu_{F^2\rho}-\nu_{\rho})F~,~~~~~~~~~~~~~~\epsilon(F)~=~0~.
\label{deltaviadeltae}
\eeq
The nilpotency of \mb{\Delta_{E}} implies
\bea
 (\Delta_{\rho}+\nu_{\rho})^{2}&=&0~, \label{deltanunilp} \\
   (\Delta_{\rho}\nu_{\rho})&=&0~, \label{deltaannihilatesnu} \\
\Delta_{\rho}^{2}&=&\left(\nu_{\rho}~,~\cdot~\right)~.\label{absnil}
\eea
We summarize the above information in Fig.~\ref{statementtable}. 

\begin{figure} 
\caption{The following diagram holds for an arbitrary pair of measure density
$\rho$ and possibly degenerate antisymplectic structure $E$, cf.~\cite{K02}:}
\label{statementtable}
\begin{center}
\beq
\begin{array}{cccc}
\exists~{\rm Darboux~coordinate} &&&
\exists~{\rm Darboux~coordinate~system~and} \cr
{\rm system~such~that}~\rho=1. &~~~\Leftrightarrow~~~&&
\exists~{\rm anticanonical~transformation} \cr
&&&{\rm such~that}~\rho=J,~{\rm the~Jacobian}.\cr \cr
&&&\Downarrow~ \cr
&& \cr
 \Delta_{\rho}^{2}=0~~~~~~~~~~\Leftrightarrow&
 \nu_{\rho}~{\rm is~a~Casimir}~~~~~~~&\Leftarrow& \nu_{\rho}=0   \cr\cr
&\Updownarrow& \cr \cr
&\exists~{\rm Darboux~coordinate~system}  \cr
&{\rm such~that}~\sqrt{\rho}~{\rm eigenvector} \cr
&{\rm for}~\Delta_{0}~{\rm with~eigenvalue}~\nu_{\rho}.&
\end{array}
\eeq
\end{center}
The above eigenvalue is constant within an antisymplectic leaf/orbit. In the 
non-degenerate case the Casimir \mb{\nu_{\rho}} is a Grassmann-odd constant.
Evidently Grassmann-odd constants cannot be non-zero, if the theory does not
have any external Grassmann-odd parameters. In practice, this is the case.
The diagram contains two implication arrows $\Downarrow$, that are
not bi-implications $\Updownarrow$. One is the possibility of a non-zero
Casimir \mb{\nu_{\rho}}; the other is a non-trivial $\Delta$-cohomology
obstruction, cf.\ Subsection~\ref{secvaryingsol}.
\end{figure}

\subsection{Anticanonical Transformations}
\label{secanticanon}

\noi
An anticanonical transformation preserves by definition the antisymplectic 
structure $E$. Infinitesimally, it is generated by a bosonic vector field $X$
such that
\beq
   X(F,G)~=~(X(F),G)+(F,X(G))~.
\label{exe}
\eeq
A Hamiltonian vector field \mb{X\!=\!\ad\Psi}, where 
\mb{\ad\Psi\!\equiv\!(\Psi,\cdot)} denotes the ``adjoint action'' \wrt the
antibracket, and where \mb{\Psi} is a Grassmann-odd generator, is an example
of an infinitesimal anticanonical transformations \e{exe}. This follows
directly from the Jacobi identity \e{jacid}. It is natural to call an
infinitesimal anticanonical transformation $X$ in \eq{exe} for 
\mb{\ad}-{\em closed}, and a Hamiltonian vector field \mb{X\!=\!\ad\Psi} 
for \mb{\ad}-{\em exact}. If $E$ is non-degenerate, then all \mb{\ad}-closed 
vector fields are locally of the \mb{\ad}-exact type. Here we shall 
elaborate on \mb{\ad}-closed vector fields in a possibly degenerate 
antisymplectic manifold. To this end, let 
\beq
{\rm div}_{\rho}X~\equiv~
\frac{(-1)^{\epsilon_{A}}}{\rho}\papal{\Gamma^{A}}(\rho X^{A}) 
\eeq
denote the divergence of a vector field $X$ \wrt a measure density \mb{\rho}.
Then
\beq
X~\ad{\rm -closed}~~~~~~~\Rightarrow~~~~~~~ 
[ \stackrel{\rightarrow}{\Delta}_{\rho}, X]
~=~-\Hf ({\rm div}_{\rho}X,~\cdot~)~,
\label{adclosedleibnitz}
\eeq
and
\beq
X~\ad{\rm -closed}~~~~~~~\Rightarrow~~~~~~~
\Hf (\Delta_{\rho}~{\rm div}_{\rho}X)~=~X(\nu_{\rho})~.
\label{adclosedabsnil}
\eeq
Equations \es{adclosedleibnitz}{adclosedabsnil} are \mb{\ad}-closed 
versions of the Leibniz rule \e{deltarholeibnitz} and the relation 
\e{absnil}, respectively. They reduce to those relations,
if $X$ is \mb{\ad}-exact, because the odd Laplacian is the divergence of a 
Hamiltonian vector field \cite{schwarz93,KN93}
\beq
  \Delta_{\rho}\Psi~=~-\Hf {\rm div}_{\rho}(\ad \Psi)
~,~~~~~~~~~~~~~~~~~\epsilon(\Psi)~=~1~.
\label{deltadivham}
\eeq
In Darboux coordinates with \mb{\rho\!=\!1} the \eq{adclosedabsnil} becomes
\beq
X~\ad{\rm -closed}~~~~~~~\Rightarrow~~~~~~~
(\Delta_{1}~{\rm div}_{1}X)~=~0~.
\label{adclosedabsnil1}
\eeq
This non-covariant result will be needed for the Lemma~\ref{thebvlemma}
below, which in turn is used to justify the definition \e{absmass} of 
\mb{\nu_{\rho}}. To avoid circular logic, we mention that the special case
\eq{adclosedabsnil1} can also be proven directly without relying on the 
concept of \mb{\nu_{\rho}}.

\noi
Consider now a one-parameter family of (not necessarily anticanonical)
passive coordinate transformations \mb{\Gamma^{A}(t)} for some parameter 
\mb{t \in [t_i,t_f]}, and governed by a one-parameter generating vector
field \mb{X_{t}\!=\!X_{t}(\Gamma)},
\beq
{d\Gamma^{A}(t) \over dt}~=~X_{t}^{A}~.
\label{evolutioneq}
\eeq
We are here and below guilty of infusing some active picture language into
a passive picture, \ie properly speaking, the active vector field is 
{\em minus} $X$, and so forth. The solution 
\beq
\Gamma^{A}(t)~=~U(t;t_i) \Gamma^{A}(t_i)
\label{canontransf}
\eeq
can be expressed with the help of a path-ordered exponential
\beq
U(t_f;t_i) ~\equiv~ {\cal P}\exp \int_{t_i}^{t_f}\! dt~X_{t}~.
\label{pathorderedU}
\eeq
The Jacobian
\beq
J(t_f;t_i) ~\equiv~ \sdet
\left(\frac{\partial \Gamma^{A}(t_f)}{\partial \Gamma^{B}(t_i)}\right)
\label{jacobiandef}
\eeq
is given by
\beq
\ln J(t_f;t_i)
~=~\int_{t_i}^{t_f}\! dt~ U(t_f;t) {\rm div}_{1}^{(t_i)}X_{t}~,
\label{jacobiandiv1}
\eeq 
where \mb{{\rm div}_{1}^{(t)}} refers to the divergence with 
\mb{\rho\!=\!1} in the coordinates \mb{\Gamma^{A}(t)}.
The formula \e{jacobiandiv1} can be deduced from the differential equation
\beq
{d \over dt}\ln J(t;t_i)
~=~(-1)^{\epsilon_{A}}\papal{\Gamma^{A}(t)}{d \Gamma^{A}(t) \over dt} 
~=~{\rm div}_{1}^{(t)}X_{t}  
~=~{\rm div}_{1}^{(t_i)}X_{t} + X_{t}\left(\ln J(t;t_i)\right)~.
\label{jacdiffeq}
\eeq
The measure density \mb{\rho} transforms in the passive picture with the 
Jacobian
\beq
\rho(t_f)~=~\frac{\rho(t_i)}{ J(t_f;t_i)}~.\label{rhotransf}
\eeq
Therefore \mb{\rho} satisfies the following differential equation
\beq
{d \over dt}\ln\rho(t)~=~-{\rm div}_{1}^{(t)} X_{t}~=~
X_{t}(\ln\rho(t))-{\rm div}_{\rho} X_{t}~.
\label{rhodiffeq}
\eeq
Next put \mb{\rho} back into the divergence in \eq{jacobiandiv1}. Then
\beq
\frac{J(t_f;t_i)}{\rho(t_i)}U(t_f;t_i)\rho(t_i)
~=~\exp\left[\int_{t_i}^{t_f}\! dt~ U(t_f;t){\rm div}_{\rho} X_{t}\right]~.
\label{jacobiandivrho}
\eeq 
This equation can be understood both in the active and the passive picture, 
and it will play a key r\^ole later on. In deriving it we have used that
\beq
\int_{t_i}^{t_f}\! dt~U(t_f;t) X_{t}\left(\ln\rho(t_i)\right)
~=~-\int_{t_i}^{t_f}\! dt~ {d \over dt}
\left[U(t_f;t)\ln\rho(t_i)\right]
~=~\left[U(t_f;t_i)-{\bf 1}\right]\ln\rho(t_i)~.
\eeq
{}Finally, recall the remarkable fact that the $\Delta$-operator 
-- despite being a second order differential operator -- 
has a Leibniz type interplay with \mb{\ad}-closed vector fields, cf.\ 
\eq{adclosedleibnitz}.
{}For this reason we can form identities that look very similar to 
well-known identities in ordinary Quantum Mechanics. {}For instance,
\beq
X_{t}~\ad{\rm -closed}~~~~~~~\Rightarrow~~~~~~~
 [ \stackrel{\rightarrow}{\Delta}_{\rho}, U(t_f;t_i)]~=~- \Hf
\int_{t_i}^{t_f} \! dt~U(t_f;t)~\ad ( {\rm div}_{\rho} X_{t} )~U(t;t_i)~.
\label{entanglementformula}
\eeq
We may now give a short direct proof of a result used in \eq{bvlemma0}:

\begin{lemma}  
Let \mb{\Gamma^{A}(t_{i}) \to \Gamma^{A}(t_{f})} be a finite 
anticanonical transformation between Darboux coordinates in a possibly 
degenerate antisymplectic manifold. Then the Jacobian \mb{J(t_f;t_i)} 
satisfies
\beq
\Delta_{1}^{(t_i)} \sqrt{J(t_f;t_i)}~=~0~.
\label{bvlemma}
\eeq
Here \mb{\Delta_{1}^{(t_i)}} refers to the odd Laplacian with \mb{\rho\!=\!1}
in the initial Darboux coordinates \mb{\Gamma^{A}(t_{i})}.
\label{thebvlemma}
\end{lemma}

\noi
{\sc Proof of Lemma~\ref{thebvlemma}:}
\bea
-\Delta_{1}^{(t_i)}\ln \sqrt{J(t_f;t_i)}
&=&-\Hf \int_{t_i}^{t_f}\! dt~ [\Delta_{1}^{(t_i)},U(t_f;t)] 
{\rm div}_{1}^{(t_i)}X_{t} \cr
&=& \frac{1}{4} \int_{t_i}^{t_f}\! dt~
\int_{t}^{t_f}\! dt^{\prime}~ U(t_f;t^{\prime})
\left({\rm div}_{1}^{(t_i)}X_{t^{\prime}},
~U(t^{\prime};t){\rm div}_{1}^{(t_i)}X_{t}\right) \cr
&=&\frac{1}{4} \left(\int_{t_i}^{t_f}\! dt^{\prime}~
U(t_f;t^{\prime}){\rm div}_{1}^{(t_i)}X_{t^{\prime}},
\int_{t_i}^{t_f}\! dt~U(t_f;t){\rm div}_{1}^{(t_i)}X_{t}\right)~
\theta(t^{\prime}\!-\!t) \cr
&=& \Hf \left(\ln \sqrt{J(t_f;t_i)},~
\ln \sqrt{J(t_f;t_i)} \right)~,
\label{bvlemmaproof}
\eea
where use has been made of the equations \e{adclosedabsnil1}, 
\es{jacobiandiv1}{entanglementformula}. \proofbox

\noi
Lemma~\ref{thebvlemma} is a degenerate generalization of a well-known 
result \cite{BV84,K02} for the non-degenerate case.
The covariant version of Lemma~\ref{thebvlemma} reads
\beq
\left(\frac{U(t_f;t_i)\rho(t_i)}{\rho(t_f)}\right)^{-\Hf}
\Delta_{\rho}
\left(\frac{U(t_f;t_i)\rho(t_i)}{\rho(t_f)}\right)^{\Hf}
~=~\nu\left(U(t_f;t_i)\rho(t_i);~\rho(t_f),E(t_f) \right)
~=~\left[U(t_f;t_i)-{\bf 1}\right]\nu_{\rho}~,
\label{covbvlemma}
\eeq
where
\mb{\Gamma^{A}(t_{i})\to\Gamma^{A}(t_{f})\!
\equiv\!U(t_f;t_i)\Gamma^{A}(t_{i})}
is a finite anticanonical transformation.

\noi 
We observe that it was never necessary in this Subsection to assume that
\mb{\rho} and $E$ are compatible,  \ie that \mb{\Delta_{\rho}} is nilpotent.
In the remaining part of the paper we shall assume that the
\mb{\Delta}-operator is nilpotent, except for a subtlety \e{nilpotencydo}
concerning second-class constraints.

\subsection{Varying the Solutions to the Quantum Master Equation}
\label{secvaryingsol}

\noi
Let us now consider solutions to the Quantum Master Equation. 

\begin{itemize}
\item
We are mainly interested in deformations of the quantum master action $W$ 
in \mb{\Delta_{\rho}e^{\Ih W}=0} for nilpotent \mb{\Delta_{\rho}}, where $W$
satisfies certain rank and boundary conditions \cite{BV81,BV83,BV84}. 
Here both $\rho$ and $E$ are kept fixed.

\item 
As a precursor for the above problem, it is of interest to vary the 
semi-density \mb{\sqrt{\rho}} in \mb{\Delta_{1}^{(t_i)}\sqrt{\rho}=0} for 
nilpotent \mb{\Delta_{1}^{(t_i)}}. Here the antisymplectic structure $E$ is
kept fixed.
\end{itemize}

\noi
Let us collectively write \mb{(\Delta_{\rho}\sigma)=0} to represent both
types of problems, so that \mb{\sigma\in \Sigma} denotes a solution in the
space $\Sigma$ of solutions to the Quantum Master Equation. Now consider
a one-parameter family of solutions $\sigma(t)$, where \mb{t \in [t_i,t_f]}.
Obviously, the difference of neighboring solutions is $\Delta$-closed: 
\mb{\Delta_{\rho} (d\sigma/dt)=0}. If the difference is furthermore 
\mb{\Delta}-exact, we may write\footnote{In general, there is non-trivial 
\mb{\Delta}-cohomology. In finite dimensions, for a constant non-degenerate
antisymplectic matrix $E^{AB}$, whose fermionic blocks vanish, the non-trivial
$\Delta_{0}$-cohomology is one-dimensional, generated by  the {\em fermionic
top-monomial}, \ie the monomial of all fermionic and no bosonic 
variables \cite{deltacohomology}. In local field theory, cohomology may 
arise from locality requirements. Furthermore, our treatment obviously 
only applies to a path-connected solution space.}
\beq
{d \sigma(t) \over dt}
~=~-[\stackrel{\rightarrow}{\Delta}_{\rho},\Psi(t)]\sigma(t)~,
\label{deltaexact}
\eeq
where \mb{\Psi(t)} is a one-parameter family of fermionic functions.
Next introduce a path-ordered exponential
\beq
V(t_f;t_i) ~\equiv~ {\cal P}\exp\left[-\int_{t_i}^{t_f}\! dt~
[\stackrel{\rightarrow}{\Delta}_{\rho},\Psi(t)] \right] ~.
\label{pathorderedV}
\eeq
Integrating \eq{deltaexact} along the path, we find
\beq
\sigma(t_f)~=~V(t_f;t_i)\sigma(t_i)
~=~V(t_f;t_i)\sigma(t_i)V(t_i;t_f)V(t_f;t_i)1 
~=~U(t_f;t_i)\sigma(t_i) \cdot v(t_f;t_i)~,
\label{solsigma}
\eeq
where we have used the identity 
\beq
e^{-[\stackrel{\rightarrow}{\Delta},\Psi]}\sigma
e^{[\stackrel{\rightarrow}{\Delta},\Psi]}
~=~e^{-[[\stackrel{\rightarrow}{\Delta},\Psi],~\cdot~]}\sigma
~=~e^{{\rm ad} \psi}\sigma~,
\eeq
and defined 
\beq
v(t_f;t_i) ~\equiv~ V(t_f;t_i)1
~=~\exp\left[-\int_{t_i}^{t_f}\! dt~ U(t_f;t) \Delta_{\rho}\Psi(t). \right]
~=~ \sqrt{\frac{J(t_f;t_i)}{\rho(t_i)}U(t_f;t_i)\rho(t_i)}~.
\label{littlev}
\eeq
Use has been made of  \eq{deltadivham}, \eq{jacobiandivrho} and the 
differential equation
\beq
{d \over dt} v(t;t_i)
~=~-[\stackrel{\rightarrow}{\Delta}_{\rho},\Psi(t)]v(t;t_i)
~=~(\Psi(t),v(t;t_i))-(\Delta_{\rho}\Psi(t))v(t;t_i)~.
\label{vdiffeq}
\eeq
In the third equality of \e{littlev} we re-interpret the \mb{\Psi(t)} family,
which originates from a \mb{\Delta}-exact variation, as a generator of an
\mb{\ad}-exact anticanonical transformation. There is thus a one-to-one
correspondence between \mb{\ad}-exact anticanonical transformations and
\mb{\Delta}-exact variations. Moreover, as anticanonical transformations
can be understood passively, one may also give \mb{\Delta}-exact variations
a passive interpretation, \ie one is not changing the solution \mb{\sigma};
only the coordinates \mb{\Gamma^{A}}. The detailed mechanism for this
one-to-one correspondence is of great interest, both conceptionally and in 
practice.

\begin{definition}
We shall say that an anticanonical transformation \mb{U(t_f;t_i)}
acts on a pair \mb{\sigma(t_i)} and \mb{\rho(t_i)}
according to the following ``twisted'' transformation rules:
\bea
\sigma(t_i)~\longrightarrow~\sigma(t_f)&=&\sqrt{\frac{J(t_f;t_i)}{\rho(t_i)}}
U(t_f;t_i)\left[\sigma(t_i)\sqrt{\rho(t_i)}\right]~, \label{canontransfrule}\\
\rho(t_i)~\longrightarrow~\rho(t_f)&=&\frac{\rho(t_i)}{J(t_f;t_i)}~.
\eea
\end{definition}

\noi
To be more precise, it is the stabilizer subgroup
\mb{\{ U\! \in\! {\cal G} ~|~ U(\nu_{\rho})\!=\!\nu_{\rho} \}}
that acts on solutions to the Quantum Master Equation
\mb{\Delta_{\rho}\sigma\!=\!0}. The full group \mb{{\cal G}} of anticanonical
transformations acts on solutions to the modified Quantum Master Equation
\mb{(\Delta_{\rho}\!+\!\nu_{\rho})\sigma\!=\!0}.
 
\noi
Letting the anticanonical generator \mb{\Psi(t)} depend on $t$ is somewhat 
academic, because one may always find an equivalent constant generator $\Psi$
(and choose the parameter interval to be \mb{[t_i,t_f]=[0,1]}), such that 
\mb{U(t_f\!=\!1;t_i\!=\!0)=e^{\ad \Psi}}.
While $t$-dependent $\Psi$'s provide a deeper theoretical understanding, 
it is preferred in practice to work with such $t$-independent $\Psi$'s
where path-ordering issues are absent. In the latter case the above
one-parameter solution is of the form
\beq
\sigma(t)~=~e^{-t [\stackrel{\rightarrow}{\Delta},\Psi]}\sigma_{i}~,
\eeq
and \eq{jacobiandivrho} reduces to 
\beq
\ln\sqrt{\frac{J(t_f;t_i)}{\rho(t_i)}U(t_f;t_i)\rho(t_i)}~=~
-E(\ad \Psi) \Delta_{\rho}\Psi~,
\label{jacobiandive}
\eeq
where
\beq
 E(x)~=~\int_{t_i=0}^{t_f=1} \!dt~e^{xt}~=~\frac{e^x-1}{x}~.
\eeq
It follows from \eqs{solsigma}{littlev} that 

\begin{proposition}:~~A finite \mb{\Delta}-exact transformation
\beq
e^{\Ih W_f}~=~e^{-[\stackrel{\rightarrow}{\Delta},\Psi]}e^{\Ih W_i}
\label{boltzmannexactvar}
\eeq
deforms the quantum action $W$ according to
\beq
 W_f ~=~ e^{\ad \Psi}W_i +(i\hbar) E(\ad \Psi) \Delta\Psi
~=~ e^{\ad \Psi}W_i +(i\hbar) \frac{e^{\ad\Psi}-1}{\ad\Psi}\Delta\Psi~.
\label{Wexactvar}
\eeq
\label{propWexactvar}
\end{proposition}

\noi
This important deformation formula will be used repeatedly throughout the
remainder of the paper. By expanding in Planck's constant,
\beq
 W~=~S+\sum_{n=1}^{\infty} (i \hbar)^n W_{n}~,~~~~~~~~~~~~~~~~~~~~~~
\Psi~=~\sum_{n=0}^{\infty} (i \hbar)^n \Psi_{n}~,
\label{WPsi}
\eeq
one sees that the classical action \mb{S} undergoes a classical anticanonical
transformation
\beq
 S_f ~=~ e^{\ad \Psi_{0}}S_i~,
\label{Wexactvar0}
\eeq
while the leading quantum correction \mb{W_{1}} transforms as
\beq
W_{1,f} ~=~ e^{\ad \Psi_{0}}W_{1,i}+ E(\ad \Psi_{0}) \Delta\Psi_{0}
+\left(E(\ad \Psi_{0})\Psi_{1},S_f\right)~.
\label{Wexactvar1}
\eeq
To summarize, the deformations of the {\em classical} solutions \mb{S} to the 
Classical Master Equation \mb{(S,S)=0} are generated by the group of 
anticanonical transformations \mb{e^{\ad \Psi_{0}}}, cf.~\e{Wexactvar0}.
This should be compared to the {\em quantum} situation, where deformations 
of solutions $W$ to the Quantum Master Equation \mb{\Delta e^{\Ih W}=0} are 
similarly generated by the group of quantum anticanonical transformations 
\mb{e^{\ad \Psi}}. Here $\Psi$ depends on $\hbar$ in accordance with 
\e{WPsi}, but with the important difference that the group action is 
applied in a non-standard way, twisted by the semi-density \mb{\sqrt{\rho}},
cf.~\eq{canontransfrule}.
This twisting effect is not felt at the classical level.

\noi
There are important exceptions where the above twisting is not present at 
all. This happens for instance in Darboux coordinates 
\mb{\Gamma^{A}=\left\{\phi^{\alpha};\phi^{*}_{\alpha}\right\}} when
\mb{\rho=\rho(\phi)} and \mb{\Psi=\Psi(\phi)} are independent of the 
antifields \mb{\phi^{*}_{\alpha}}, so that \mb{\Delta\Psi=0}. Then the 
formula \e{Wexactvar} reduces to a purely anticanonical transformation
\beq
W_{f}~=~ e^{\ad \Psi}W_i
~=~W_i(\phi;\phi^{*}\!+\!\frac{\partial \Psi}{\partial \phi})~,
\eeq
a formula that is intimately tied to the original way of gauge-fixing 
in the field-antifield formalism \cite{BV81}.

\noi
The dilation transformation \mb{e^{\Ih W}\to Ce^{\Ih W}} (or
\mb{e^{\Ih X}\to Ce^{\Ih X}}), where $C$ is a constant factor, is clearly a
symmetry of the Quantum Master Equation. 
{}From a mathematical standpoint, granted that the action satisfies 
pertinent rank conditions, the scaling represents non-trivial 
\mb{\Delta}-cohomology, which is excluded from our reasoning \e{expmaxvar}
of gauge independence for \mb{{\cal Z}}. It obviously does change the 
partition function \mb{{\cal Z}\to C{\cal Z}}. On the other hand, from a 
physics perspective such an overall constant rescaling is totally trivial
and plays no r\^ole whatsoever.

\noi
As another simple application, let us briefly mention the second type of 
problem. We have a nilpotent $\Delta$-operator \mb{\Delta_{1}^{(t_i)}} with
\mb{\rho(t_i)\!=\!1} in the coordinates \mb{\Gamma^{A}(t_i)}.
The constant semi-density \mb{\sigma(t_i)\!=\!1} is a trivial solution
to  \mb{(\Delta_{1}^{(t_i)}\sigma)=0}. Now act with an anticanonical 
transformation $U(t_f;t_i)$ on \mb{\sigma(t_i)\!=\!1} according to the
transformation rule \eq{canontransfrule}. Then
\beq
\rho(t_f)~=~\frac{\rho(t_i)}{J(t_f;t_i)}~=~\frac{1}{J(t_f;t_i)}~,
\eeq 
so the transformed semi-density becomes 
\beq
\sigma(t_f)~=~\frac{U(t_f;t_i)\left[\sigma(t_i)\sqrt{\rho(t_i)}\right]}
{\sqrt{\rho(t_f)}}~=~\sqrt{J(t_f;t_i)}~. 
\eeq 
This in turn provides a descriptive proof of Lemma~\ref{thebvlemma}.

\setcounter{equation}{0}
\section{Irreducible First-Level Gauge-Fixing Formalism}
\label{secirr}

\subsection{Review of Original Gauge-Fixing Formalism}

\noi
Our starting point is an action \mb{W=W(\Gamma;\hbar)} that possesses $N$
gauge symmetries that should be fixed. In the original gauge-fixing 
prescription of the field-antifield formalism \cite{BV81,BV83,BV84} there 
is no $X$-part. In non-degenerate Darboux coordinates 
\mb{\Gamma^{A}=\left\{\phi^{\alpha};\phi^{*}_{\alpha}\right\}} with a density
\mb{\rho} that is independent of the antifields 
\mb{\phi^{*}_{\alpha}}, we may reformulate gauge-fixing in the following
way that is easy to generalize later:

\begin{itemize}
\item[1a.]
{}First change the Boltzmann factor \mb{e^{\Ih W}} with a $\Delta$-exact 
transformation generated by a gauge fermion function \mb{\Psi},
\beq
e^{\Ih W^{\Psi}}~=~e^{-[\stackrel{\rightarrow}{\Delta},\Psi]}e^{\Ih W}~.
\label{boltzmannexactvar0}
\eeq
This formula obviously preserves the Quantum Master Equation, and it 
generalizes readily to a gauge fermion {\em operator} \mb{\hat{\Psi}}, but
we shall not pursue such a generalization here.
According to \eq{Wexactvar} the transformed action \mb{W^{\Psi}} becomes
\beq
W^{\Psi}~=~ e^{\ad \Psi}W +(i\hbar) E(\ad \Psi) \Delta\Psi~.
\label{zerothlevelWexactvar}
\eeq

\item[1b.]
Next put the antifields \mb{\phi^{*}_{\alpha}\to 0} to zero,
\beq
{\cal Z}^{\Psi}
~=~\int\! [d\phi]\left.\rho~e^{\Ih W^{\Psi}}\right|_{\phi^{*}=0}
~=~\int\! [d\Gamma]~\rho~\delta(\phi^{*})~e^{\Ih W^{\Psi}}~.
\label{zerothlevelza}
\eeq

\item[1c.]
The partition function \mb{{\cal Z}^{\Psi}} defined this way does not 
depend on the gauge fermion \mb{\Psi}. {\sc Proof}: \ \ 
Start with exponentiating the \mb{\delta}-function in \eq{zerothlevelza}:
\beq 
\delta(\phi^{*}) ~=~\int\! [d\lambda] e^{\Ih X}
\eeq 
with a trivial action 
\mb{X=\phi^{*}_{\alpha} \lambda^{\alpha}} that obviously satisfies the Quantum
Master Equation \mb{\Delta_{\hl}e^{\frac{i}{\hbar}X}=0}, where
\mb{\Delta_{\hl}\equiv \Delta + (-1)^{\epsilon_{\alpha}}
(\partial  / \partial \lambda^{\alpha})
(\partial  / \partial \lambda^{*}_{\alpha})} 
is the suitably extended \mb{\Delta}-operator.
Then the partition function 
\beq
{\cal Z}^{\Psi}=\int\! [d\Gamma][d\lambda] \rho~e^{\Ih X}
e^{-[\stackrel{\rightarrow}{\Delta}_{\hl},\Psi]}e^{\Ih W}
\eeq 
becomes of the $W$-$X$--form discussed in the Introduction. The independence
of $\Psi$ follows straightforwardly from the symmetry \e{deltasym} of the 
$\Delta$-operator,
\beq
{\cal Z}^{\Psi+\delta\Psi}-{\cal Z}^{\Psi} 
~=~\int\! [d\Gamma][d\lambda] \rho \int_{0}^{1} \! dt~
e^{-(1-t)[\stackrel{\rightarrow}{\Delta}_{\hl},\Psi]}e^{\Ih W}~ 
[\stackrel{\rightarrow}{\Delta}_{\hl},\delta\Psi]
e^{t[\stackrel{\rightarrow}{\Delta}_{\hl},\Psi]}e^{\Ih X}~=~0~.
\eeq
The \mb{\lambda^{\alpha}}'s and the \mb{\Delta_{\hl}} can be viewed as part of
the so-called first-level formalism, 
cf.~Subsection~\ref{secfirstlevelformalism}. \proofbox

\item[2a.]
One may reach an alternative version of the partition function 
\e{zerothlevelza} by using the symmetry \e{deltasym} of the $\Delta$-operator
to write \eq{zerothlevelza} as
\beq
{\cal Z}^{\Psi}~=~\int\! [d\Gamma]~ \rho~ e^{\Ih W}
e^{[\stackrel{\rightarrow}{\Delta},\Psi]} \delta(\phi^{*}) 
~=~\int\! [d\Gamma]~ \rho~ e^{\Ih W} 
\delta\left(e^{-\ad \Psi}\phi^{*}\right)
e^{E(-\ad \Psi) \Delta\Psi}~.
\label{zerothlevelz}
\eeq

\item[2b.]
If furthermore the gauge fermion \mb{\Psi} is independent of the 
antifields \mb{\phi^{*}_{\alpha}}, as is normally assumed, the transformation
\e{zerothlevelWexactvar} reduces to a purely anticanonial transformation,
\beq
 W^{\Psi} ~=~ e^{\ad \Psi} W~,
\eeq
and one arrives at the familiar prescription of the original field-antifield
formalism, where gauge-fixing is done by an explicit substitution of the
antifields \mb{\phi^{*}_{\alpha}\to\partial \Psi / \partial \phi^{\alpha}}
with a field gradient of a gauge fermion \mb{\Psi}:
\beq
{\cal Z}^{\Psi}~=~\int\! [d\phi]\left.\rho~
e^{\Ih W}\right|_{\phi^{*}=\frac{\partial \Psi}{ \partial \phi}}~.
\label{familiarbv}
\eeq
\end{itemize}

\noi
The above gauge-fixing procedure with explicit removal of the $N$ antifields
obviously refers to a particular set of coordinates on the supermanifold, 
and is therefore not covariant. The $X$-part of the new formulation is 
precisely introduced \cite{BMS} as a covariantization of the gauge-fixing
prescription.

\subsection{First-Level Formalism}
\label{secfirstlevelformalism}

\noi
The gauge-fixing procedure was considerably generalized in the nineties into
a so-called {\em multi-level formalism} \cite{BT1,BT,BMS} to allow for 
covariant and more flexible gauge-fixing choices. {}For systematic reasons we
shall retroactively call the original non-degenerate antisymplectic
phase space variables \mb{\Gamma^{A}=\{\phi^{\alpha};\phi^{*}_{\alpha}\}}
for {\em zeroth-level fields}, the expansion parameter 
\mb{\hbar\!\equiv\!\hbar_{\bo}} for the {\em zeroth-level Planck constant},
and the gauge-fixing  procedure of the last subsection
for {\em zeroth-level gauge-fixing}. 

\noi
In the (irreducible) {\em first-level} formalism one introduces $N$ Lagrange
multipliers \mb{\lambda^{\alpha}\!\equiv\!\lambda_{\bl}^{\alpha}} of 
Grassmann parity \mb{\epsilon_{\alpha}\!\equiv\!\epsilon_{\alpha}^{\bl}} 
and $N$ antifields \mb{\lambda^{*}_{\alpha}\!\equiv\!\lambda^{*}_{\bl\alpha}},
which we collectively call the {\em first-level fields}. The phase space 
variables
\beq
\Gamma^{A}_{\hl} ~\equiv~ \left\{\Gamma^{A};\lambda^{\alpha},
\lambda^{*}_{\alpha} \right\}
\eeq
for the first-level formalism thus consist of the zeroth and the first-level
fields\footnote{Notation: We use capital roman letters  $A$, $B$, $C$, 
$\ldots$ from the beginning of the alphabet as upper index for both 
\mb{\Gamma^{A}} and \mb{\Gamma^{A}_{\hl}}, respectively. Usually a quantity
\mb{Q_{\bn}} with a soft-bracket index \mb{\bn} is associated with the 
\nth-level only, while a quantity \mb{Q_{\hn}} with a 
hard-bracket index \mb{\hn} accumulates all the levels \mb{\leq n}.}.
The first-level odd Laplacian
\beq
\Delta_{\hl}~\equiv~\Delta + (-1)^{\epsilon_{\alpha}}
\papal{\lambda^{\alpha}}\papal{\lambda^{*}_{\alpha}}
\label{Deltaaug}
\eeq
gives rise to an extended antisymplectic structure in the standard way. We
shall always assume there is a trivial measure density associated with the
first-level sector\footnote{In the multi-level formalism the previous levels
are treated covariantly and the present level non-covariantly. This means at
the first-level that general zeroth-level coordinate transformations 
\mb{\Gamma^{A} \to \Gamma^{\prime A}} are allowed, while the first-level 
fields \mb{\lambda^{\alpha}} and \mb{\lambda^{*}_{\alpha}} are considered to
be fixed from the onset. This implies for instance that it is 
consistent to choose a trivial measure in the first-level sector.}.

\noi
{}Furthermore, one introduces a {\em first-level Planck constant} 
\mb{\hbar_{\bl}} as a new expansion parameter for the quantum action
\beq
X~=~\Omega + (i\hbar_{\bl})\Xi + (i\hbar_{\bl})^2\tilde{\Omega}
+ {\cal O}(\hbar_{\bl}^3)~, \label{Xplanckexp}
\eeq
where the dependence of the \mb{\lambda^{\alpha}}'s, the 
\mb{\lambda^{*}_{\alpha}}'s and the previous (zeroth) level objects is
implied. At the end of the calculations one substitutes back 
\mb{\hbar_{\bl} \to \hbar}. The (first-level) Planck number grading 
\mb{\pl\!\equiv\!\pl_{\bl}} is defined as \cite{BT1} 
\beq
\pl(F\cdot G)~=~\pl(F)+\pl(G)~~,~~~~ \pl(\hbar_{\bo})~=~\pl(\Gamma^{A})~=~0
~~,~~~~
\pl(\hbar_{\bl})~=~\pl(\lambda^{\alpha})~=~-\pl(\lambda^{*}_{\alpha})~=~1~.
\label{Planck}
\eeq
One may compactly write the Planck number grading as a Planck number operator
\beq
\pl~=~- \left(\lambda^{\alpha}\lambda^{*}_{\alpha},~\cdot~\right)_{\hl}
+\hbar_{\bl} \frac{\partial }{\partial \hbar_{\bl}}~.
\eeq
We remark that 
the introduction of two different expansion parameters \mb{\hbar_{\bo}} and
\mb{\hbar_{\bl}} is spurred on one hand by the wish to limit the number of 
terms in $X$ by imposing (first-level) Planck number conservation, and on the
other hand to allow $\Omega$, $\Xi$, $\tilde{\Omega}$, etc., to depend on 
\mb{\hbar_{\bo}}.  If the latter is not an issue, one only needs one Planck
constant.

\noi
At the first level one is guided by the following

\begin{principle}
The gauge-fixing action $X$ satisfies three requirements:
\begin{enumerate}
\item
Planck number conservation:
\mb{\pl (\frac{X}{\hbar_{\bl}})=0}.
\item
The Quantum Master Equation: 
\mb{\Delta_{\hl} \exp\left[\frac{i}{\hbar_{\bl}}X\right]=0}.
\item
The Hessian of $X$ should have rank equal to half the number of fields
\mb{\Gamma^{A}_{\hl}}, \ie \mb{2N} in the irreducible case.
\end{enumerate}
\label{principle3}
\end{principle}

\noi
Strickly speaking, it is enough that the rank conditions are met only
on stationary field configurations, a technicality we shall assume 
implicitly from now on.
Planck number conservation limits the lowest-order terms in $X$ to
\beq
  X ~=~ G_{\alpha}\lambda^{\alpha} + (i\hbar_{\bl})H
-\lambda^{*}_{\alpha}R^{\alpha} + {\cal O}((\lambda^{*})^{2})~,
~~~~~~~~~~~~~~~~~~~~~~\pl(X)~=~1~,
\label{X1stlevel}
\eeq
where
\beq 
-R^{\alpha}~=~ \Hf U^{\alpha}_{\beta \gamma}~\lambda^{\gamma}\lambda^{\beta}
 (-1)^{\epsilon_{\beta}+1}
+ (i\hbar_{\bl})V^{\alpha}{}_{\beta}\lambda^{\beta} 
+ (i\hbar_{\bl})^2 ~\tilde{G}^{\alpha}~,~~~~~~~~~~\pl(R^{\alpha})~=~2~.
\eeq
The Quantum Master Equation generates a tower of equations;
the first few read 
\bea
 (G_{\alpha},G_{\beta})&=& G_{\gamma} U^{\gamma}_{\alpha \beta}~,
\label{nonabelinvo} \\
 (\Delta G_{\beta})-(H,G_{\beta})
&=&(-1)^{\epsilon_{\alpha}}U^{\alpha}_{\alpha\beta}
+  G_{\alpha}V^{\alpha}{}_{\beta}~, \label{weakeq2} \\
e^{H} (\Delta  e^{-H}) ~=~- (\Delta H)+\Hf (H,H)
&=&V^{\alpha}{}_{\alpha}-G_{\alpha}\tilde{G}^{\alpha}~.
\label{weakeq3}
\eea
The most important \eq{nonabelinvo} is a non-Abelian involution of 
the $N$ gauge-fixing constraints $G_{\alpha}$.

\noi
There are essentially two equivalent ways of performing {\em first-level
gauge-fixing}: One may gauge-fix either the $X$ or the $W$-part. 

\begin{itemize}
\item[1a.]
To gauge-fix the $X$-part, first change the Boltzmann factor \mb{e^{\Ih X}}
with a \mb{\Delta_{\hl}}-exact transformation generated by a gauge fermion 
\mb{\Psi},
\beq
e^{\Ih X^{\Psi}}~=~e^{[\stackrel{\rightarrow}{\Delta}_{\hl},\Psi]}e^{\Ih X}~.
\label{boltzmannexactvar1}
\eeq

\item[1b.]
Next put the Lagrange multiplier antifields \mb{\lambda^{*}_{\alpha} \to 0}
to zero,
\beq
{\cal Z}^{\Psi}_{\hl}
~=~\int\! [d\Gamma][d\lambda]\left.\rho~
e^{\Ih (W+X^{\Psi})}\right|_{\lambda^{*}=0}
~=~\int\! [d\Gamma_{\hl}]~\rho~\delta(\lambda^{*})~e^{\Ih (W+X^{\Psi})}~.
\label{firstlevelza}
\eeq

\item[1c.]
The partition function \mb{{\cal Z}^{\Psi}_{\hl}} defined this way does not
depend on the gauge fermion \mb{\Psi}. {\sc Proof}: \ \ 
Use second-level techniques, cf.\ Section~\ref{sechigherlevels}, and 
exponentiate the \mb{\delta}-function in \eq{firstlevelza}: 
\beq
\delta(\lambda^{*})e^{\Ih W}~=~\int\! [d\lambda_{\bz}] e^{\Ih W_{\hz}}
\eeq
with an action \mb{W_{\hz}=\lambda^{*}_{\alpha}\lambda^{\alpha}_{\bz}+W}
that satisfies the Master Equation \mb{\Delta_{\hz}e^{\Ih W_{\hz}}=0}.
Then the partition function  
\beq
{\cal Z}^{\Psi}_{\hl}~=~\int\! [d\Gamma_{\hl}][d\lambda_{\bz}] \rho~
e^{\Ih W_{\hz}} e^{[\stackrel{\rightarrow}{\Delta}_{\hz},\Psi]}e^{\Ih X}
\eeq
becomes of the $W$-$X$--form discussed in the Introduction. 
\proofbox

\item[1d.]
If furthermore the gauge fermion \mb{\Psi} is independent 
of the Lagrange multiplier antifields \mb{\lambda^{*}_{\alpha}}, 
as is normally assumed, the gauge-fixing action \e{boltzmannexactvar1}
reduces to
\beq
X^{\Psi}~=~e^{-\ad_{\hl} \Psi}X -(i\hbar) E(-\ad \Psi) \Delta\Psi~,
\label{firstlevelXexactvarprime}
\eeq
where in detail,
\beq
e^{-\ad_{\hl} \Psi}X
~=~X\left(e^{-\ad \Psi}\Gamma; \lambda, 
\lambda^{*}\!-\!E(-\ad \Psi)\frac{\partial \Psi}{ \partial \lambda}\right)~.
\label{firstlevelXgaugefixing}
\eeq

\item[2a.]
To alternatively  gauge-fix the $W$-part, one may use the 
symmetry \e{deltasym} of the \mb{\Delta_{\hl}}-operator to re-write 
\eq{firstlevelza} as
\beq
{\cal Z}^{\Psi}_{\hl}
~=~\int\! [d\Gamma_{\hl}] ~\rho~ e^{\Ih X}
e^{-[\stackrel{\rightarrow}{\Delta}_{\hl},\Psi]}
\delta(\lambda^{*})e^{\Ih W} 
~=~\int\! [d\Gamma_{\hl}] ~\rho~ e^{\Ih X}~
\delta\left(e^{\ad_{\hl}\Psi}\lambda^{*}\right)
e^{\Ih W^{\Psi}_{\hl}}~,
\label{firstlevelz}
\eeq
where we have defined 
\beq
e^{\Ih W^{\Psi}_{\hl}}
~=~e^{-[\stackrel{\rightarrow}{\Delta}_{\hl},\Psi]}e^{\Ih W}~.
\label{Wboltzmannexactvar1}
\eeq

\item[2b.]
If the gauge fermion \mb{\Psi} is independent 
of the Lagrange multiplier antifields \mb{\lambda^{*}_{\alpha}},
the action \mb{W^{\Psi}_{\hl}} in \eq{Wboltzmannexactvar1} reduces to
\mb{W^{\Psi}} in \eq{zerothlevelWexactvar}, and
the Lagrange multiplier antifields are gauge-fixed as
\beq 
\lambda^{*}_{\alpha}
~=~-E(\ad\Psi)\frac{\partial \Psi}{ \partial \lambda^{\alpha}}~,
\label{gf1}
\eeq
\ie the partition function reads
\beq
{\cal Z}^{\Psi}_{\hl}
~=~\int\! [d\Gamma] [d\lambda] \left. \rho~ e^{\Ih (W^{\Psi}+ X)}\right|_{
\lambda^{*}=-E(\ad\Psi)\frac{\partial \Psi}{ \partial \lambda}}~.
\label{firstlevelzprime}
\eeq
\end{itemize}

\subsection{Going On-shell \wrt the Constraints}
\label{Going}

\noi
A tractable gauge is the \mb{\lambda^{*}_{\alpha}\!\equiv\! 0} gauge,
\ie a trivial gauge fermion \mb{\Psi\!\equiv\!0}.
Then $X$ reduces to only two terms 
\beq 
\left. X\right|_{\lambda^{*}=0}
~=~G_{\alpha}\lambda^{\alpha} + (i\hbar_{\bl})H~.
\eeq
The \mb{\lambda^{\alpha}}'s becomes Lagrange multipliers for the 
constraints $G_{\alpha}$, which are in turn enforced directly through 
$\delta$-functions in the path integral -- hence the name. 
Moreover, the set of constraints $G_{\alpha}$ has to be irreducible
in order for the rank condition on the Hessian of $X$ to be met, \ie
\beq
\forall X^{\alpha}~:~ G_{\alpha} X^{\alpha}~=~0 ~~~\Rightarrow~~~ \exists
A^{\alpha\beta}=-(-1)^{\epsilon_{\alpha} \epsilon_{\beta}}
A^{\beta\alpha}~:~
X^{\alpha}~=~G_{\beta} A^{\beta\alpha}~.
\eeq
Let \mb{F^{\alpha} = F^{\alpha}(\Gamma;\hbar)} be arbitrary zeroth-level 
coordinate functions of statistics 
\mb{\epsilon({F^{\alpha}})=\epsilon_{\alpha}\!+\!1}
such that \mb{\bar{\Gamma}^{A} \equiv \{F^{\alpha};G_{\alpha} \}} forms a 
coordinate system in the zeroth-level sector, and let 
\mb{J=\sdet (\frac{\partial \bar{\Gamma}^A}{\partial \Gamma^B })}
denote the Jacobian of the transformation \mb{\Gamma^{A}\to\bar{\Gamma}^{A}}.
Then

\begin{theorem}:~~The Quantum Master Equation implies that the 
quantum-correction $H$ depends on the constraints \mb{G_{\alpha}}
modulo terms that vanish on-shell \wrt the \mb{G_{\alpha}}'s according to the
following square root formula \cite{BT,K94}
\beq
H ~=~-\ln\sqrt{\frac{J~\sdet(F^{\alpha},G_{\beta})}{\rho}}+ {\cal O}(G)
\label{khudaverdianh}
\eeq
up to an overall unphysical integration constant that may be discarded. 
Moreover in the Abelian case \mb{U^{\gamma}_{\alpha \beta}=0}, 
it is possible to solve off-shell as well:
\beq
H ~=~-\ln\sqrt{\frac{J~\sdet(F^{\alpha},G_{\beta})}{\rho}}
 -G_{\alpha} \int_{0}^{1} \!V^{\alpha}{}_{\beta} 
\left. \left((F,G)^{-1}\right)^{\beta}{}_{\gamma}
\right|_{(F;G) \to (tF;G)}\! \! dt~F^{\gamma}~.
\label{abeliankhudaverdianh}
\eeq
\label{theorema}
\end{theorem}

\noi
{\sc Proof of Theorem~\ref{theorema}}:~~
{}First note that the \mb{N \times N} matrix 
\mb{\Lambda^{\alpha}{}_{\beta}\equiv(F^{\alpha},G_{\beta})} is invertible, at
least in a neighborhood of the constrained surface \mb{G_{\alpha}\approx 0}.
Use the Jacobi identity and the involution \e{nonabelinvo}
\beq
((F^{\gamma},G_{\alpha}),G_{\beta})
-(-1)^{(\epsilon_{\alpha}+1)(\epsilon_{\beta}+1)}
(\alpha\leftrightarrow\beta)
~=~(F^{\gamma},(G_{\alpha},G_{ \beta}))
~=~(F^{\gamma},G_{\delta}U^{\delta}_{\alpha \beta})
\label{jacfgg1}
\eeq
to deduce that
\beq
(\matrisse{\Lambda}^{-1})^{\gamma}{}_{\delta}
(\matrisse{\Lambda}^{\delta}{}_{\alpha},G_{\beta})
-(-1)^{(\epsilon_{\alpha}+1)(\epsilon_{\beta}+1)}
(\alpha\leftrightarrow\beta)
~=~U^{\gamma}_{\alpha \beta} 
+(-1)^{\epsilon_{\gamma}\epsilon_{\delta}}
G_{\delta}(\matrisse{\Lambda}^{-1})^{\gamma}{}_{\epsilon}
(F^{\epsilon},U^{\delta}_{\alpha \beta})~.
\label{jacfgg2}
\eeq
Now supertrace to get
\beq
(\str\ln\matrisse{\Lambda},G_{\beta})
+(-1)^{(\epsilon_{\alpha}+1)\epsilon_{\beta}}
(\matrisse{\Lambda}^{\alpha}{}_{\beta},G_{\gamma})
(\matrisse{\Lambda}^{-1})^{\gamma}{}_{\alpha}
~=~(-1)^{\epsilon_{\alpha}}U^{\alpha}_{\alpha \beta}
+(-1)^{\epsilon_{\alpha}(\epsilon_{\gamma}+1)}
G_{\gamma}(\matrisse{\Lambda}^{-1})^{\alpha}{}_{\delta}
(F^{\delta},U^{\gamma}_{\alpha \beta})~.
\label{jacfgg3}
\eeq
Next use the new coordinates 
\mb{\bar{\Gamma}^{A} \equiv \{F^{\alpha};G_{\alpha} \}} to rewrite
\bea
(\Delta G_{\beta})
&=&\frac{(-1)^{\epsilon_{A}}}{2\bar{\rho}}\papal{\bar{\Gamma}^{A}}
\bar{\rho}(\bar{\Gamma}^{A},G_{\beta}) \cr
&=&(\ln\sqrt{\bar{\rho}},G_{\beta})
+\frac{(-1)^{\epsilon_{\alpha}+1}}{2}\papal{F^{\alpha}}
(F^{\alpha},G_{\beta})
+\frac{(-1)^{\epsilon_{\alpha}}}{2}\papal{G_{\alpha}}
(G_{\alpha},G_{\beta}) \cr
&=&(\ln\sqrt{\bar{\rho}},G_{\beta})
+\frac{(-1)^{(\epsilon_{\alpha}+1)\epsilon_{\beta}}}{2}
(\Lambda^{\alpha}{}_{\beta}\papar{F^{\alpha}})
+\frac{(-1)^{\epsilon_{\alpha}}}{2}\papal{G_{\alpha}}
[G_{\gamma} U^{\gamma}_{\alpha \beta}] \cr
&=&(\ln\sqrt{\frac{\bar{\rho}}{\sdet\Lambda}},G_{\beta})
+(-1)^{\epsilon_{\alpha}}U^{\alpha}_{\alpha \beta}
+(-1)^{\epsilon_{\alpha}(\epsilon_{\gamma}+1)}G_{\gamma}\papal{G_{\alpha}}
 U^{\gamma}_{\alpha \beta}\cr
&&-\frac{(-1)^{(\epsilon_{\alpha}+1)\epsilon_{\beta}}}{2}
(\Lambda^{\alpha}{}_{\beta}\papar{G_{\gamma}})(G_{\gamma},G_{\delta})
(\matrisse{\Lambda}^{-1})^{\delta}{}_{\alpha} \cr
&&+\frac{(-1)^{\epsilon_{\alpha}(\epsilon_{\gamma}+1)}}{2}
G_{\gamma}(\matrisse{\Lambda}^{-1})^{\alpha}{}_{\delta}
(F^{\delta},F^{\epsilon})\papal{F^{\epsilon}}U^{\gamma}_{\alpha \beta}
~,\label{weakeq2a}
\eea
where \mb{\bar{\rho}=\frac{\rho}{J}} denotes the transformed density, and
where \eq{jacfgg3} has been used in the last equality. The last three terms 
are of order \mb{{\cal O}(G)}. Then one of the consequences of the Master
Equation \e{weakeq2} is
\beq
\left[ \ln\sqrt{\frac{\sdet\Lambda}{\bar{\rho}}}+H\right]
\papar{F^{\beta}}~=~{\cal O}(G)~,
\label{weakeq2b} 
\eeq
and the square root formula \e{khudaverdianh} follows by integration. 
The integration constant represents the trivial
dilation symmetry \mb{H\to H+c},
where $c$ is a constant. Clearly $H$ only appears in 
differentiated form in the Quantum Master Equation, so any integration
constant is a priori allowed. Restricting ourselves to consider only classes
of solutions that are \mb{\Delta_{\hl}}-exactly connected, it is consistent
to always set this integration constant to zero. 
\proofbox

\noi
The square root formula \e{khudaverdianh} gives $H$ as a function of the 
\mb{F^{\alpha}}'s. But since the \mb{F^{\alpha}}'s are arbitrary, this
dependence must be trivial. This fact may also be shown directly:

\begin{lemma} \cite{BT,K94}:~~The factor 
\mb{J~\sdet(F^{\alpha},G_{\beta})} is independent of the \mb{F^{\alpha}}'s up
to terms that vanish on-shell \wrt \mb{G_{\alpha}}, if the \mb{G_{\alpha}}'s
satisfy the involution \eq{nonabelinvo}.
\label{lemmaa}
\end{lemma}

\noi
{\sc Proof of Lemma~\ref{lemmaa}}:~~Exponentiate the two determinants by 
introducing a ghost pair $\bar{C}_{A}$ and $C^{A}$ of statistics 
\mb{\epsilon_{A}\!+\!1}, and another ghost pair \mb{\bar{B}_{\alpha}} and 
$B^{\alpha}$ of statistics \mb{\epsilon_{\alpha}\!+\!1}, so that the product
of the two determinants inside the square root can be written 
as a partition function
\beq
{\cal Z}_{\rm det}~=~ J~\sdet(F^{\alpha},G_{\beta})
~=~\int\![d\bar{C}][dC][d\bar{B}][dB]~e^{\Ih S_{\rm det}}
\label{toyz}
\eeq
with a determinant action given as
\beq 
S_{\rm det}~=~\bar{C}_A (\bar{\Gamma}^{A}\papar{\Gamma^{B}}) C^{B}
+ \bar{B}_{\alpha}(F^{\alpha},G_{\beta}) B^{\beta}~.
\label{toydetaction}
\eeq 
Now there are several ways to proceed. Perhaps the most enlightening treatment
is to rewrite this as a mini field-antifield system within our theory. 
Consider a ``classical'' action
\beq
 S_{0} ~=~ \bar{C}^{\alpha} (G_{\alpha}\papar{\Gamma^{A}}) C^{A}~,
\label{toyclasaction}
\eeq
where we have split \mb{\bar{C}_{A}=\{\bar{C}_{\alpha};\bar{C}^{\alpha}\}} of
Grassmann parity \mb{\epsilon({\bar{C}_{\alpha}}) = \epsilon_{\alpha}} and
\mb{\epsilon({\bar{C}^{\alpha}})=\epsilon_{\alpha}\!+\!1}, respectively.
On-shell \wrt \mb{G_{\alpha}} the classical action \mb{S_{0}} is 
invariant \mb{\delta S_{0} \approx 0} under the following BRST-like symmetry
\beq
\delta C^{A}  ~=~ (\Gamma^{A},G_{\alpha}) B^{\alpha} \mu~,
\label{toybrst}
\eeq
because of the involution \e{nonabelinvo}. Here $\mu$ is a Grassmann-odd
parameter. The standard field-antifield recipe \cite{BV81,BV83} now instructs
us to construct a minimal proper action as
\beq
S_{\min}~=~S_{0}+C_{A}^{*} (\Gamma^{A},G_{\alpha}) B^{\alpha}~,
\label{toysmin}
\eeq
and a non-minimal proper action
\beq
S ~=~ S_{\min} + \bar{C}_{\alpha} \bar{B}^{*\alpha}~,
\label{toys}
\eeq
where we have identified \mb{\bar{C}_{\alpha}} with Nakanishi-Lautrup
auxiliary fields. It is easy to check that \mb{(S,S)_{BC}\!\approx\!0}
and \mb{\Delta_{BC} S \!=\! 0},
where
\beq
\Delta_{BC}~\equiv~(-1)^{\epsilon_{\alpha}+1}
\papal{B^{\alpha}}\papal{B^{*}_{\alpha}}
+(-1)^{\epsilon_{\alpha}+1}
\papal{\bar{B}_{\alpha}}\papal{\bar{B}^{*\alpha}}
+(-1)^{\epsilon_{A}+1}
\papal{C^{A}}\papal{C^{*}_{A}}
+(-1)^{\epsilon_{A}+1}
\papal{\bar{C}_{A}}\papal{\bar{C}^{*A}}~.
\label{deltabc}
\eeq
So the Quantum Master Equation for $S$ is satisfied on-shell. Now choose a
gauge fermion as
\beq
\Psi~=~\bar{B}_{\alpha} (F^{\alpha}\papar{\Gamma^{A}}) C^{A}~.
\label{toypsi}
\eeq
The gauge-fixed action
\beq
S\left(C_{A}^{*}\!=\!\frac{\partial \Psi}{\partial C^{A}};
\bar{B}^{*\alpha}\!=\!\frac{\partial \Psi}{\partial \bar{B}_{\alpha}}\right)
~=~\bar{C}^{\alpha} (G_{\alpha}\papar{\Gamma^{A}}) C^{A}
+\bar{C}_{\alpha} (F^{\alpha} \papar{\Gamma^{A}}) C^{A}
+\bar{B}_{\alpha}(F^{\alpha},G_{\beta}) B^{\beta}
~=~S_{\rm det}
\label{toygfaction}
\eeq
is precisely the determinant action $S_{\rm det}$. 
Hence the partition function \e{toyz} does not depend on $\Psi$, and
since the $F^{\alpha}$'s only appear inside the gauge fermion $\Psi$,
we conclude that the partition function \e{toyz} does not depend on 
the $F^{\alpha}$'s as well.
\proofbox

\noi
The following Theorem~\ref{theoremb} is in some respect a reversed statement
of Theorem~\ref{theorema}:

\begin{theorem} :~~{}For an arbitrary set of irreducible constraints 
\mb{G_{\alpha}} and structure functions \mb{U^{\gamma}_{\alpha \beta}} that
satisfy the involution \e{nonabelinvo}, there exist functions \mb{H}, 
\mb{V^{\alpha}{}_{\beta}}, \mb{\tilde{G}^{\alpha}}, etc., 
such that $X$ is a solution to the Quantum Master Equation.
\label{theoremb}
\end{theorem}

\noi
{\sc Proof}:~~This relies on Abelianization, \ie
there exist Abelian constraints \mb{G_{\alpha}^{0}} with
\mb{(G_{\alpha}^{0},G_{\beta}^{0})=0}, and an invertible 
rotation matrix \mb{\Lambda^{\alpha}{}_{\beta}}, such that
\mb{G_{\beta}=G_{\alpha}^{0}\Lambda^{\alpha}{}_{\beta}}.
\proofbox
  
\begin{corollary} \cite{BT,K94}:~~The partition function\footnote{To recover
the zeroth-level gauge-fixing \e{familiarbv} one first goes to Darboux 
coordinates \mb{\Gamma^{A}=\{\phi^{\alpha};\phi^{*}_{\alpha}\}} 
with $\rho\!=\!1$, and then substitute \mb{F^{\alpha} \to \phi^{\alpha}} and
\mb{G_{\alpha} \to \phi^{*}_{\alpha}-\partial \Psi / \partial \phi^{\alpha}}.
In our conventions the Lagrange multipliers $\lambda^{\alpha}$ and the 
antifields $\phi^{*}_{\alpha}$ (of the previous level) carry the {\em same}
Grassmann parity. In detail, we define
\beq
\epsilon\left(\phi^{\alpha}\right)~\equiv~\epsilon_{\alpha}^{\bo}~,~~~~~~
\epsilon\left(\phi^{*}_{\alpha}\right)~\equiv~\epsilon_{\alpha}^{\bo}+1
~\equiv~\epsilon_{\alpha}^{\bl}~,~~~~~~
\epsilon\left(\lambda^{\alpha}\right)~
\equiv~\epsilon_{\alpha}^{\bl}~\equiv~\epsilon_{\alpha}~,
\eeq
and so forth.}
\beq
{\cal Z}^{G}_{\hl}~=~\int\! [d\Gamma]~e^{\Ih W} \delta(G)
\sqrt{\rho~J~\sdet(F^{\alpha},G_{\beta})}
\label{khudaverdianz}
\eeq
is independent of the \mb{G_{\alpha}}'s satisfying the involution 
\eq{nonabelinvo}.
\label{corollaryb}
\end{corollary}

\noi
{\sc Sketched Proof}:~~
The Corollary does not explicitly refer to an $X$-part, but we may always 
assume an underlying $X$-part because of Theorem~\ref{theoremb}. Therefore 
the broad strategies concerning independence of the gauge-fixing $X$-part 
mentioned in the Introduction apply. 
\proofbox

\noi
Another interesting result is the following

\begin{theorem}:~~The on-shell square root formula \eq{khudaverdianh} for $H$,
viewed as part of $X$, is form invariant under finite
\mb{\Delta_{\hl}}-exact deformations of $X$, even if $X$ does not solve the 
Quantum Master Equation.
\label{theoremc}
\end{theorem}

\noi
{\sc Proof of Theorem~\ref{theoremc}}:~~
Assume that the gauge-fixing action $X_{f}$ that contains the investigated
quantum correction $H_{f}$, is a \mb{\Delta_{\hl}}-exact deformation
\beq
X_{f}~=~e^{\ad_{\hl} \Psi}X_{i}
+(i\hbar_{\bl})E(\ad_{\hl}\Psi)\Delta_{\hl}\Psi~,
\label{Xexactvar}
\eeq
of an initial gauge-fixing action $X_{i}$ with a one-loop correction $H_{i}$
that obeys the on-shell square root formula \eq{khudaverdianh}, \ie
\beq
H_{i} ~=~c-\ln \sqrt{ \sdet (\frac{\partial \{F_i;G_i\}}{\partial \Gamma})
~\frac{\sdet(F_{i}^{\alpha},G_{i,\beta})}{\rho}}+ {\cal O}(G_{i})~,
\label{khudaverdianhi}
\eeq
where we included the integration constant $c$.
We have to show that a similar formula holds for $H_{f}$.
Order by order in \mb{\hbar_{\bl}} the \eq{Xexactvar}  implies that
\bea
 \Omega_{f} &=&e^{\ad_{\hl} \Psi_{0}}\Omega_{i}~,  \label{Xexactvar0} \\
\Xi_{f} &=&e^{\ad_{\hl} \Psi_{0}}\Xi_{i}
+ E(\ad_{\hl} \Psi_{0}) \Delta_{\hl}\Psi_{0}
+\left(E(\ad_{\hl} \Psi_{0})\Psi_{1},~\Omega_{f}\right)_{\hl}~,
\label{Xexactvar1} \\
\tilde{\Omega}_{f}&=&{\cal O}(\lambda^{*})~.\label{Xexactvar2}
\eea
The generator 
\mb{\Psi=\Psi_{0}+(i\hbar_{\bl}) \Psi_{1}+ {\cal O}(\hbar_{\bl}^2)} 
conserves the Planck number,  \mb{\pl(\Psi)=0}. 
This restricts the possible lowest terms to
\bea
\Psi_{0}&=&\Psi_{\bo} +(-1)^{\epsilon_{\alpha}+\epsilon_{\beta}}
\lambda^{\alpha}\Psi_{\bl\alpha}{}^{\beta} 
\lambda^{*}_{\beta}+{\cal O}((\lambda^{*})^{2})~,\label{psiexpansion0} \\
\Psi_{1}&=&{\cal O}(\lambda^{*})~, \label{psiexpansion1}
\eea
where the sign factors in front of the matrix \mb{\Psi_{\bl\alpha}{}^{\beta}}
of Grassmann grading \mb{\epsilon_{\alpha}\!+\!\epsilon_{\beta}} are 
introduced for later convenience. 
{}First we look at the \eq{Xexactvar0} for $\Omega_{f}$:
\beq
G_{f,\alpha} \lambda^{\alpha}~=~\left. \Omega_{f} \right|_{\lambda^{*}=0}
~=~\exp\left[\ad\Psi_{\bo}-(-1)^{\epsilon_{\alpha}+\epsilon_{\beta}}
\lambda^{\alpha}\Psi_{\bl\alpha}{}^{\beta} 
\papal{\lambda^{\beta}}\right]G_{i,\gamma} \lambda^{\gamma}~.
\label{omegalambdastar}
\eeq
The constraints \mb{G_{f,\alpha}} are a composition of a rotation and
an anticanonical transformation,
\beq
G_{f,\alpha}~=~\Lambda_{\alpha}{}^{\beta}~\tilde{G}_{i,\beta}~,
\label{gsolution}
\eeq
where tilde $\sim$ denotes the anticanonical transformation
\beq
\Gamma^{A}~~~~~\longrightarrow~~~~~
\tilde{\Gamma}^{A}~\equiv~\exp\left[\ad\Psi_{\bo}\right]\Gamma^{A}~.
\eeq
We shall later need an expression for the superdeterminant of the rotation
matrix \mb{\Lambda_{\alpha}{}^{\beta}},
\beq
\ln\sdet\matrisse{\Lambda}~=~\str\ln \matrisse{\Lambda}
~=~-E(\ad\Psi_{\bo})\str\matrisse{\Psi}_{\bl}~.
\label{sdetlambdasolution}
\eeq
To prove the \eqs{gsolution}{sdetlambdasolution} one may use one-parameter
techniques, \ie let the generator \mb{\Psi\to t\Psi} be proportional to a 
parameter \mb{t\in[0,1]} to study the transition
\beq
G_{\alpha}(t\!=\!0)~\equiv~G_{i,\alpha}~~~~~\longrightarrow~~~~~
G_{\alpha}(t\!=\!1)~\equiv~G_{f,\alpha}~.
\eeq
It follows from \eq{omegalambdastar} that the constraints obey the 
differential equation
\beq
{d G_{\alpha}(t) \over dt}~=~(\Psi_{\bo},G_{\alpha}(t)) 
-\Psi_{\bl}{}_{\alpha}{}^{\beta} G_{\beta}(t)~.\label{gdiffeq}
\eeq
The first term on the \rhs represents an anticanonical transformation, while
the second term is a rotation. The solution to \eq{gdiffeq} is 
\beq
G_{\alpha}(t)~=~
\Lambda_{\alpha}{}^{\beta}(t)~
\exp\left[ t~ \ad\Psi_{\bo}\right]G_{i,\beta}~,
\eeq
where the rotation matrix \mb{\Lambda(t)} is a path-ordered matrix 
expression in the parameter \mb{t^{\prime}\in[0,t]},
\beq
 \Lambda(t)~=~{\cal P} \exp\left[-\int_{0}^{t} \!dt^{\prime}~
e^{(t-t^{\prime})\ad\Psi_{\!\bo}}\Psi_{\bl} \right]~.
\eeq
This leads immediately to \eqs{gsolution}{sdetlambdasolution}.

\noi
Next we look at the \rhs of the \eq{Xexactvar1} for \mb{\Xi_{f}}.
The third term is proportional to either the constraints \mb{G_{f,\alpha}}
or to 
\mb{\lambda^{*}_{\alpha}} because of \eqs{psiexpansion1}{omegalambdastar},
so only the first two terms contribute on-shell:
\beq
H_{f}~=~\left. \Xi_{f} \right|_{\lambda^{*}=0} 
~=~ \exp\left[\ad\Psi_{\bo}\right]H_{i}
+E(\ad \Psi_{\bo})\left( \Delta \Psi_{\bo} 
+ \str\matrisse{\Psi}_{\bl}\right)+ {\cal O}(G_{f})~.
\label{xilambdastar}
\eeq
Combining \eq{xilambdastar}, \e{khudaverdianhi}, \e{jacobiandive},
\e{sdetlambdasolution} and \e{gsolution}, one deduces the Theorem:
\bea
-H_{f}&=&-\tilde{c}+\ln \sqrt{ \sdet 
(\frac{\partial \{ \tilde{F}_i;\tilde{G}_i\}}{\partial \tilde{\Gamma}})
~\frac{\sdet(\tilde{F}_{i}^{\alpha},\tilde{G}_{i,\beta})}
{\rho(\tilde{\Gamma})}}
+{\cal O}(\tilde{G}_{i}) \cr
&&+\ln \sqrt{\frac{\rho(\tilde{\Gamma})}{\rho} 
~\sdet (\frac{\partial \tilde{\Gamma}^{A}}
{\partial \Gamma^{B} })} +  \ln \sdet\matrisse{\Lambda}
+ {\cal O}(G_{f}) \cr
&=&-c+\ln \sqrt{ \sdet (\frac{\partial \{ \tilde{F}_i;\tilde{G}_i\}}
{\partial \Gamma})~\sdet\matrisse{\Lambda}^2 \frac{\sdet
(\tilde{F}_{i}^{\alpha},\tilde{G}_{i,\beta})}{\rho}} + {\cal O}(G_{f})\cr
&=&-c+\ln \sqrt{ \sdet (\frac{\partial \{ F_{f};G_f\}}{\partial \Gamma})
~\frac{\sdet(F_{f}^{\alpha},G_{f,\beta})}{\rho}} + {\cal O}(G_{f})~,
\eea  
with \mb{F_{f}^{\alpha}=\tilde{F}_{i}^{\alpha}}. 
\proofbox

\noi
Since one may in principle create every $X$-solution through
\mb{\Delta_{\hl}}-exact deformations of some trivial $X$-action like
that of the \mb{\phi^{*}_{\alpha}\!=\!0} gauge, one may interpret 
Theorem~\ref{theoremc} as generating the square root formula \e{khudaverdianh}
via first-level anticanonical transformations \mb{e^{\ad\Psi}}.
Theorem~\ref{theoremc} also shows that the integration constant from 
Theorem~\ref{theorema} is invariant under \mb{\Delta_{\hl}}-exact 
deformations, so one may consistently discard it.
The proof shows that only the classical part $\Psi_{0}$ 
of the underlying generator $\Psi$ plays an active r\^ole in the 
transformation of  $H$ on-shell. Here the word {\em classical} is used 
in the first-level sense, \ie for objects independent of \mb{\hbar_{\bl}}.
Moreover, we have seen that $\Psi_{0}$ generates rotations 
and zeroth-level anticanonical transformations of the $G_{\alpha}$'s.

\setcounter{equation}{0}
\section{Second-Class Constraints}
\label{secsecondclass}

\noi
It is of interest to extend the irreducible first-level construction of a 
gauge-fixed Lagrangian path integral to include antisymplectic 
second-class constraints \cite{BT1,BBD1}. Consider therefore a set of 
\mb{2\ND} second-class constraints \mb{\Theta^{a}} with Grassmann parity 
\mb{\epsilon({\Theta^a})\!=\!\epsilon_{a}} that reduce the \mb{2N}-dimensional
antisymplectic manifold down to a physical submanifold of dimension 
\mb{2(N \!-\!\ND)}. This proceeds quite analogous to the Poisson-bracket 
treatment of second-class constraints in the Hamiltonian formalism\footnote{
Here, we work partly at the ``gauge-generating'' zeroth-level and partly at
the ``gauge-fixing'' first-level. Therefore the second-class constraints
\mb{\Theta^{a}\!=\!\Theta^{a}(\Gamma;\hbar_{(-1)})} and several of the Dirac
constructions to be introduced below could in principle depend on a Planck
expansion parameter \mb{\hbar_{(-1)}}, which we assign to a previous ``minus
first'' level. Moreover, we postpone for simplicity the issue of 
reparametrizations of the constraints
\mb{\Theta^{a} \to \Theta^{\prime a}=\Lambda^{a}{}_{b}(\Gamma)~\Theta^{b}}
to Section~\ref{reparam}, \ie the {\em defining} set of constraints are kept
fixed for now. In fact we derive in Section~\ref{reparam} that a
reparametrization invariant formulation {\em necessitates} off-shell
corrections to eqs.\ \e{thetavanish}, \e{nilpotencyd} and \e{delta2com}.
{}Finally, let us mention that antisymplectic {\em first-class} constraints,
and moreover, the conversion from second to first-class antisymplectic
constraints, have been addressed in \cite{BM}.}.
The antibracket matrix 
\beq
E^{ab} ~\equiv~ (\Theta^{a},\Theta^{b}) \label{definingthetatheta}
\eeq
of the second-class constraints \mb{\Theta^{a}} has by definition an inverse
matrix \mb{E_{ab}},
\beq
E_{ab}E^{bc} ~=~ \delta^{c}_{a}~,
\eeq
so that one can introduce a Dirac antibracket completely analogous to the
Dirac Poisson bracket \cite{BT1}
\beq
(F,G)_{D} ~\equiv~ (F,G) - (F,\Theta^{a})E_{ab}(\Theta^{b},G)~,
\label{Diracbracket}
\eeq
where \mb{F\!=\!F(\Gamma)} and \mb{G\!=\!G(\Gamma)} are arbitrary functions.
The bracket satisfies a Jacobi identity
\beq
\sum_{F,G,H~{\rm cycl.}}(-1)^{(\epsilon_{F}+1)(\epsilon_{H}+1)}
((F,G)_{D},H)_{D}~=~0
\label{jacidd}
\eeq
everywhere in the extended phase space \mb{\Gamma^{A}}.
The projection property ensures that the Dirac antibracket vanishes,
\beq
(\ad_{D}\Theta^{a})F~\equiv~(\Theta^{a},F)_{D} ~=~ 0 ~, \label{thetavanish}
\eeq  
when taken of any function \mb{F} with any of the constraints \mb{\Theta^{a}}.
In addition, there exists a nilpotent Dirac $\Delta$-operator $\Delta_{D}$, 
\beq
   \Delta_{D}^{2}~=~0~,\label{nilpotencyd}
\eeq
so that the Dirac antibracket \e{Diracbracket} equals the failure of 
\mb{\Delta_{D}} to act as a derivation, in complete analogy with the usual
$\Delta$-operator. It reads
\beq
\Delta_{D} ~=~ \frac{(-1)^{\epsilon_{A}}}{2\rho_{D}}
\papal{\Gamma^{A}}\rho_{D}E^{AB}_{D}\papal{\Gamma^{B}}~,
\label{diracdelta}
\eeq
with a degenerated antisymplectic metric
\beq
E_{D}^{AB} ~\equiv~ (\Gamma^{A},\Gamma^{B})_{D} ~,
\eeq
and with a compatible Dirac measure density \mb{\rho_{D}=\rho_{D}(\Gamma)}.
By definition the Dirac measure density \mb{\rho_{D}} transforms as
\beq
\rho^{\prime}_{D}~=~ \frac{\rho_{D}}
{\sdet (\frac{\partial \Gamma^{\prime A}}{\partial \Gamma^{B} })}
\label{rhodtransformationrule0}
\eeq
under change of coordinates \mb{\Gamma^{A}\to\Gamma^{\prime A}},
while the Dirac measure density \mb{\rho_{D}} is required to transform as
\beq
\rho^{\prime}_{D}~=~ \rho_{D}~\sdet (\Lambda^{a}{}_{b})+{\cal O}(\Theta)
\label{rhodtransformationrule1}
\eeq
under reparametrization of the constraints
\mb{\Theta^{a} \to \Theta^{\prime a}=\Lambda^{a}{}_{b}(\Gamma)~\Theta^{b}}.
{}Finally, the $\Delta_{D}$-operator annihilates the constraints
\beq
(\Delta_{D}\Theta^{a}) ~=~ 0~,  \label{delta2com}
\eeq
because of \eq{thetavanish}, and independently of the choice of \mb{\rho_{D}}.
In the case of higher-order $\Delta$-operators the \mb{\Theta^{a}}'s 
become {\em operators}, and the condition \e{delta2com} should be replaced
with \mb{[\Delta_{D},\Theta^{a}]=0}, cf.~\cite{BBD1}.

\subsection{First-Level Partition Function}
\label{dirac1pf}

\noi
With the above ingredients, the corresponding irreducible first-level
Lagrangian path integral formulation can be carried out very analogous to
the case without second-class constraints \cite{BT1}. The appropriate path
integral in the \mb{\lambda^{*}_{\alpha}=0} gauge, is, 
\beq
{\cal Z}_{\hl D} ~=~ \int \! [d\Gamma][ d\lambda]~\rho_{D}~
e^{\Ih (W_{D} + X_{D})}\prod_{a}\delta(\Theta^{a}) ~,
\label{2nd2nd}
\eeq
with both $W_{D}$ and $X_{D}$ satisfying the corresponding Quantum Master
Equations
\beq
\Delta_{D} \exp\left[\frac{i}{\hbar_{\bo}}W_{D}\right]~=~0
~,~~~~~~~~~~~~~~~~~~~~~~~~
\Delta_{\hl D} \exp\left[\frac{i}{\hbar_{\bl}}X_{D}\right]~=~0~.
\label{theDmastereq}
\eeq
At the first-level there are \mb{N\!-\!\ND} Lagrange multipliers
\mb{\lambda^{\alpha}} and \mb{N\!-\!\ND} corresponding antifields
\mb{\lambda^{*}_{\alpha}}.
One may again expand the action
\beq
X_{D}~=~G_{\alpha}\lambda^{\alpha}+(i\hbar_{\bl})H+{\cal O}(\lambda^{*})
\label{XDirac}
\eeq
in terms allowed by the Planck number conservation.
The Quantum Master Equation for \mb{X_{D}} shows that the \mb{N\!-\!\ND}
gauge-fixing functions \mb{G_{\alpha}} are in involution \wrt the 
Dirac antibracket,
\beq
  (G_{\alpha},G_{\beta})_{D} ~=~ G_{\gamma} U^{\gamma}_{\alpha \beta}~.
\label{nonabelinvoD}
\eeq
As in the case with no second-class constraints, an on-shell closed-form
expression for the one-loop correction $H$ has been found \cite{BBD1}.
Let \mb{F^{\alpha}\!=\!F^{\alpha}(\Gamma;\hbar)} be 
arbitrary zeroth-level coordinate functions such that 
\beq
\bar{\Gamma}^{A} ~\equiv~ \{F^{\alpha}; G_{\alpha}; \Theta^{a} \}
\label{3splitD}
\eeq 
forms a coordinate system in the zeroth-level sector, and let
\mb{J_{D}=\sdet (\frac{\partial \bar{\Gamma}^{A}}{\partial \Gamma^{B}})}
denote the Jacobian of the transformation \mb{\Gamma^{A}\to\bar{\Gamma}^{A}}.
Then

\begin{theorem}:~~The Quantum Master Equation implies that the 
one-loop correction $H$ depends on the constraints \mb{G_{\alpha}}
modulo terms that vanish on-shell \wrt the \mb{G_{\alpha}}'s 
and the \mb{\Theta^{a}}'s according to the following square root formula
\beq
H ~=~-\ln\sqrt{\frac{J_{D}~\sdet(F^{\alpha},G_{\beta})_{D}}{\rho_{D}}}
+ {\cal O}(G;\Theta)~.
\label{khudaverdianhD}
\eeq
\label{theoremad}
\end{theorem}

\noi
One may check that 

\begin{lemma}:~~The factor 
\mb{J_{D}~\sdet(F^{\alpha},G_{\beta})_{D}} is independent of the 
\mb{F^{\alpha}}'s up to terms that vanish on-shell \wrt \mb{G_{\alpha}}, if
the \mb{G_{\alpha}}'s satisfy the involution \e{nonabelinvoD} \wrt the Dirac
antibracket.
\label{lemmaad}
\end{lemma}

\noi
{\sc Proof of Lemma~\ref{lemmaad}}:~~ We may also this time build an auxiliary
field-antifield system. It is almost identical to the case without 
second-class constraints, so we shall only point out some of the differences.
The ``classical'' action $S_{0}$ now reads
\beq 
S_{0} ~=~ \bar{C}^{\alpha} (G_{\alpha}\papar{\Gamma^{A}}) C^{A}+
\bar{C}_{a} (\Theta^{a}\papar{\Gamma^{A}}) C^{A}~,
\eeq
where we have split the antighost 
\mb{\bar{C}_{A}=\{\bar{C}_{\alpha};\bar{C}^{\alpha};\bar{C}_{a}\}}
in three parts that reflects the splitting in \e{3splitD}. On-shell
\wrt $G_{\alpha}$ the classical action $S_{0}$ is invariant 
\mb{\delta S_{0} \approx 0} under the following BRST-like symmetry
\beq
\delta C^{A}  ~=~ (\Gamma^{A},G_{\alpha})_{D} B^{\alpha} \mu~,
\label{toybrstD}
\eeq
because of the involution \e{nonabelinvoD}.
Note that in the Lemma we can work off-shell \wrt the  $\Theta^{a}$'s.
Hence the minimal proper action of this auxiliary field-antifield
system is
\beq
S_{\min}~=~S_{0}+C_{A}^{*} (\Gamma^{A},G_{\alpha})_{D} B^{\alpha}~,
\label{toysminD}
\eeq
and the non-minimal proper action is
\beq
S ~=~ S_{\min} + \bar{C}_{\alpha} \bar{B}^{\alpha}_{*}~.
\label{toysD}
\eeq
\proofbox

\noi
There are at least three very good reasons to impose the \mb{\rho_{D}}
transformation rule \e{rhodtransformationrule1}. First of all, it is precisely
what is needed to make the partition function \e{2nd2nd} invariant under 
reparametrization of the constraints 
\mb{\Theta^{a} \to \Theta^{\prime a}=\Lambda^{a}{}_{b}(\Gamma)~\Theta^{b}}.
Secondly, note that the rule \e{rhodtransformationrule1} also 
render the expression inside the square root of \e{khudaverdianhD} 
reparametrization invariant, up to terms that vanish on-shell \wrt the 
$\Theta^{a}$'s. Thirdly, we shall show in Section~\ref{reparam} below that
the rule \e{rhodtransformationrule1} is needed to make the Dirac odd Laplacian
\mb{\Delta_{D}} reparametrization invariant on-shell.

\subsection{Unitarizing Coordinates}

\noi
The $2N$ zeroth-level variables \mb{\Gamma^{A}=\Gamma^{A}(\gamma;\Theta)} can
be viewed as functions of \mb{2(N\!-\!\ND)} physical variables \mb{\gamma^{A}}
and \mb{2\ND} second-class variables $\Theta^{a}$ such that the second-class
constraints satisfy \mb{\Theta^{a}(\Gamma(\gamma;\Theta))=\Theta^{a}}.
In other words, we may choose so-called {\em unitarizing} coordinates 
$\Gamma^{A}$ that split \mb{\Gamma^{A}=\{\gamma^{A};\Theta^{a}\}} directly 
into a physical and a second-class subsector.
We use capital roman letters  $A$, $B$, $C$, $\ldots$
from the beginning of the alphabet as upper index for both the full and the
reduced variables \mb{\Gamma^{A}} and \mb{\gamma^{A}}, respectively. 
A change of unitarizing coordinates
\beq
\Gamma^{A}~=~\{\gamma^{A};\Theta^{a}\}~~~\longrightarrow~~~ 
\Gamma^{\prime A}~=~\{\gamma^{\prime A};\Theta^{\prime a}\}
\eeq
has in general the form
\beq
\gamma^{\prime A}~=~\gamma^{\prime A}(\Gamma)~,~~~~~~~~~~~~
 \Theta^{\prime a}=\Lambda^{a}{}_{b}(\Gamma)~\Theta^{b}~,
\eeq
where the matrix \mb{\Lambda^{a}{}_{b}} is invertible. This causes the
Jacobian $J$ of the coordinate transformation to factorize on-shell,
\beq
J~=~\sdet (\frac{\partial \Gamma^{\prime A}}{\partial \Gamma^{B}})
~=~\sdet (\frac{\partial \gamma^{\prime A}}{\partial \gamma^{B}})~
\sdet (\frac{\partial \Theta^{\prime a}}{\partial \Theta^{b} })
+{\cal O}(\Theta)~.\label{jfac}
\eeq 
The Dirac antibracket becomes
\beq
 (F,G)_{D}
~=~(F \papar{\gamma^{A}})(\gamma^{A},\gamma^{B})_{D}(\papal{\gamma^{B}}G)
\eeq
in unitarizing coordinates.

\subsection{Reduction to Physical Submanifold}

\noi
In unitarizing coordinates \mb{\Gamma^{A}=\{\gamma^{A};\Theta^{a}\}}, 
we may assign reduced ``tilde'' objects that live on the physical submanifold.
In order of appearance,
\bea
\phd{E}^{AB}&\equiv&\left.(\gamma^{A},\gamma^{B})_{D} \right|_{\Theta=0}~,\cr
\phd{\rho}&\equiv&\left. \rho_{D}\right|_{\Theta=0}~,\cr
\phd{\Delta}&\equiv&\frac{(-1)^{\epsilon_{A}}}{2\phd{\rho}}
\papal{\gamma^{A}}\phd{\rho} \phd{E}^{AB}\papal{\gamma^{B}}~=~
\left. \Delta_{D} \right|_{\Theta=0}~,\cr
\phd{W}&\equiv&\left. W_{D}\right|_{\Theta=0}~,\cr
\phd{X}&\equiv&\left. X_{D}\right|_{\Theta=0}~,\cr
\phu{G}_{\alpha}&\equiv&\left. G_{\alpha}\right|_{\Theta=0}~,\cr
\phd{H}&\equiv&\left. H \right|_{\Theta=0}~,\cr
\phd{F}^{\alpha}&\equiv&\left. F^{\alpha} \right|_{\Theta=0}
~,~~~~~~~~{\rm etc.}
\eea
The antibracket 
\beq
\ph{(F,G)}~\equiv~ (F\papar{\gamma^{A}})\phd{E}^{AB}(\papal{\gamma^{B}}G)~,
\eeq 
the measure \mb{\phd{\rho}[d\gamma]}, the odd Laplacian
\mb{\phd{\Delta}}, the actions \mb{\phd{W}} and \mb{\phd{X}}, etc., 
are all independent of the defining set of constraints \mb{\Theta^{a}} 
(and of the unitarizing coordinates) used in the Dirac construction, cf.\
Section~\ref{reparam}.
The odd Laplacian \mb{\phd{\Delta}} is nilpotent, \mb{\phd{\Delta}^2=0},
because \mb{\Delta_{D}} does not contain \mb{\Theta}-derivatives (when 
using unitarizing coordinates).
{}Furthermore, the actions $\phd{W}$ and $\phd{X}$ satisfy the Quantum 
Master Equations,
\beq
\phd{\Delta} \exp\left[\frac{i}{\hbar_{\bo}}\phd{W}\right]~=~0
~,~~~~~~~~~~~~~~~~~~~~~~~~
\phu{\Delta}_{\hl} \exp\left[\frac{i}{\hbar_{\bl}}\phd{X}\right]~=~0~.
\label{thephmastereq}
\eeq
The first-level partition function \e{2nd2nd} reduces to
\beq
{\cal Z}_{\hl D}~=~\int\![d\gamma][ d\lambda]~\phd{\rho}~
e^{\Ih(\phd{W}+\phd{X})}
\label{2nd2ndph}
\eeq
on the physical submanifold. The resulting partition function
is precisely of the general form \e{partition} for fields entirely
living on the physical submanifold. Moreover, the coordinates 
\mb{\bar{\Gamma}^{A}\equiv\{F^{\alpha}; G_{\alpha}; \Theta^{a} \}}
used in \e{3splitD}, and in particular the coordinates
\mb{\{\phd{F}^{\alpha}; \phu{G}_{\alpha}; \Theta^{a} \}}
are both examples of unitarizing coordinates.
Therefore the square root formula \e{khudaverdianhD} reduces to
\beq
\phd{H} ~=~-\ln\sqrt{\frac{\phd{J}~
\sdet\ph{(\phd{F}^{\alpha},\phu{G}_{\beta})}}{\phd{\rho}}}
+ {\cal O}(\phd{G})~.
\label{khudaverdianhph}
\eeq
Here we have used that the Jacobian
\beq
\left. J_{D} \right|_{\Theta=0}
~=~\left. \sdet (\frac{\partial \bar{\Gamma}^{A}}
{\partial \Gamma^{B}})\right|_{\Theta=0}
~=~\sdet(\frac{\partial 
\{\phd{F}^{\alpha};\phu{G}_{\alpha}\}}
{\partial \gamma^{B}})
~\equiv~\phd{J}
\eeq 
satisfies the factorization property \e{jfac}. We summarize the above
observations in the following

\begin{theorem}
{\bf -- Reduction Theorem}:~~
A first-level field-antifield theory \e{2nd2nd} with second-class constraints
\mb{\Theta^{a}} may always be written in a set of unitarizing coordinates 
\mb{\Gamma^{A}=\{\gamma^{A};\Theta^{a}\}}. In these coordinates the theory
reduces to a physical theory \e{2nd2ndph} with physical coordinates 
\mb{\gamma^{A}} on the physical submanifold. The reduction is independent of
the parametrization of the constraints \mb{\Theta^{a}} and the choice of
unitarizing coordinates \mb{\Gamma^{A}=\{\gamma^{A};\Theta^{a}\}}.
\label{theoremrd}
\end{theorem}

\subsection{Transversal Coordinates}

\noi
Let us define {\em transversal} coordinates as unitarizing coordinates  
\mb{\Gamma^{A}=\{\gamma^{A};\Theta^{a}\}} with the additional property that
\beq
(\gamma^{A},\Theta^{a})~=~0~,
\eeq
so that the second-class variables \mb{\Theta^{a}} and the physical variables
\mb{\gamma^{A}} are perpendicular to each other in the antibracket sense.
{}For every system of unitarizing coordinates 
\mb{\Gamma^{A}=\{\gamma^{A};\Theta^{a}\}} there exist unique deformation 
functions \mb{X^{A}_{a}=X^{A}_{a}(\Gamma)} such that a unique ``primed'' set
of coordinates \mb{\Gamma^{\prime A}=\{\gamma^{\prime A};\Theta^{\prime a}\}},
defined as
\beq
\left\{
\begin{array}{rcl}
\gamma^{\prime A}&=&\gamma^{A}-X^{A}_{a}\Theta^{a}~, \cr
\Theta^{\prime a}&=&\Theta^{a}~,
\end{array} 
\right.
\eeq 
is a set of transversal coordinates: 
\mb{(\gamma^{\prime A},\Theta^{\prime a})\!=\!0}. 
In fact, \mb{X^{A}_{a}} satisfy the following fixed-point equation,
\beq
X^{A}_{a}~=~(\gamma^{A},\Theta^{b})E_{ba}
-(-1)^{\epsilon_{c}(1+\epsilon_{A})}\Theta^{c}(X^{A}_{c},\Theta^{b})E_{ba}
\eeq
that may be solved recursively 
\mb{X^{A}_{a}=(\gamma^{A},\Theta^{b})E_{ba}+{\cal O}(\Theta)}
to all orders in $\Theta$. 

\noi
We conclude that each set of second-class constraints \mb{\Theta^{a}} may be
complemented with variables \mb{\gamma^{A}} into a system of transversal
coordinates \mb{\Gamma^{A}=\{\gamma^{A};\Theta^{a}\}}. In transversal 
coordinates the Dirac antibracket matrix
\beq
 (\gamma^{A},\gamma^{B})_{D}~=~(\gamma^{A},\gamma^{B})
\eeq
becomes the original antibracket matrix.  This
may be used to give a short proof of the remarkable fact that the Jacobi
identity \e{jacidd} for the Dirac antibracket holds everywhere in the 
extended phase space \mb{\Gamma^{A}}. Clearly, for all physical purposes
it would have been enough to have the Jacobi identity \e{jacidd} satisfied
just on the physical submanifold. Nevertheless, the Dirac construction
\e{Diracbracket} provides the stronger Jacobi identity \e{jacidd} for free.

\noi
Similarly, we may always impose strong nilpotency \e{nilpotencyd} of
\mb{\Delta_{D}} when considering an arbitrary but fixed set of second-class
constraints \mb{\Theta^{a}}. However, we shall see in the next 
Section~\ref{reparam} that strong nilpotency \e{nilpotencyd} and the 
transformation rule \e{rhodtransformationrule1} cannot both be maintained
under reparametrization of the second-class constraints \mb{\Theta^{a}}. We
have already seen the necessity of the transformation rule 
\e{rhodtransformationrule1}, so instead we would surprisingly have to relax
the nilpotency requirement \e{nilpotencyd} for \mb{\Delta_{D}}. A manifestly
reparametrization invariant Ansatz turns out to be that the square of the
Dirac odd Laplacian,
\beq
\Delta_{D}^{2}~=~O^{A} \papal{\Gamma^{A}}~,
\label{nilpotencydo}
\eeq
is a first order differential operator with coefficient functions
\mb{O^{A}={\cal O}(\Theta)} that vanish on-shell \wrt the second-class
constraints \mb{\Theta^{a}}.

\subsection{Reparametrization of the Second-Class Constraints}
\label{reparam}

\noi
Let us now reparametrize the defining set of second-class constraints
\beq
\Theta^{a}~~~~~\longrightarrow~~~~~~ 
\Theta^{\prime a}~=~\Lambda^{a}{}_{b}(\Gamma)~\Theta^{b}~
\eeq
in \eq{definingthetatheta}, and build the Dirac antibracket 
\mb{(\cdot,\cdot)^{\prime}_{D}} and odd Laplacian \mb{\Delta^{\prime}_{D}}
from the primed set of constraints \mb{\Theta^{\prime a}}.
Our aim is dual: First of all, we must check that the different choices of
the second-class constraints do not lead to different physical quantities
on the physical submanifold. Secondly, it is of interest to know whether a
relation can be maintained strongly everywhere in the extended phase space, or
whether there appear additional contributions of order \mb{{\cal O}(\Theta)}.

\noi
The Dirac antibracket does not transform on-shell under reparametrization,
\beq
(F,G)_{D}~~~~~\longrightarrow~~~~~~
(F,G)_{D}^{\prime}~=~(F,G)_{D}+{\cal O}(\Theta)~.
\label{diracantibrackettransf0}
\eeq
In fact, one may calculate the above transformation to any order of precision
in \mb{\Theta}. This is important because higher order terms in 
\e{diracantibrackettransf0} that naively appear to play no physical r\^ole, 
can be exposed by \mb{\Theta}-differentiations. 
The calculations are simplified by choosing coordinates \mb{\gamma^{A}} such
that \mb{\Gamma^{A}=\{\gamma^{A};\Theta^{a}\}} are transversal coordinates.
To second order in \mb{\Theta} one finds
\bea
(F,G)_{D}^{\prime}-(F,G)_{D}
&=&-(F\papar{\Theta^{a}})(\Theta^{a}\papar{\Theta^{\prime b}}) 
(\Theta^{\prime b},G)_{D}
-(F,\Theta^{\prime a})_{D} (\papal{\Theta^{\prime a}}\Theta^{b}) 
(\papal{\Theta^{b}}G) \cr
&&-(F,\Theta^{\prime a})_{D} (\papal{\Theta^{\prime a}}\Theta^{b}) 
E_{bc}(\Theta^{c}\papar{\Theta^{\prime d}}) 
(\Theta^{\prime d},G)_{D} \cr
&&+(F\papar{\Theta^{a}})(\Theta^{a}\papar{\Theta^{\prime b}}) 
(\Theta^{\prime b},\Theta^{\prime c})_{D}(\papal{\Theta^{\prime c}}\Theta^{d})
(\papal{\Theta^{d}}G)+{\cal O}(\Theta^{3})~. 
\label{diracantibrackettransf2n}
\eea
In particular, the analogue of \eq{thetavanish} becomes
\bea
(\Theta^{a},F)^{\prime}_{D} 
&=&-(\Theta^{a}\papar{\Theta^{\prime b}}) (\Theta^{\prime b},F)_{D}
+(\Theta^{a}\papar{\Theta^{\prime b}})
(\Theta^{\prime b},\Theta^{\prime c})_{D}(\papal{\Theta^{\prime c}}\Theta^{d})
(\papal{\Theta^{d}}F)+{\cal O}(\Theta^{3}) \cr
&=&{\cal O}(\Theta)~.
\label{thetavanishprime}
\eea 
Similarly, to first order in \mb{\Theta}, the Dirac odd Laplacian transforms as
\bea
(\Delta^{\prime}_{D}F)
&=&\frac{(-1)^{\epsilon_{A}}}{2}\papal{\gamma^{A}} (\gamma^{A},F)_{D}^{\prime}
+\frac{(-1)^{\epsilon_{a}}}{2}\papal{\Theta^{a}} (\Theta^{a},F)_{D}^{\prime}
+\Hf \left(\ln\rho^{\prime}_{D} ,F\right)_{D}^{\prime} \cr
&=&(\Delta_{D}^{\prime\prime}F)
-\frac{(-1)^{\epsilon_{A}}}{2\rho_{D}^{\prime\prime}}\papal{\gamma^{A}} 
\left[\rho_{D}^{\prime\prime}\left(\gamma^{A},\Theta^{\prime a}\right)_{D}
(\papal{\Theta^{\prime a}}\Theta^{b}) (\papal{\Theta^{b}}F)\right]\cr
&&-\frac{(-1)^{\epsilon_{A}}}{2\rho_{D}^{\prime\prime}}
(\papal{\gamma^{A}}\Theta^{\prime a})
(\papal{\Theta^{\prime a}}\Theta^{b})\papal{\Theta^{b}}
\left[\rho_{D}^{\prime\prime}(\gamma^{A},F)_{D}\right]\cr
&&-\frac{(-1)^{\epsilon_{a}}}{2}(\Theta^{\prime a},~
\papal{\Theta^{\prime a}}\Theta^{b})_{D} (\papal{\Theta^{b}}F)+R(F)~,
\label{diracdeltatransf1}
\eea
where the remainder \mb{R(F)={\cal O}(\Theta^2)} is a second order 
differential operator consisting of terms that contain at least as many 
powers of \mb{\Theta}'s as $\Theta$-derivatives. It vanishes to the second 
order \mb{{\cal O}(\Theta^2)} in \mb{\Theta} when it is normal-ordered.
We have furthermore defined
\beq
\rho_{D}^{\prime\prime}~\equiv~ \frac{\rho_{D}^{\prime}}
{\sdet(\frac{\partial \Theta^{\prime}}{\partial \Theta})}~,
\eeq
and 
\beq
\Delta_{D}^{\prime\prime}~\equiv~
\frac{(-1)^{\epsilon_{A}}}{2\rho_{D}^{\prime\prime}}\papal{\gamma^{A}}
 \rho_{D}^{\prime\prime}(\gamma^{A},~\cdot~)_{D}
\eeq
is the Dirac odd Laplacian in ``unprimed'' transversal coordinates 
\mb{\Gamma^{A}=\{\gamma^{A};\Theta^{a}\}} and equipped with 
\mb{\rho_{D}^{\prime\prime}} as Dirac measure. To make sure that the Dirac 
odd Laplacian \mb{\Delta_{D}} does not transform on-shell,
\beq
(\Delta_{D}F)~~~~~\longrightarrow~~~~~~
(\Delta^{\prime}_{D}F)~=~(\Delta_{D}F)+{\cal O}(\Theta)~,
\label{diracdeltatransf0}
\eeq
we would clearly have to impose 
\mb{\rho_{D}^{\prime\prime}=\rho_{D}+{\cal O}(\Theta)}, which is just the 
\mb{\rho_{D}} transformation rule \e{rhodtransformationrule1}. In general, the
Dirac odd Laplacian \mb{\Delta_{D}} does change when we leave the physical 
submanifold. 

\noi
On the other hand, the transformation rule \e{rhodtransformationrule1}
provides us with a limited freedom in choosing \mb{\rho_{D}^{\prime\prime}},
or rather, in choosing \mb{\rho_{D}^{\prime}}. It is by construction clear 
that the squares \mb{\Delta_{D}^{2}}, \mb{\Delta_{D}^{\prime 2}} and 
\mb{\Delta_{D}^{\prime\prime 2}} are all first order differential operators,
and let us a priori assume that \mb{\Delta_{D}} is strongly nilpotent,
\ie \eq{nilpotencyd}. Then the rule 
\mb{\rho_{D}^{\prime\prime}=\rho_{D}+{\cal O}(\Theta)} implies that
\mb{\Delta_{D}^{\prime\prime}} is at least nilpotent on-shell, because
\mb{\Delta_{D}} does not contain $\Theta$-derivatives (in transversal
coordinates). Also \mb{\Delta^{\prime}_{D}} becomes nilpotent on-shell,
\beq
(\Delta_{D}^{2}F)~~~~~\longrightarrow~~~~~~
(\Delta^{\prime 2}_{D}F)~=~(\Delta_{D}^{2}F)
+O^{\prime A} (\papal{\Gamma^{A}}F)~,
~~~~~~~~~~~~O^{\prime A}~=~{\cal O}(\Theta)~,
\label{diracdelta2transf}
\eeq
because each \mb{\Theta}-derivative in \e{diracdeltatransf1} is accompanied
with at least one power of $\Theta$. 

\noi
The analogue of \eq{delta2com}, derived using transversal coordinates, becomes
\bea
(\Delta^{\prime}_{D}\Theta^{a})&=&
-\frac{(-1)^{\epsilon_{A}}}{2\rho_{D}}\papal{\gamma^{A}} 
\left[\rho_{D}\left(\gamma^{A},\Theta^{\prime b}\right)_{D}
(\papal{\Theta^{\prime b}}\Theta^{a}) \right]
-\frac{(-1)^{\epsilon_{b}}}{2}(\Theta^{\prime b},~
\papal{\Theta^{\prime b}}\Theta^{a})_{D}+{\cal O}(\Theta^2) \cr
&=&-(\Delta_{D}\Theta^{\prime b})(\papal{\Theta^{\prime b}}\Theta^{a})
-(-1)^{\epsilon_{b}}(\Theta^{\prime b},~
\papal{\Theta^{\prime b}}\Theta^{a})_{D}+{\cal O}(\Theta^2) \cr
&=&(-1)^{\epsilon_{b}}\Theta^{\prime b}
\Delta_{D}(\papal{\Theta^{\prime b}}\Theta^{a})+{\cal O}(\Theta^2)
~=~{\cal O}(\Theta)~.
\label{delta2comprime}
\eea
Applying \mb{\Delta^{\prime}_{D}} one more time one gets
\beq
(\Delta^{\prime 2}_{D}\Theta^{a})
~=~(-1)^{\epsilon_{b}}\Delta_{D}(\Theta^{\prime b}
\Delta_{D}(\papal{\Theta^{\prime b}}\Theta^{a}))
+{\cal O}(\Theta^2) 
~=~{\cal O}(\Theta)~.
\label{delta2comprime2}
\eeq
{}From this we conclude somewhat surprisingly that \mb{\Delta^{\prime}_{D}} is
in general {\em not} nilpotent away from the physical submanifold, 
independently of the choice of 
\mb{\rho_{D}^{\prime\prime}=\rho_{D}+{\cal O}(\Theta)}. 
A manifestly reparametrization invariant formulation is to assume the weaker
nilpotency \e{nilpotencydo} from the beginning. This has of course no 
consequences for the physics, which only lives on-shell.

\setcounter{equation}{0}
\section{Reducible Gauge-Fixing}
\label{secreduc}

\noi
We consider in this Section an interesting generalization, where the 
gauge-fixing functions \mb{G_{\alpha_{0}}} in the $X$-part become reducible.
This is quite analogous to reducibility among the gauge-generators in the
zeroth-level $W$-part. Recall that originally  the stage of reducibility
in the $W$-sector was introduced so that a zeroth-stage
gauge theory corresponds to an irreducible gauge algebra, \ie if the ghosts 
do not carry gauge symmetry.  Similarly, first-stage gauge theories 
have ghosts-for-ghosts that do not carry gauge symmetry, and so 
forth \cite{BV83}. We shall here adjust this terminology to the gauge-fixing 
$X$-part in the first-level formalism.

\noi
The motivation to work with an overcomplete set of constraints is a 
well-known theme in the theory of constrained dynamics: Often the independent
set of constraints breaks symmetries (such as, \eg, Lorentz covariance)
or locality that one would like to preserve during the quantization process.
Here an overcomplete set of constraints can provide an immediate remedy.

\noi
One starts as usual with a zeroth-level theory \mb{W=W(\Gamma;\hbar)} that
has \mb{N} gauge symmetries that should be fixed. Next one introduces 
\mb{N_{0}} Lagrange multipliers \mb{\lambda^{\alpha_{0}}} and \mb{N_{0}}
antifields \mb{\lambda^{*}_{\alpha_{0}}}. {}For each positive integer $i$ 
one chooses a number \mb{N_{i}} of so-called (first-level) $i$'{\em th-stage}
ghosts \mb{\gh^{\alpha_{i}}\equiv\gh_{\bl i}^{\alpha_{i}}}, 
with Grassmann parity \mb{\epsilon_{\alpha_{i}}\!+\!i} where 
\mb{\epsilon_{\alpha_{i}}\!\equiv\!\epsilon_{\alpha_{i}}^{\bl i}},
and \mb{N_{i}} antifields 
\mb{\gh^{*}_{\alpha_{i}}\!\equiv\!\gh^{i*}_{ \bl\alpha_{i}}}, 
of opposite statistics, where the index \mb{\alpha_{i}} runs through 
\mb{\alpha_{i}=1, \ldots, N_{i}}.
The integers \mb{N_{0}, N_{1}, N_{2}, \ldots}, can be chosen at will, as long
as all of the following alternating sums are non-negative:
\beq
\forall i \ge -1:~~ \sum_{j=-1}^{i} (-1)^{i-j} N_{j}~\ge~ 0~,
\eeq
where \mb{N_{-1} \!\equiv \!N}.
In particular, one may easily check from the above inequalities that 
\beq
\forall i \ge -1:~~ N_{i}~\ge~ 0~,
\eeq
as it should be. The {\em stage $s$ of reducibility} is defined as the maximum
\beq
 s~\equiv~\max\left(\{i \ge 0 | N_i>0\}\cup \{-1\} \right)~,
\eeq
over the non-empty set \mb{\{i \ge 0 | N_i>0\}\cup \{-1\} }. 
A {\em zeroth-stage theory} with \mb{s\!=\!0} requires 
\beq
 N~=~N_{0}~>~0~=~N_{1}~=~N_{2}~=~N_{3}~=~\ldots~.
\eeq
This is precisely the irreducible case of Section~\ref{secirr}.
Similarly, a {\em first-stage theory} with \mb{s\!=\!1} requires 
\beq
N ~\geq~ 0~,~~~~~~N_{0}-N~=~N_{1}~>~0~=~N_{2}~=~N_{3}~=~N_{4}~=~\ldots~, 
\eeq
while a {\em higher stage theory}  requires 
\bea
s=2:&&N_{0}\geq N\geq 0~,~~ N_{1}-N_{0}+N=N_{2}>0=N_{3}=N_{4}=N_{5}=\ldots~,\cr
s=3:&& N_{3}-N_{2}+N_{1}+N=N_{0}\geq N \geq 0~,~~
N_{2}\geq N_{3}>0=N_{4}=N_{5}=\ldots~, \cr
s=4:&& N_{4}-N_{3}+N_{2}=N_{1}-N_{0}+N\geq 0~,~~N_{0}\geq N\geq 0~,~~
N_{3}\geq N_{4}>0=N_{5}=\ldots~, \cr
\ldots&&\ldots~,
\eea
and so forth. The stage $s$ of reducibility could be \mb{\infty}. 
If \mb{s <\infty }, then 
\mb{\sum_{i=0}^{\infty}N_{2i-1}=\sum_{i=0}^{\infty}N_{2i}}.

\noi
Summarizing, the minimal field content in the first-level formalism is
\beq
\Gamma^{A}_{\hl\min}\equiv\left\{\Gamma^{A};
\lambda^{\alpha_{0}},\lambda^{*}_{\alpha_{0}};
\gh^{\alpha_{1}},\gh^{*}_{\alpha_{1}};
\gh^{\alpha_{2}},\gh^{*}_{\alpha_{2}};\ldots ;
\gh^{\alpha_{s}},\gh^{*}_{\alpha_{s}}\right\}~.
\eeq
In addition to the \mb{2\sum_{i=-1}^{s}N_{i}} minimal fields and antifields,
there is a triangular tower of \mb{4\sum_{i=1}^{s}iN_{i}} non-minimal fields
and antifields, in complete analogy with reducibility at the zeroth-level,
cf.\ \Ref{BV83} and Subsection~\ref{finitestagedirac}. The first-level minimal
odd Laplacian becomes
\beq
\Delta_{\hl\min}~\equiv~\Delta+(-1)^{\epsilon_{\alpha_{0}}}
\papal{\lambda^{\alpha_{0}}}\papal{\lambda^{*}_{\alpha_{0}}} 
+\sum_{i=1}^{s}(-1)^{\epsilon_{\alpha_{i}}+i}
\papal{\gh^{\alpha_{i}}}\papal{\gh^{*}_{\alpha_{i}}}~,
\label{reduDeltaaug}
\eeq
while the first-level minimal Planck-number operator is
\beq
\pl_{\min}~=~-\left(\lambda^{*}_{\alpha_{0}}\lambda^{\alpha_{0}}
+\sum_{i=1}^{s}(i\!+\!1) \gh^{*}_{\alpha_{i}}\gh^{\alpha_{i}},~
\cdot~\right)_{\hl}+\hbar_{\bl} \frac{\partial }{\partial \hbar_{\bl}}~.
\eeq
The gauge-fixing action \mb{X_{\min}} should again satisfy 
the Principle~\ref{principle3}, \ie 1) Planck number conservation,
2) the Quantum Master Equation and 3) rank requirements. 
Although it is straightforward to expand \mb{X_{\min}} in action 
terms allowed by Planck number, it quickly becomes space consuming.
Instead we shall focus on a few important terms
\bea
X_{\min}&=&G_{\alpha_{0}}~\lambda^{\alpha_{0}} + (i\hbar_{\bl})~H 
+ \lambda^{*}_{\alpha_{0}}~Z^{\alpha_{0}}{}_{\alpha_{1}}~\gh^{\alpha_{1}}
+\sum_{i=1}^{s-1} \gh^{*}_{\alpha_{i}}~ 
Z^{\alpha_{i}}{}_{\alpha_{i+1}}~\gh^{\alpha_{i+1}}\cr
&&+\lambda^{*}_{\alpha_{0}}\left[\Hf U^{\alpha_{0}}_{\beta_{0}\gamma_{0}}~
\lambda^{\gamma_{0}}(-1)^{\epsilon_{\beta_{0}}+1}
+(i\hbar_{\bl})~V^{\alpha_{0}}{}_{\beta_{0}}\right]\lambda^{\beta_{0}}
+(i\hbar_{\bl})^2~\lambda^{*}_{\alpha_{0}}\tilde{G}^{\alpha_{0}} \cr
&&+\sum_{i=1}^{s}\gh^{*}_{\alpha_{i}}\left[
U^{\alpha_{i}}_{\beta_{i}\alpha_{0}}~
\lambda^{\alpha_{0}}(-1)^{\epsilon_{\beta_{i}}+i+1}
+(i\hbar_{\bl})~V^{\alpha_{i}}{}_{\beta_{i}} \right]\gh^{\beta_{i}}
+\ldots~.\label{X0}
\eea
In particular, the Faddeev-Popov term 
\mb{\lambda^{*}_{\alpha_{0}}Z^{\alpha_{0}}{}_{\alpha_{1}}\gh^{\alpha_{1}}}
and its higher-stage counterparts 
\mb{\gh^{*}_{\alpha_{i}}Z^{\alpha_{i}}{}_{\alpha_{i+1}}\gh^{\alpha_{i+1}}}
will be important new ingredients 
(as compared to the irreducible case). 
The structure functions 
\mb{Z^{\alpha_{i-1}}{}_{\alpha_{i}}
\equiv Z_{i}^{\alpha_{i-1}}{}_{\alpha_{i}}(\Gamma;\hbar)} will carry
Grassmann parity \mb{\epsilon_{\alpha_{i}}\!+\!\epsilon_{\alpha_{i-1}}}.
We stress that the \eq{X0} should not be read as a systematic expansion of
the action \mb{X_{\min}}. Rather the terms in \eq{X0} were selected simply
because they enter the first few consequences of the Quantum Master Equation
for \mb{X_{\min}},
\bea
(G_{\alpha_{0}},G_{\beta_{0}})
&=& G_{\gamma_{0}} U^{\gamma_{0}}_{\alpha_{0} \beta_{0}}~,
\label{redunonabelinvo} \\
G_{\alpha_{0}}Z^{\alpha_{0}}{}_{\alpha_{1}}&=&0~, \label{Greduc} \\
Z^{\alpha_{i-1}}{}_{\alpha_{i}}Z^{\alpha_{i}}{}_{\alpha_{i+1}}&=&{\cal O}(G)~,
\label{Zreduc} \\
(Z^{\alpha_{i}}{}_{\beta_{i+1}},G_{\beta_{0}})
&=&Z^{\alpha_{i}}{}_{\alpha_{i+1}}
U^{\alpha_{i+1}}_{\beta_{i+1} \beta_{0}}
-(-1)^{(\epsilon_{\beta_{0}}+1)(\epsilon_{\gamma_{i}}+\epsilon_{\beta_{i+1}})}
U^{\alpha_{i}}_{\gamma_{i}\beta_{0}}
Z^{\gamma_{i}}{}_{\beta_{i+1}}
+{\cal O}(G)~,\label{reduweakeq5} \\
(\Delta G_{\beta_{0}})-(H,G_{\beta_{0}})
&=&\sum_{i=0}^{s}(-1)^{\epsilon_{\alpha_{i}}+i}
U^{\alpha_{i}}_{\alpha_{i}\beta_{0}}
 +G_{\alpha_{0}}V^{\alpha_{0}}{}_{\beta_{0}}~, \label{reduweakeq2} \\
 -(\Delta H)+\Hf (H,H)
&=&\sum_{i=0}^{s}V^{\alpha_{i}}{}_{\alpha_{i}}
-G_{\alpha_{0}}\tilde{G}^{\alpha_{0}}~.
\label{reduweakeq3}
\eea
The first \eq{redunonabelinvo} is just the usual non-Abelian involution of
the gauge-fixing functions \mb{G_{\alpha_{0}}}. The second \eq{Greduc} and
the third \eq{Zreduc} show that \mb{G_{\alpha_{0}}} and 
\mb{Z^{\alpha_{i-1}}{}_{\alpha_{i}}}, respectively, are in general reducible.
They imply that the action $X$ exhibit  gauge symmetries on-shell \wrt the
\mb{G_{\alpha_{0}}}'s,
\beq
\delta\gh^{\alpha_{i}}
~=~Z^{\alpha_{i}}{}_{\alpha_{i+1}}\xi^{\alpha_{i+1}}~,
\eeq
where \mb{\xi^{\alpha_{i+1}}} are gauge parameters.

\noi
Note that an irreducible gauge-fixing action $X$ corresponding to \mb{s=0}
has {\em no} gauge symmetry at the first-level\footnote{ 
Compare this with the zeroth-level terminology, where a {\em trivial} gauge
algebra, \ie with {\em no} gauge symmetry, 
is strictly speaking of stage ``$-1$''. In other words, the definition 
of stage of the first-level $X$-action has been shifted by one unit as 
compared to the original definition of stage of the zeroth-level $W$-action
\cite{BV83}. This shift is introduced to avoid speaking of negative stages.}.
This is why we could choose a trivial first-level gauge \mb{\Psi\equiv 0} in
Subsection~\ref{Going}. In the reducible case the gauge fermion \mb{\Psi}
should meet certain rank requirements.

\subsection{First-Stage Reducibility}
\label{secfirststagereduc}

\noi
In Subsections~\ref{secfirststagereduc}-\ref{secnonminappr} we shall work 
out the simplest case of reducible gauge-fixing in detail, namely first-stage
reducibility. In this setting the gauge-fixing action \mb{X} has \mb{N_{1}}
gauge symmetries in the \mb{\lambda^{\alpha_{0}}}-variables due to 
reducibility \eq{Greduc} among the gauge-fixing functions \mb{G_{\alpha_{0}}}.
Besides the two minimal first-level fields \mb{\lambda^{\alpha_{0}}} and
\mb{\gh^{\alpha_{1}}}, there are also two non-minimal fields,
\mb{\bgh_{\alpha_{1}}} of statistics \mb{\epsilon_{\alpha_{1}}\!+\!1},
and \mb{\pi_{\alpha_{1}}} of statistics \mb{\epsilon_{\alpha_{1}}}, \ie
\beq
\Gamma^{A}_{\hl}\equiv\left\{\Gamma^{A};
\lambda^{\alpha_{0}},\lambda^{*}_{\alpha_{0}};
\gh^{\alpha_{1}},\gh^{*}_{\alpha_{1}};
\bgh_{\alpha_{1}},\bgh^{*\alpha_{1}};
\pi_{\alpha_{1}},\pi^{*\alpha_{1}}\right\}~.
\eeq
The Planck-number operator is chosen to be
\beq
\pl~=~-\left(\lambda^{*}_{\alpha_{0}}\lambda^{\alpha_{0}}+2
\gh^{*}_{\alpha_{1}}\gh^{\alpha_{1}}-
\bgh_{\alpha_{1}}\bgh^{*\alpha_{1}},~
\cdot~\right)_{\hl}+\hbar_{\bl} \frac{\partial }{\partial \hbar_{\bl}}~,
\eeq
or equivalently, 
\beq
\begin{array}{rclcrclcrclcrcl}
\pl(\lambda^{\alpha_{0}})&=&1 &,&
\pl(\gh^{\alpha_{1}})&=&2 &,&
\pl(\bgh_{\alpha_{1}})&=&-1 &,&
\pl(\pi_{\alpha_{1}})&=&0~, \\ \\
\pl(\lambda^{*}_{\alpha_{0}})&=&-1 &,&
\pl(\gh^{*}_{\alpha_{1}})&=&-2 &,&
\pl(\bgh^{*\alpha_{1}})&=&1 &,&
\pl(\pi^{*\alpha_{1}})&=&0~,  \\ \\
\pl(\Gamma^{A})&=&0 &,& && & &
\pl(\hbar_{\bl})&=&1 &,&
\pl(\hbar)&=&0~.
\end{array}
\eeq
Note in particular that \mb{\pi_{\alpha_{1}}} and \mb{\pi^{*\alpha_{1}}} have
vanishing Planck number, so they appear on the same footing as the original
variables \mb{\Gamma^{A}} in a Planck number expansion. This is just 
one of many reasons to enlarge the \mb{2N}-dimensional zeroth-level phase
space \mb{\Gamma^{A}} into a \mb{2N_{0}}-dimensional phase space
\beq
\Gamma^{A_{\ext}}~\equiv~\{\Gamma^{A};\pi_{\alpha_{1}},\pi^{*\alpha_{1}}\}~,
\eeq
where \mb{N_{0}=N+N_{1}}.
We shall see in Subsection~\ref{backtoreduc} that this space plays a
profound r\^ole. The odd Laplacian reads
\beq
\Delta_{\hl}~\equiv~\Delta_{\ext} +(-1)^{\epsilon_{\alpha_{0}}}
\papal{\lambda^{\alpha_{0}}}\papal{\lambda^{*}_{\alpha_{0}}}
+(-1)^{\epsilon_{\alpha_{1}}+1}
(\papal{\gh^{\alpha_{1}}}\papal{\gh^{*}_{\alpha_{1}}}
+\papal{\bgh_{\alpha_{1}}}\papal{\bgh^{*\alpha_{1}}})~,
\label{reduDeltaaug1}
\eeq
where
\beq
\Delta_{\ext}~\equiv~\Delta+(-1)^{\epsilon_{\alpha_{1}}}
\papal{\pi_{\alpha_{1}}}\papal{\pi^{*\alpha_{1}}}~.
\eeq 
The subscript ``$\ext$'' will everywhere in this Section refer to the
extended space \mb{\Gamma^{A_{\ext}}}.

\noi
However let us first take a more traditional route. Recall that in the
original field-antifield approach the minimal sector is introduced to 
obtain solutions to the Master Equation satisfying the appropriate rank
condition, and the non-minimal sector is {\em only} added to have 
well-defined gauge-fixing choices at hand \cite{BV81,BV83}. So ignoring for 
the moment the non-minimal fields  
\mb{\{\bgh_{\alpha_{1}},\bgh^{*\alpha_{1}};
\pi_{\alpha_{1}},\pi^{*\alpha_{1}}\}}, 
the rank of the Hessian of \mb{X_{\min}} in the minimal sector 
\beq
\Gamma^{A}_{\hl\min}~\equiv~\{ \Gamma^{A};
\lambda^{\alpha_{0}},\lambda^{*}_{\alpha_{0}};
\gh^{\alpha_{1}},\gh^{*}_{\alpha_{1}}\}
\eeq
should be \mb{2N_{0}}, where \mb{N_{0}=N+N_{1}}. This implies that the two
rectangular matrices \mb{\partial G_{\alpha_{0}} / \partial \Gamma^{A}}
and \mb{Z^{\alpha_{0}}{}_{\alpha_{1}}} have maximal rank, \ie
\beq
\rank (G_{\alpha_{0}}\papar{\Gamma^{A}})~=~N~,~~~~~~~~~~~~~~~~~~~
\rank (Z^{\alpha_{0}}{}_{\alpha_{1}})~=~N_{1}~,
\eeq
and therefore the \mb{Z^{\alpha_{0}}{}_{\alpha_{1}}} matrix does not have
zero-eigenvalue right eigenvectors.

\subsection{First-Level Gauge-Fixing}

\noi
The standard Ansatz for the non-minimal action is
\beq
X~=~X_{\min}(\Gamma^{A}_{\hl\min};\hbar_{\hl})
+ \pi_{\alpha_{1}}\bgh^{*\alpha_{1}}~,\label{minansatz}
\eeq
while a simple choice for \mb{\Psi} reads
\beq
\Psi ~=~ -\bgh_{\alpha_{1}}\chi^{\alpha_{1}}
~,~~~~~~~~~~~~~~~~~\pl(\Psi)~=~0~,\label{reducPsi}
\eeq
with \mb{N_{1}} first-level gauge-fixing conditions 
\mb{\chi^{\alpha_{1}}=\chi^{\alpha_{1}}(\Gamma;\lambda_{0};\hbar)}
of Grassmann parity \mb{\epsilon_{\alpha_{1}}}.
Planck number conservation restricts us to a linear dependence of 
\mb{\lambda^{\alpha_{0}}},
\beq
\chi^{\alpha_{1}}~=~\omega^{\alpha_{1}}{}_{\alpha_{0}}\lambda^{\alpha_{0}}
\label{reducchi}
\eeq
(up to an inessential constant proportional to \mb{\hbar_{\bl}}).
This gauge-fixing choice is based on a matrix 
\mb{\omega^{\alpha_{1}}{}_{\alpha_{0}}
=\omega^{\alpha_{1}}{}_{\alpha_{0}}(\Gamma;\hbar)}
of Grassmann parity 
\mb{\epsilon_{\alpha_{0}}\!+\!\epsilon_{\alpha_{1}}},
such that the Faddeev-Popov matrix
\beq
\Delta^{\alpha_{1}}{}_{\beta_{1}}~\equiv~
\omega^{\alpha_{1}}{}_{\alpha_{0}}Z^{\alpha_{0}}{}_{\beta_{1}}
\label{fpmatrix}
\eeq
 is invertible, \ie
\beq
\rank\left(\Delta^{\alpha_{1}}{}_{\beta_{1}}\right)~=~N_{1}~.
\label{fpdetrank}
\eeq
One could in principle let \mb{\omega^{\alpha_{1}}{}_{\alpha_{0}}}
depend on \mb{\pi_{\alpha_{1}}}, but one may prove a constraint 
\mb{\pi_{\alpha_{1}}\approx 0} that would turn this idea into a
vacuous exercise. According to the general theory outlined in 
Subsection~\ref{secfirstlevelformalism}
the first-level partition function is given by 
\beq
{\cal Z}^{\Psi}_{\hl} 
~=~ \int \! d\mu \left. e^{\Ih (W + X^{\Psi})} 
\right|_{\lambda^{*}_{0},\gh_{1}^{*},\bgh_{1}^{*},\pi^{*}_{1}=0}
~=~ \int \! d\mu \left. e^{\Ih (W^{\Psi} + X)} \right|_{\Sigma}~,
\label{reduWXpathint}
\eeq
with a measure 
\beq
d\mu ~=~ \rho [d\Gamma] [d\lambda_{0}] [d\gh_{1}] [d\bgh_{1}] [d\pi_{1}]~,
\label{firststagemeasure}
\eeq
and a gauge-fixing surface \mb{\Sigma} specified by
\bea
\lambda^{*}_{\alpha_{0}}
&=&-E(\ad\Psi)\frac{\partial \Psi}{\partial\lambda^{\alpha_{0}}}
~=~\bgh_{\alpha_{1}}E(\ad\Psi)\omega^{\alpha_{1}}{}_{\alpha_{0}}~,
\label{lambda0star} \\
\gh^{*}_{\alpha_{1}}
&=&-E(\ad\Psi)\frac{\partial \Psi}{\partial\gh^{\alpha_{1}}}~=~0~,
\label{lambda1star} \\
\bgh^{*\alpha_{1}}
&=&-E(\ad\Psi)\frac{\partial \Psi}{\partial \bgh_{\alpha_{1}}}
~=~E(\ad\Psi)\chi^{\alpha_{1}}~, \\ 
\pi^{*\alpha_{1}}
&=&-E(\ad\Psi)\frac{\partial \Psi}{\partial\pi_{\alpha_{1}}}~=~0~,
\label{reducantifield1s}
\eea
cf.\ the prescription \e{gf1}. 

\begin{lemma}:~~
The path integrations over the Faddeev-Popov ghost pair 
\mb{\{\gh^{\alpha_{1}};\bgh_{\alpha_{1}}\}} can be performed explicitly.
The first-level partition function \e{reduWXpathint} thereby simplifies to
\beq
{\cal Z}_{\hl}
~=~\int \![d\Gamma][d\lambda_{0}][d\pi_{1}]~\rho~
e^{\Ih W+\Ih G_{\omega,\alpha_{0}}\lambda^{\alpha_{0}}-H}~
\sdet(\Delta^{\alpha_{1}}{}_{\beta_{1}})~,  
\label{partitionfunctionreduc}  
\eeq
where 
\beq
 G_{\omega,\alpha_{0}}~\equiv~G_{\alpha_{0}}
+\pi_{\alpha_{1}}\omega^{\alpha_{1}}{}_{\alpha_{0}}
\eeq
are \mb{\omega}-deformed gauge-fixing constraints, and 
\mb{\Delta^{\alpha_{1}}{}_{\beta_{1}}} 
is the Faddeev-Popov matrix \e{fpmatrix}.
\label{lemmareducpathint}
\end{lemma}

\noi
{\sc Proof of Lemma~\ref{lemmareducpathint}}:~~It is convenient to split
the gauge-fixed action 
\beq
S~=~(\frac{\hbar_{\bl}}{\hbar}W^{\Psi}+X)|_{\Sigma}~=~S_{0}+S_{FP}+V~,
\eeq
into a part
\beq 
S_{0}~\equiv~\frac{\hbar_{\bl}}{\hbar}W+G_{\alpha_{0}}\lambda^{\alpha_{0}}
+(i\hbar_{\bl})H
+\pi_{\alpha}\chi^{\alpha}~
\eeq
that is independent of the ghosts and antighosts 
\mb{\{\gh^{\alpha_{1}};\bgh_{\alpha_{1}}\}}, a Faddeev-Popov term
\mb{S_{FP}\equiv\bgh_{\alpha_{1}}
\Delta^{\alpha_{1}}{}_{\beta_{1}}\gh^{\beta_{1}}}
that is quadratic in \mb{\{\gh^{\alpha_{1}};\bgh_{\alpha_{1}}\}}, and a
part $V$ that contains all interaction terms, tadpole terms and terms 
quadratic in the antighost \mb{\bgh_{\alpha_{1}}}. At this point the 
only quantities left that carry Planck number, 
are \mb{\hbar_{\bl}}, \mb{\lambda^{\alpha_{0}}}, \mb{\gh^{\alpha_{1}}} and
\mb{\bgh_{\alpha_{1}}}. Since the action \mb{S} has Planck number 
\mb{\pl(S)=1}, the multiplicities \mb{m_{\hbar}}, \mb{m_{\lambda}},
\mb{m_{1}}, and \mb{\bar{m}_{1}} of
\mb{\hbar_{\bl}}, \mb{\lambda^{\alpha_{0}}}, \mb{\gh^{\alpha_{1}}} and
\mb{\bgh_{\alpha_{1}}}, respectively, must obey  
\beq
 m_{\hbar} +m_{\lambda}+ 2m_{1}-\bar{m}_{1} ~=~1
\eeq
in any given term in the action. Equivalently,
\beq
\bar{m}_{1}-m_{1}~=~m_{\hbar}+m_{\lambda}+ m_{1}-1~\equiv~RHS~.
\eeq
Clearly the \rhs \mb{RHS\geq -1}. There are no terms with \mb{RHS\!=\!-1},
and the terms with \mb{RHS\!=\!0} are precisely the free part 
\mb{S_{0}+S_{FP}}. This implies that all the terms in the $V$-part have fewer
ghosts \mb{\gh^{\alpha_{1}}} than antighosts \mb{\bgh_{\alpha_{1}}}, \ie
\beq
 m_{1}~<~\bar{m}_{1} ~. \label{ruleofthumb}
\eeq 
Next one scales the ghosts and antighosts,
\beq
\gh^{\alpha_{1}} ~\to~ \frac{1}{\varepsilon}\gh^{\alpha_{1}}~,~~~~~~~~~~~
\bgh_{\alpha_{1}} ~\to~ \varepsilon\bgh_{\alpha_{1}}~,
\eeq
and let \mb{\varepsilon\to 0}. The $V$-term drops out because of the rule 
\e{ruleofthumb}, while the free part \mb{S_{0}+S_{FP}} and the path integral 
measure are unchanged. Hence the integration over 
\mb{\{\gh^{\alpha_{1}};\bgh_{\alpha_{1}}\}} can be explicitly performed.
\proofbox

\subsection{A Square Root Formula for $H$}

\noi
The \mb{\omega}-deformed constraints \mb{G_{\omega,\alpha_{0}}} have two very
important properties:
\begin{enumerate}

\item
{}First, the \mb{N_{0}} constraints \mb{G_{\omega,\alpha_{0}}} are
{\em irreducible} on the \mb{2N_{0}}-dimensional extended space
\beq
\Gamma^{A_{\ext}}~\equiv~\{\Gamma^{A};\pi_{\alpha_{1}},\pi^{*\alpha_{1}}\}~.
\eeq
Besides containing the original reducible constraints 
\mb{G_{\alpha_{0}}\approx 0} of rank \mb{N}, the \mb{\omega}-deformed 
constraints \mb{G_{\omega,\alpha_{0}}\approx 0} in addition contain \mb{N_{1}}
conditions \mb{\pi_{\alpha_{1}}\approx 0}, as is clear from the formula
\beq
\pi_{\beta_{1}}~=~G_{\omega,\alpha_{0}}Z^{\alpha_{0}}{}_{\alpha_{1}}
(\Delta^{-1})^{\alpha_{1}}{}_{\beta_{1}}~,
\eeq
where use has been made of \eq{Greduc}.
See Subsection~\ref{backtoirr} below for more details.

\item
Second, the \mb{G_{\omega,\alpha_{0}}} are in non-Abelian involution
\beq
 (G_{\omega,\alpha_{0}},G_{\omega,\beta_{0}})_{\ext}
~=~G_{\omega,\gamma_{0}} U^{\gamma_{0}}_{\omega,\alpha_{0}\beta_{0}}~,
\label{omeganonabelinvo}
\eeq
\wrt the extended \mb{(\cdot,\cdot)_{\ext}} bracket. The \mb{\omega}-deformed
structure functions read
\bea
U^{\gamma_{0}}_{\omega,\alpha_{0}\beta_{0}}
&=&\Hf \left[\delta^{\gamma_{0}}_{\delta_{0}}-
Z^{\gamma_{0}}{}_{\alpha_{1}}(\Delta^{-1})^{\alpha_{1}}{}_{\beta_{1}}
\omega^{\beta_{1}}{}_{\delta_{0}}\right]
U^{\delta_{0}}_{\alpha_{0}\beta_{0}} 
+Z^{\gamma_{0}}{}_{\alpha_{1}}(\Delta^{-1})^{\alpha_{1}}{}_{\beta_{1}}
(\omega^{\beta_{1}}{}_{\alpha_{0}},G_{\omega,\beta_{0}})_{\ext} \cr
&&-(-1)^{(\epsilon_{\alpha_{0}}+1)(\epsilon_{\beta_{0}}+1)}
(\alpha_{0}\leftrightarrow\beta_{0})+{\cal O}(G_{\omega})
\label{uomega}
\eea
to the zeroth order in \mb{G_{\omega}}.
\end{enumerate}

\noi
In the following Theorem~\ref{theoremareduc} the fundamental r\^ole played
by the \mb{2N_{0}}-dimensional extended space \mb{\Gamma^{A_{\ext}}} is
displayed further. Let
\beq
 F_{\bar{\omega}}^{\alpha_{0}}~\equiv~F^{\alpha_{0}}
+\bar{\omega}^{\alpha_{0}}{}_{\alpha_{1}}\pi^{*\alpha_{1}}
+\tilde{\omega}^{\alpha_{0}\alpha_{1}}\pi_{\alpha_{1}}+\ldots
\eeq
be arbitrary zeroth-level coordinate functions of statistics
\mb{\epsilon_{\alpha_{0}}\!+\!1}, where ``\mb{\ldots}'' denotes terms that
are at least quadratic in  \mb{\pi_{\alpha_{1}}} and \mb{\pi^{*\alpha_{1}}},
and where
\beq
 F^{\alpha_{0}}~=~F^{\alpha_{0}}(\Gamma;\hbar)~,~~~~~~~~
\bar{\omega}^{\alpha_{0}}{}_{\alpha_{1}}
~=~\bar{\omega}^{\alpha_{0}}{}_{\alpha_{1}}(\Gamma;\hbar)
~~~~~~~{\rm and}~~~~~~~~
\tilde{\omega}^{\alpha_{0}\alpha_{1}}
~=~\tilde{\omega}^{\alpha_{0}\alpha_{1}}(\Gamma;\hbar)~, 
\eeq
such that the transformation
\beq
\Gamma^{A_{\ext}}~\equiv~\{\Gamma^{A};\pi_{\alpha_{1}},\pi^{*\alpha_{1}} \}
~~\longrightarrow~~
\bar{\Gamma}^{A_{\ext}}_{\omega,\bar{\omega}} 
~\equiv~ \{ F^{\alpha_{0}}_{\bar{\omega}};G_{\omega,\alpha_{0}} \}
\eeq
is a coordinate transformation of the \mb{2N_{0}}-dimensional space 
\mb{\Gamma^{A_{\ext}}}, and let
\mb{J_{\omega,\bar{\omega}}
=\sdet (\frac{\partial \bar{\Gamma}^{A_{\ext}}_{\omega,\bar{\omega}}}
{\partial \Gamma^{B_{\ext}}})} denote the Jacobian. Then

\begin{theorem}:~~The Quantum Master Equation for \mb{X_{\min}} implies 
that the quantum correction $H$ is given the following square root formula 
\beq
H ~=~-\left. \ln \sqrt{\frac{J_{\omega,\bar{\omega}}~
    \sdet( F_{\bar{\omega}}^{\alpha_{0}}, G_{\omega,\beta_{0}})_{\ext}}
{\rho~\sdet(\Delta^{\alpha_{1}}{}_{\beta_{1}})^2}} 
\right|_{\pi_{1},\pi_{1}^{*}=0}+ {\cal O}(G)~. 
\label{khudaverdianreduc}
\eeq
\label{theoremareduc}
\end{theorem}

\noi
There are essentially two new features in the reducible case that we would
like to emphasize as compared to the irreducible case, cf.\ 
Theorem~\ref{theorema}. 
{}First and most pronounced, almost all reference to the original space
\mb{\Gamma^{A}} has been replaced with the extended space 
\mb{\Gamma^{A_{\ext}}}.  The two conditions \mb{G_{\alpha_{0}}\approx 0} and
\mb{\pi_{\alpha_{1}}\approx 0} enter slightly asymmetrically, because  
\mb{H=H(\Gamma;\hbar)} conventionally does not depend on 
\mb{\pi_{\alpha_{1}}}, and hence this asymmetry is due to notation
rather than substance, cf.\ Subsection~\ref{backtoreduc}. 
Secondly, the Faddeev-Popov determinant 
\mb{\sdet(\Delta^{\alpha_{1}}{}_{\beta_{1}})} makes an interesting
appearance in formula \e{khudaverdianreduc}.

\noi
In more details the Jacobian is given as
\beq
J_{\omega,\bar{\omega}}~\equiv~
\sdet(\bar{\Gamma}^{A_{\ext}}_{\omega,\bar{\omega}}\papar{\Gamma^{B_{\ext}}})
~=~\int  \! [d\bar{C}_{\ext}][dC_{\ext}]~e^{\Ih S_{C}}~, 
\label{Jomega}
\eeq 
with a Jacobian action
\bea
S_{C}&=&\bar{C}_{A_{\ext}}(\bar{\Gamma}^{A_{\ext}}_{\omega,\bar{\omega}}
\papar{\Gamma^{B_{\ext}}})C^{B_{\ext}} \cr
&=&(-1)^{\epsilon_{A_{\ext}}+1}C^{A_{\ext}}
(\papal{\Gamma^{A_{\ext}}}G_{\omega,\alpha_{0}})
\bar{C}^{\alpha_{0}} 
+\bar{C}_{\alpha_{0}}(F_{\bar{\omega}}^{\alpha_{0}}
\papar{\Gamma^{A_{\ext}}})C^{A_{\ext}}\cr
&=&\left[(-1)^{\epsilon_{A}+1}C^{A}
(\papal{\Gamma^{A}}G_{\alpha_{0}})
+(-1)^{\epsilon_{\alpha_{1}}+1}C_{\alpha_{1}}
\omega^{\alpha_{1}}{}_{\alpha_{0}} \right]\bar{C}^{\alpha_{0}} \cr
&& +\bar{C}_{\alpha_{0}}\left[(F^{\alpha_{0}}
\papar{\Gamma^{A}})C^{A}
+ \bar{\omega}^{\alpha_{0}}{}_{\alpha_{1}}C^{\alpha_{1}}
+ \tilde{\omega}^{\alpha_{0}\alpha_{1}}C_{\alpha_{1}}\right]
+{\cal O}(\pi_{1};\pi_{1}^{*})~. 
\label{actionc}
\eea
Similarly, one may exponentiate the other superdeterminant
\beq
\sdet( F_{\bar{\omega}}^{\alpha_{0}}, G_{\omega,\beta_{0}})_{\ext}~=~
\int \! [d\bar{B_{0}}][dB_{0}]~e^{\Ih S_{B}} 
\label{sdetomega}
\eeq
with action
\bea
S_{B}&=&\bar{B}_{\alpha_{0}}(F_{\bar{\omega}}^{\alpha_{0}},
G_{\omega,\beta_{0}})_{\ext} B^{\beta_{0}}  \cr
&=&\bar{B}_{\alpha_{0}}\left[ (F^{\alpha_{0}},G_{\beta_{0}})
-\bar{\omega}^{\alpha_{0}}{}_{\alpha_{1}}
\omega^{\alpha_{1}}{}_{\beta_{0}}\right]B^{\beta_{0}}
+{\cal O}(\pi_{1};\pi_{1}^{*})~.
\label{actionb}
\eea
Here the ghost pair \mb{(C^{A_{\ext}};\bar{C}_{A_{\ext}})} has statistics
\mb{\epsilon_{A_{\ext}}\!+\!1} and the ghost pair 
\mb{(B^{\alpha_{0}};\bar{B}_{\alpha_{0}})} has statistics 
\mb{\epsilon_{\alpha_{0}}\!+\!1}. We have split 
\mb{\bar{C}_{A_{\ext}}=\{\bar{C}_{\alpha_{0}};\bar{C}^{\alpha_{0}}\}} of
Grassmann parity 
\mb{\epsilon({\bar{C}_{\alpha_{0}}}) = \epsilon_{\alpha_{0}}} and
\mb{\epsilon({\bar{C}^{\alpha_{0}}})=\epsilon_{\alpha_{0}}\!+\!1}, 
respectively. The ghosts \mb{C^{A_{\ext}}} of the \mb{2N_{0}}-dimensional
extended space decompose as 
\mb{C^{A_{\ext}}=\{C^{A};C_{\alpha_{1}},C^{\alpha_{1}}\}}, with
Grassmann parity \mb{\epsilon({C^{A}}) = \epsilon_{A}\!+\!1}, 
\mb{\epsilon(C_{\alpha_{1}}) = \epsilon_{\alpha_{1}}\!+\!1} and
\mb{\epsilon(C^{\alpha_{1}}) = \epsilon_{\alpha_{1}}}, respectively.

\noi
Due to the above mentioned properties of \mb{G_{\omega,\alpha_{0}}},
it is clear that one may re-use the Lemma~\ref{lemmaa} from 
Section~\ref{Going} for this situation:

\begin{lemma}:~~The factor
\mb{J_{\omega,\bar{\omega}}~
\sdet(F_{\bar{\omega}}^{\alpha_{0}},G_{\omega,\beta_{0}})_{\ext}}
is independent of \mb{F^{\alpha_{0}}_{\bar{\omega}}}
up to terms that vanish on-shell \wrt \mb{G_{\omega,\alpha_{0}}},
if the \mb{G_{\omega,\alpha_{0}}}'s are in involution \wrt the 
\mb{(\cdot,\cdot)_{\ext}} bracket.
\label{lemmaareduc0}
\end{lemma}

\noi
{\sc Proof of Theorem~\ref{theoremareduc}}:~~
We would like to derive an analogue of \eq{weakeq2} with \mb{G_{\alpha}}
replaced with the \mb{\omega}-deformed \mb{G_{\omega,\alpha_{0}}} in the
extended space \mb{\Gamma^{A_{\ext}}}. From the Master Equation, we are
provided with a reducible version \e{reduweakeq2}. Let us first contract
an index on the \mb{\omega}-deformed structure functions \e{uomega},
\bea
(-1)^{\epsilon_{\alpha_{0}}}
U^{\alpha_{0}}_{\omega,\alpha_{0}\beta_{0}}
&=& (-1)^{\epsilon_{\alpha_{0}}}
\left[\delta^{\alpha_{0}}_{\gamma_{0}}-
Z^{\alpha_{0}}{}_{\alpha_{1}}(\Delta^{-1})^{\alpha_{1}}{}_{\beta_{1}}
\omega^{\beta_{1}}{}_{\gamma_{0}}\right]
U^{\gamma_{0}}_{\alpha_{0}\beta_{0}} \cr
&& +(-1)^{\epsilon_{\alpha_{0}}}
Z^{\alpha_{0}}{}_{\alpha_{1}}(\Delta^{-1})^{\alpha_{1}}{}_{\beta_{1}}
(\omega^{\beta_{1}}{}_{\alpha_{0}},G_{\omega,\beta_{0}})_{\ext} 
+{\cal O}(G_{\omega})~,\label{uomega2}
\eea
where use has been made of \eq{Greduc}. Next one derives
\bea
(-1)^{\epsilon_{\alpha_{1}}}
U^{\alpha_{1}}_{\alpha_{1} \beta_{0}}
&=&(-1)^{\epsilon_{\alpha_{1}}}
(\Delta^{-1})^{\alpha_{1}}{}_{\beta_{1}}
\omega^{\beta_{1}}{}_{\alpha_{0}}
(Z^{\alpha_{0}}{}_{\alpha_{1}},G_{\omega,\beta_{0}})_{\ext}\cr
&&+(-1)^{\epsilon_{\alpha_{0}}}Z^{\alpha_{0}}{}_{\alpha_{1}}
(\Delta^{-1})^{\alpha_{1}}{}_{\beta_{1}}
\omega^{\beta_{1}}{}_{\gamma_{0}}
U^{\gamma_{0}}_{\alpha_{0}\beta_{0}}
+{\cal O}(G_{\omega})\label{omegaweakeq5}
\eea
from \eq{reduweakeq5}. Combining equation \e{reduweakeq2}, 
\es{uomega2}{omegaweakeq5}, one gets the sought-for equation
\beq
 (\Delta_{\ext} G_{\omega,\beta_{0}})
-\left(H-\ln\sdet(\Delta^{\alpha_{1}}{}_{\beta_{1}}),~
G_{\omega,\beta_{0}}\right)_{\ext}
~=~(-1)^{\epsilon_{\alpha_{0}}}
U^{\alpha_{0}}_{\omega,\alpha_{0}\beta_{0}}
+{\cal O}(G_{\omega})~.\label{omegaweakeq2}
\eeq
Besides working in the extended space \mb{\Gamma^{A_{\ext}}},
the only real difference from \eq{weakeq2} is that \mb{H} has been shifted
by the Faddeev-Popov determinant. Hence one may proceed as in the proof of
Theorem~\ref{theorema}.
\proofbox

\subsection{From Irreducible to Reducible Constraints}
\label{backtoreduc}

\noi
The partition function for reducible gauge-fixing constraints becomes
\beq
{\cal Z}_{\hl}^{G_{\omega}} ~=~ \int \! [d\Gamma][d \pi_{1}]~
e^{\Ih W}~ \delta(G_{\omega,\alpha_{0}})
 \left.\sqrt{\rho~J_{\omega,\bar{\omega}}~
\sdet( F_{\bar{\omega}}^{\alpha_{0}}, 
G_{\omega,\beta_{0}})_{\ext}}\right|_{\pi_{1}^{*}=0}
\label{1stlevelreduc}
\eeq
by combining the two previous results 
\es{partitionfunctionreduc}{khudaverdianreduc}.
Note that the Faddeev-Popov determinant have completely cancelled out 
from the partition function \e{1stlevelreduc}! This suggests that there is
a much simpler and broader approach as follows:  
If you want \mb{N_{0}} reducible gauge-fixing constraints in the original
\mb{2N}-dimensional zeroth-level phase space \mb{\Gamma^{A}}, 
then introduce first-level variable 
\mb{\{\lambda^{\alpha_{0}},\lambda^{*}_{\alpha_{0}};
\pi_{\alpha_{1}},\pi^{*\alpha_{1}}\}},
and an action
\beq
X~=~{\cal G}_{\alpha_{0}}\lambda^{\alpha_{0}}+(i\hbar_{\bl}){\cal H}
-\lambda^{*}_{\alpha_{0}}{\cal R}^{\alpha_{0}}
+{\cal O}\left((\lambda^{*})^2\right),
\eeq
where all the structure functions 
\mb{{\cal G}_{\alpha_{0}}={\cal G}_{\alpha_{0}}(\Gamma_{\ext};\hbar)},
\mb{{\cal H}={\cal H}(\Gamma_{\ext};\hbar)}, etc, are allowed to depend
on \mb{\pi_{\alpha_{1}}} and \mb{\pi^{*\alpha_{1}}} as well.
Hence the full constraints \mb{{\cal G}_{\alpha_{0}}} live in the 
extended \mb{\Gamma^{A_{\ext}}} space and should be irreducible, 
because of rank requirements, while the reducible constraints are simply
the restriction 
\beq
   G_{\alpha_{0}}~=~
\left. {\cal G}_{\alpha_{0}}\right|_{\pi_{1},\pi_{1}^{*}=0}
\eeq
into the original \mb{\Gamma^{A}} space. According to Theorem~\ref{theorema}
applied on the extended space \mb{\Gamma^{A_{\ext}}}, the Quantum Master
Equation for $X$ will carve out the square root measure factor directly,
\beq
 {\cal H} ~=~ -\ln\sqrt{\frac{{\cal J}~
\sdet( {\cal F}^{\alpha_{0}}, {\cal G}_{\beta_{0}} )_{\ext}}{\rho}}
+{\cal O}({\cal G})~,
\label{khudaverdianho}
\eeq
with \mb{{\cal J}\equiv\sdet (\frac{\partial \bar{\Gamma}^{A_{\ext}}}
{\partial \Gamma^{B_{\ext}}})} and 
\mb{\bar{\Gamma}^{A_{\ext}}\equiv
\{ {\cal F}^{\alpha_{0}};{\cal G}_{\alpha_{0}}\}}.
There is thus no need to introduce a Faddeev-Popov ghost pair 
\mb{\{\gh^{\alpha_{1}};\bgh_{\alpha_{1}}\}}.
Therefore by appealing to the first-level irreducible theory from
Subsection~\ref{secfirstlevelformalism}-\ref{Going}, one derives the 
following version of Corollary~\ref{corollaryb}, written in the so-called
\mb{\lambda^{*}_{\alpha_{0}}=0=\pi^{*\alpha_{1}}} gauge.

\begin{corollary}:~~The partition function
\beq
{\cal Z}_{\hl}^{{\cal G}} ~=~ \int \! [d\Gamma_{\ext}]~e^{\Ih W}~ 
\delta(\pi^{*\alpha_{1}})~ \delta({\cal G}_{\alpha_{0}}) \sqrt{\rho~{\cal J}~
\sdet( {\cal F}^{\alpha_{0}}, {\cal G}_{\beta_{0}})_{\ext}}
\label{1stlevelreduco}
\eeq
is independent of  \mb{{\cal G}_{\alpha_{0}}}'s that are in involution
\wrt the \mb{(\cdot,\cdot)_{\ext}}-bracket.
\label{corollarybreduco}
\end{corollary}

\noi
We conclude the following

\begin{theorem}
{\bf -- Reduction Theorem}:~~
The irreducible partition function \e{1stlevelreduco} on the 
\mb{2N_{0}}-dimensional extended space \mb{\Gamma^{A_{\ext}}}
reduces to the reducible partition function \e{1stlevelreduc}
on the \mb{2N}-dimensional original space \mb{\Gamma^{A}}, if
one chooses the gauge-fixing conditions \mb{{\cal G}_{\alpha_{0}}}
to be the \mb{\omega}-deformed constraints
\mb{ G_{\omega,\alpha_{0}}\equiv G_{\alpha_{0}}
+\pi_{\alpha_{1}}\omega^{\alpha_{1}}{}_{\alpha_{0}}}.
\label{theoremrr0}
\end{theorem}

\noi
In the next Subsection we will present a Reduction Theorem~\ref{theoremrr}
that in some respect is opposite to the above Reduction 
Theorem~\ref{theoremrr0}.

\subsection{From Reducible to Irreducible Constraints}
\label{backtoirr}

\noi
One may always assume that the reducible constraints
\mb{G_{\alpha_{0}}}, \mb{\alpha_{0}=1,\ldots,N_{0}}, can be written as
linear combinations of irreducible constraints 
\mb{G^{\prime}_{\alpha}}, \mb{\alpha=1,\ldots,N},
\beq
G_{\alpha_{0}} ~=~ G^{\prime}_{\alpha}P^{\alpha}{}_{\alpha_{0}}~,
\label{reducformirreduc}
\eeq
such that the irreducible constraints \mb{G^{\prime}_{\alpha}} are in
involution, cf.\ \eq{nonabelinvo}.
Therefore the theory can be set up purely within the irreducible
framework of Section~\ref{secirr}. In this Subsection we check that the
reducible and the irreducible approach agree.

\begin{theorem}
{\bf -- Reduction Theorem}:~~
The reducible partition function \e{1stlevelreduc} coincides with the
irreducible partition function \e{khudaverdianz}, when re-writing
the reducible quantities in their irreducible counterparts.
\label{theoremrr}
\end{theorem}

\noi
{\sc Proof}:~~First of all, let us note that the reducible gradients
\beq
(\papal{\Gamma^{A}}G_{\alpha_{0}})
 ~=~ (\papal{\Gamma^{A}}G^{\prime}_{\alpha})P^{\alpha}{}_{\alpha_{0}}
+{\cal O}(G^{\prime})
\eeq
are linear combinations of irreducible gradients on-shell due to 
\eq{reducformirreduc}. The rectangular matrix \mb{P^{\alpha}{}_{\alpha_{0}}}
has \mb{\rank(P^{\alpha}{}_{\alpha_{0}})=N}. It follows from 
\eqs{Greduc}{reducformirreduc} that there exists an antisymmetric matrix
\mb{A^{\alpha\beta}{}_{\alpha_{1}}
=-A^{\beta\alpha}{}_{\alpha_{1}}(-1)^{\epsilon_{\alpha}\epsilon_{\beta}}}
such that
\beq
 X^{\alpha}{}_{\beta_{1}}~\equiv~ 
P^{\alpha}{}_{\alpha_{0}}Z^{\alpha_{0}}{}_{\beta_{1}}
~=~G^{\prime}_{\beta}A^{\beta\alpha}{}_{\beta_{1}}~=~{\cal O}(G^{\prime})~.
\eeq
Because of the rank condition \e{fpdetrank}, one may combine
\mb{P^{\alpha}{}_{\alpha_{0}}} and 
\mb{\omega^{\alpha_{1}}{}_{\alpha_{0}}} into an invertible
\mb{N_{0}\times N_{0}} matrix
\beq
   \twobyone{P}{\omega}_{N_{0}\times N_{0}}~,
\eeq
at least in the vicinity of the constrained surface 
\mb{G^{\prime}_{\alpha}\approx 0}.
Next define a rectangular matrix \mb{\bar{P}^{\alpha_{0}}{}_{\alpha}} via
\beq
 \twobyone{P}{\omega}_{N_{0}\times N_{0}}\onebyone{\bar{P}}_{N_{0}\times N}
~=~\twobyone{{\bf 1}}{0}_{N_{0}\times N}~,
\eeq
and a square matrix \mb{R^{\alpha_{0}}{}_{\beta_{0}}} as
\beq
 \onebyone{R}_{N_{0}\times N_{0}}~\equiv~
\onebytwo{\bar{P}}{Z}_{N_{0}\times N_{0}}~.
\eeq 
It follows that
\beq
\twobyone{P}{\omega}_{N_{0}\times N_{0}} 
\onebyone{R}_{N_{0}\times N_{0}} 
~=~\twobytwo{\bf 1}{X}{0}{\Delta}_{N_{0}\times N_{0}}~,\label{rinv}
\eeq
cf.\ \e{fpmatrix}, so that \mb{R^{\alpha_{0}}{}_{\beta_{0}}} must be
invertible as well. 
Next change the coordinates
\mb{\lambda^{\alpha_{0}}=\bar{P}^{\alpha_{0}}{}_{\alpha}\lambda^{\prime\alpha}
+Z^{\alpha_{0}}{}_{\alpha_{1}}\lambda^{\prime\alpha_{1}}}, or equivalently,
\beq
\onebyone{\lambda^{\alpha_{0}}}_{N_{0}\times 1}
~=~\onebyone{R}_{N_{0}\times N_{0}}
\twobyone{\lambda^{\prime\alpha}}
{\lambda^{\prime\alpha_{1}}}_{N_{0}\times 1}~. 
\eeq
Therefore one can decompose the ``reducible'' \mb{\delta}-function 
\bea
\delta(G_{\omega,\alpha_{0}})&=&\int \! [d\lambda_{0}]~
e^{\Ih G_{\omega,\alpha_{0}}\lambda^{\alpha_{0}}}
~=~\sdet(R^{\alpha_{0}}{}_{\beta_{0}})\int \! [d\lambda^{\prime}]
[d\lambda^{\prime}_{1}]~e^{\Ih G^{\prime}_{\alpha}\lambda^{\prime\alpha}
+\pi_{\alpha_{1}}\Delta^{\alpha_{1}}{}_{\beta_{1}}
\lambda^{\prime\beta_{1}}}\cr
&=&\frac{\sdet(R^{\alpha_{0}}{}_{\beta_{0}})}
{\sdet(\Delta^{\alpha_{1}}{}_{\beta_{1}})}~
\delta(G^{\prime})~\delta(\pi_{1})
\label{deltaomega}
\eea
in its irreducible components.

\noi
There is an almost identical story for the reducible partners
\mb{F^{\alpha_{0}}}, \mb{\alpha_{0}=1,\ldots,N_{0}}, 
in terms of irreducible primed functions \mb{F^{\prime\alpha}}, 
\mb{\alpha=1,\ldots,N}, although with some important differences.
{}For instance, we shall assume directly that the reducible {\em gradients}
are linear combinations of irreducible gradients,
\beq
(F^{\alpha_{0}}\papar{\Gamma^{A}})
 ~=~ Q^{\alpha_{0}}{}_{\alpha}(F^{\prime\alpha}\papar{\Gamma^{A}})~.
\eeq
The rectangular matrix \mb{Q^{\alpha_{0}}{}_{\alpha}} has rank $N$.
This means there exists a matrix \mb{\bar{Z}^{\alpha_{1}}{}_{\alpha_{0}}}
of rank \mb{N_{1}} such that
\beq
\bar{Z}^{\alpha_{1}}{}_{\alpha_{0}}Q^{\alpha_{0}}{}_{\alpha}~=0~.
\eeq
Then \mb{\bar{\omega}^{\alpha_{0}}{}_{\alpha_{1}}} is chosen such that
\beq
 \bar{\Delta}^{\alpha_{1}}{}_{\beta_{1}}~\equiv~
 \bar{Z}^{\alpha_{1}}{}_{\alpha_{0}}\bar{\omega}^{\alpha_{0}}{}_{\beta_{1}}
\label{barfpmatrix}
\eeq
 is invertible, \ie
\beq
\rank\left(\bar{\Delta}^{\alpha_{1}}{}_{\beta_{1}}\right)~=~N_{1}~.
\label{barfpdetrank}
\eeq
One may combine \mb{Q^{\alpha_{0}}{}_{\alpha}} and 
\mb{\bar{\omega}^{\alpha_{0}}{}_{\alpha_{1}}} into an invertible
\mb{N_{0}\times N_{0}} matrix
\beq
 \onebytwo{Q}{\bar{\omega}}_{N_{0}\times N_{0}}.
\eeq
Next define a rectangular matrix \mb{\bar{Q}^{\alpha}{}_{\alpha_{0}}} via
\beq
\onebyone{\bar{Q}}_{N_\times N_{0}}
\onebytwo{Q}{\bar{\omega}}_{N_{0}\times N_{0}}
~=~\onebytwo{{\bf 1}}{0}_{N\times N_{0}}~,
\eeq
and a square matrix \mb{\bar{R}^{\alpha_{0}}{}_{\beta_{0}}} as
\beq
 \onebyone{\bar{R}}_{N_{0}\times N_{0}}~\equiv~
\twobyone{\bar{Q}}{\bar{Z}}_{N_{0}\times N_{0}}~. 
\eeq 
{}Finally change the coordinates
\bea
\onebyone{\bar{C}^{\alpha_{0}}}_{N_{0}\times 1}
&=&\onebyone{R}_{N_{0}\times N_{0}}
\twobyone{\bar{C}^{\prime\alpha}}
{\bar{C}^{\prime\alpha_{1}}}_{N_{0}\times 1}~, \cr 
\onebyone{\bar{C}_{\alpha_{0}}}_{1\times N_{0}}
&=&\onebytwo{\bar{C}^{\prime}_{\alpha}}
{\bar{C}^{\prime}_{\alpha_{1}}}_{1\times N_{0}}
\onebyone{\bar{R}}_{N_{0}\times N_{0}}~, \cr 
\onebyone{B^{\alpha_{0}}}_{N_{0}\times 1}
&=&\onebyone{R}_{N_{0}\times N_{0}}
\twobyone{B^{\prime\alpha}}{B^{\prime\alpha_{1}}}_{N_{0}\times 1}~, \cr
\onebyone{\bar{B}_{\alpha_{0}}}_{1\times N_{0}}
&=&\onebytwo{\bar{B}^{\prime}_{\alpha}}
{\bar{B}^{\prime}_{\alpha_{1}}}_{1\times N_{0}}
\onebyone{\bar{R}}_{N_{0}\times N_{0}}~. 
\eea
The two superdeterminant actions \es{actionc}{actionb} become 
\bea
S_{C}&=&(-1)^{\epsilon_{A}+1}C^{ A}
(\papal{\Gamma^{A}}G^{\prime}_{\alpha})\bar{C}^{\prime\alpha}
+(-1)^{\epsilon_{\alpha_{1}}+1}C_{\alpha_{1}}
\Delta^{\alpha_{1}}{}_{\beta_{1}} \bar{C}^{\prime\beta_{1}} \cr
&& +\bar{C}^{\prime}_{\alpha}(F^{\prime\alpha}\papar{\Gamma^{A}})C^{A}
+\bar{C}^{\prime}_{\alpha_{1}}
\bar{\Delta}^{\alpha_{1}}{}_{\beta_{1}}C^{\beta_{1}} 
+\bar{C}^{\prime}_{\alpha_{0}}\bar{R}^{\alpha_{0}}{}_{\beta_{0}}
\tilde{\omega}^{\beta_{0}\alpha_{1}}C_{\alpha_{1}}
+{\cal O}(G^{\prime};\pi_{1};\pi_{1}^{*})~,
\label{actionc2}
\eea
and
\bea
S_{B}&=&\bar{B}^{\prime}_{\alpha}
(F^{\prime\alpha},G^{\prime}_{\beta})B^{\prime\beta} 
-\bar{B}^{\prime}_{\alpha_{1}}\bar{\Delta}^{\alpha_{1}}{}_{\beta_{1}}
\Delta^{\beta_{1}}{}_{\gamma_{1}}B^{\prime\gamma_{1}}
+{\cal O}(G^{\prime};\pi_{1};\pi_{1}^{*})~,
\label{actionb2}
\eea
respectively. Integration over \mb{\bar{C}^{\prime\alpha_{1}}} yields a
\mb{\delta}-function \mb{\delta(C_{\alpha_{1}})}. Hence
one may drop the \mb{\tilde{\omega}^{\alpha_{0}\alpha_{1}}} term from
the \mb{S_{C}} action \e{actionc2}, so that the Jacobian 
\mb{J_{\omega,\bar{\omega}}} factorizes on-shell,
\beq
J_{\omega,\bar{\omega}}~=~J^{\prime}~
\frac{\sdet(\Delta^{\alpha_{1}}{}_{\beta_{1}})}
{\sdet(R^{\alpha_{0}}{}_{\beta_{0}})}~
\frac{\sdet(\bar{R}^{\alpha_{0}}{}_{\beta_{0}})}
{\sdet(\bar{\Delta}^{\alpha_{1}}{}_{\beta_{1}})}
+{\cal O}(G^{\prime};\pi_{1};\pi_{1}^{*})~,
\eeq
with \mb{J^{\prime}\equiv\sdet (\frac{\partial \bar{\Gamma}^{\prime A}}
{\partial \Gamma^{B}})} and 
\mb{\bar{\Gamma}^{\prime A}\equiv\{ F^{\prime\alpha};G^{\prime}_{\alpha}\}}.
Similarly,
\beq
\sdet( F_{\bar{\omega}}^{\alpha_{0}}, G_{\omega,\beta_{0}})_{\ext}
~=~\sdet( F^{\prime\alpha}, G^{\prime}_{\beta})~
\frac{\sdet(\Delta^{\alpha_{1}}{}_{\beta_{1}})}
{\sdet(R^{\alpha_{0}}{}_{\beta_{0}})}~
\frac{\sdet(\bar{\Delta}^{\alpha_{1}}{}_{\beta_{1}})}
{\sdet(\bar{R}^{\alpha_{0}}{}_{\beta_{0}})}
+{\cal O}(G^{\prime};\pi_{1};\pi_{1}^{*})~.
\eeq
Therefore formula \e{khudaverdianreduc} becomes
\beq
H ~=~-\ln \sqrt{\frac{J^{\prime}~\sdet( F^{\prime\alpha}, G^{\prime}_{\beta})}
{\rho~\sdet(R^{\alpha_{0}}{}_{\beta_{0}})^2}}
+ {\cal O}(G^{\prime})~, 
\label{khudaverdianreducreduc}
\eeq
and by combining equations \e{partitionfunctionreduc},
\es{deltaomega}{khudaverdianreducreduc},
one arrives at the irreducible partition function \e{khudaverdianz}
of Corollary~\ref{corollaryb}, \ie
\beq
{\cal Z}^{G^{\prime}}_{\hl} ~=~ \int \! [d\Gamma]~e^{\Ih W}\delta(G^{\prime})
\sqrt{\rho~J^{\prime}~\sdet(F^{\prime\alpha},G^{\prime}_{\beta})}~. 
\label{1stlevelreducreduc}
\eeq
\proofbox

\noi
In the notation of the above proof one may summarize the bijective
correspondence between the constraints 
\mb{\{G_{\omega,\alpha_{0}}\} \leftrightarrow 
\{G^{\prime}_{\alpha};\pi_{\alpha_{1}}\}} as follows:
\bea
 \onebyone{G_{\omega,\alpha_{0}}}_{1\times N_{0}}
&=&\onebytwo{G^{\prime}_{\alpha}}{\pi_{\alpha_{1}}}_{1\times N_{0}}
\twobyone{P}{\omega}_{N_{0}\times N_{0}}~, \\
\onebytwo{G^{\prime}_{\alpha}}{\pi_{\alpha_{1}}}_{1\times N_{0}}
&=&\onebyone{G_{\omega,\alpha_{0}}}_{1\times N_{0}}
\onebytwo{\bar{P}}{Z\Delta^{-1}}_{N_{0}\times N_{0}}~.
\eea
Other useful observations are
\bea
\onebyone{\bf 1}_{N_{0}\times N_{0}}
&=&\twobyone{P}{\omega}_{N_{0}\times N_{0}} 
\onebyone{R}_{N_{0}\times N_{0}} 
\twobytwo{\bf 1}{-X\Delta^{-1}}{0}{\Delta^{-1}}_{N_{0}\times N_{0}} \cr
&=&\twobyone{P}{\omega}_{N_{0}\times N_{0}} 
\onebytwo{\bar{P}}{(Z-\bar{P}X)\Delta^{-1}}_{N_{0}\times N_{0}} ~,
\eea
which follows from \eq{rinv}. Therefore one also has
\bea
\onebyone{\bf 1}_{N_{0}\times N_{0}}
&=&\onebytwo{\bar{P}}{(Z-\bar{P}X)\Delta^{-1}}_{N_{0}\times N_{0}}
\twobyone{P}{\omega}_{N_{0}\times N_{0}} \cr
&=&\onebyone{\bar{P}}_{N_{0}\times N} \onebyone{P}_{N\times N_{0}}
+\onebyone{(Z-\bar{P}X)\Delta^{-1}}_{N_{0}\times N_{1}}
\onebyone{\omega}_{N_{1}\times N_{0}}~,
\eea
and in particular
\beq
\onebyone{R^{-1}}_{N_{0}\times N_{0}}
~=~\twobyone{P-X\Delta^{-1}\omega}{\Delta^{-1}\omega}_{N_{0}\times N_{0}}~.
\eeq

\subsection{Non-Minimal Approach}
\label{secnonminappr}

\noi
With the standard non-minimal Ansatz \e{minansatz} it is mandatory to choose
a non-trivial gauge fermion \mb{\Psi\neq 0}. In this Subsection we shall 
study the most general non-minimal solution
\bea
X&=&{\cal G}_{A_{0}}~\lambda^{A_{0}}  + (i\hbar_{\bl})~{\cal H}
+\lambda^{*}_{A_{0}}~{\cal Z}^{A_{0}}{}_{\alpha_{1}}~\gh^{\alpha_{1}}
+\Hf\lambda^{*}_{A_{0}}~{\cal U}^{A_{0}}_{0~B_{0}C_{0}}~\lambda^{C_{0}}
\lambda^{B_{0}}(-1)^{\epsilon_{B_{0}}+1} \cr
&&+\gh^{*}_{\alpha_{1}}~{\cal U}^{\alpha_{1}}_{1~\beta_{1}A_{0}}~
\lambda^{A_{0}}\gh^{\beta_{1}}(-1)^{\epsilon_{\beta_{1}}}
+(i\hbar_{\bl})~\lambda^{*}_{A_{0}}~
{\cal V}^{A_{0}}{}_{B_{0}}~\lambda^{B_{0}}
+\ldots \label{reducXnonmin}~,
\eea
and therefore we may put \mb{\Psi=0} without loss of generality. In the
above \eq{reducXnonmin}, which should not be read as a systematic Planck
expansion, we have grouped together fields and antifields of the same Planck
number, \ie,
\beq
\Gamma^{A_{\ext}}\equiv\{\Gamma^{A};
\pi_{\alpha_{1}},\pi^{*\alpha_{1}}\}~,~~~
\lambda^{A_{0}}\equiv
\{\lambda^{\alpha_{0}}; -\bgh^{*\alpha_{1}}\}~,~~~ {\rm and} ~~~~
\lambda^{*}_{A_{0}}\equiv
\{\lambda^{*}_{\alpha_{0}};\bgh_{\alpha_{1}}\}~,\label{fieldgroup}
\eeq
of Planck number $0$, $1$ and $-1$, respectively. The minus in front of
\mb{\bgh^{*\alpha_{1}}} in \eq{fieldgroup} is introduced so that
\mb{(\lambda^{A_{0}},\lambda^{*}_{B_{0}})_{\hl}=\delta^{A_{0}}_{B_{0}}}.
The extended index \mb{A_{0}} runs over 
\mb{A_{0}\!=\!1, \ldots, N_{0}\!+\!N_{1}}.
It is interesting to see how the Quantum Master Equation determines the
one-loop correction  \mb{{\cal H}\!=\!{\cal H}(\Gamma_{\ext};\hbar)}
in this general case.
The first few consequences of the Quantum Master Equation read
\bea
 ({\cal G}_{A_{0}},{\cal G}_{B_{0}})_{\ext}
&=& {\cal G}_{C_{0}}~ {\cal U}^{C_{0}}_{0~A_{0} B_{0}}~,
\label{reducnonabelinvononmin} \\
{\cal G}_{A_{0}}{\cal Z}^{A_{0}}{}_{\alpha_{1}}&=&0~, \label{Gredunonmin} \\
({\cal Z}^{A_{0}}{}_{\beta_{1}},{\cal G}_{B_{0}})_{\ext}
&=&{\cal Z}^{A_{0}}{}_{\alpha_{1}}~{\cal U}^{\alpha_{1}}_{1~\beta_{1}B_{0}}
+(-1)^{(\epsilon_{B_{0}}+1)(\epsilon_{\beta_{1}}+1)}
{\cal U}^{A_{0}}_{0~B_{0}C_{0}}~{\cal Z}^{C_{0}}{}_{\beta_{1}}
+{\cal O}({\cal G})~,\label{reducweakeq5nonmin} \\
 (\Delta_{\ext} {\cal G}_{B_{0}})-({\cal H},{\cal G}_{B_{0}})_{\ext}
&=&(-1)^{\epsilon_{A_{0}}}{\cal U}^{A_{0}}_{0~A_{0}B_{0}}
-(-1)^{\epsilon_{\alpha_{1}}}{\cal U}^{\alpha_{1}}_{1~\alpha_{1}B_{0}}
 +{\cal G}_{A_{0}}~{\cal V}^{A_{0}}{}_{B_{0}}~. 
\label{reducweakeq2nonmin} 
\eea
The extended constraints 
\mb{{\cal G}_{A_{0}}\!=\!{\cal G}_{A_{0}}(\Gamma_{\ext};\hbar)}
and the extended generators 
\mb{{\cal Z}^{A_{0}}{}_{\beta_{1}}\!=\!
{\cal Z}^{A_{0}}{}_{\beta_{1}}(\Gamma_{\ext};\hbar)}
are both reducible sets of functions,
\beq
{\cal G}_{A_{0}}~\equiv~\{{\cal G}_{\alpha_{0}};{\cal G}_{\alpha_{1}} \}
~,~~~~~~~~~~~~~~
{\cal Z}^{A_{0}}{}_{\beta_{1}}~\equiv~
\{{\cal Z}^{\alpha_{0}}{}_{\beta_{1}};\Delta^{\alpha_{1}}{}_{\beta_{1}}\}~.
\eeq
Here \mb{\Delta^{\alpha_{1}}{}_{\beta_{1}}} is defined as the matrix from the
quadratic  \mb{\{\bgh_{\alpha_{1}};\gh^{\beta_{1}}\}} term in the action
\e{reducXnonmin}. A first-stage theory has a total of \mb{4(N_{0}\!+\!N_{1})}
fields and antifields, and we know that the Hessian for \mb{X} has half
rank on stationary field configurations. Putting all the antifields to zero,
the Hessian must have full rank \mb{(=2(N_{0}\!+\!N_{1}))} in the field-field
quadrant. Hence it follows that \mb{{\cal G}_{\alpha_{0}}} is irreducible and
\mb{{\Delta}^{\alpha_{1}}{}_{\beta_{1}}} has maximal rank. Equations 
\es{reducnonabelinvononmin}{Gredunonmin} therefore show that
\beq
{\cal G}_{\alpha_{1}} ~=~{\cal O}({\cal G}_{\alpha_{0}})
~,~~~~~~~~~~~~~~~~~~~~~~~~
{\cal Z}^{\alpha_{0}}{}_{\beta_{1}} ~=~{\cal O}({\cal G}_{\alpha_{0}})~,
\eeq
respectively. We  arrive at the following version of 
Lemma~\ref{lemmareducpathint}:

\begin{lemma}:~~
The path integrations over the Faddeev-Popov ghost pair 
\mb{\{\gh^{\alpha_{1}};\bgh_{\alpha_{1}}\}} can be performed explicitly. 
The first-level partition function \e{reduWXpathint} thereby simplifies to
\beq
{\cal Z}_{\hl}~=~\int \![d\Gamma][d\lambda_{0}][d\pi_{1}]~\rho~
e^{\Ih W+\Ih {\cal G}_{\alpha_{0}}\lambda^{\alpha_{0}}-{\cal H}}~
\sdet(\Delta^{\alpha_{1}}{}_{\beta_{1}})~.  
\label{partitionfunctionreduch}  
\eeq
\label{lemmareducpathinth}
\end{lemma}

\noi
One can also prove a version of Theorem~\ref{theorema}. For that purpose,
let \mb{\bar{\Gamma}^{A_{\ext}}\!\equiv\!
\{ {\cal F}^{\alpha_{0}}; {\cal G}_{\alpha_{0}}\}} and 
\mb{{\cal J}\!\equiv\!\sdet (\frac{\partial \bar{\Gamma}^{A_{\ext}}}
{\partial \Gamma^{B_{\ext}}})}.

\begin{theorem}:~~The Quantum Master Equation for the non-minimal \mb{X} in
\eq{reducXnonmin} implies that the quantum correction \mb{\cal H} is given
by the following square root formula 
\beq
{\cal H} ~=~- \ln \sqrt{\frac{{\cal J}~
    \sdet( {\cal F}^{\alpha_{0}}, {\cal G}_{\beta_{0}})_{\ext}}
{\rho~\sdet(\Delta^{\alpha_{1}}{}_{\beta_{1}})^2}}
+ {\cal O}({\cal G})~. 
\label{khudaverdianreduch}
\eeq
\label{theoremareduch}
\end{theorem}

\noi
{\sc Proof of Theorem~\ref{theoremareduch}}:~~
The \eq{reducweakeq5nonmin} can be rewritten as
\beq
( \ln \sdet(\Delta^{\alpha_{1}}{}_{\beta_{1}}),~{\cal G}_{B_{0}})_{\ext}
~=~(-1)^{\epsilon_{\alpha_{1}}}{\cal U}^{\alpha_{1}}_{1~\alpha_{1}B_{0}}
-(-1)^{\epsilon_{\alpha_{1}}}{\cal U}^{\alpha_{1}}_{0~\alpha_{1}B_{0}}
+{\cal O}({\cal G})~,
\eeq
which with the help of \eq{reducweakeq2nonmin} becomes
\beq
(\Delta_{\ext} {\cal G}_{B_{0}})
-\left({\cal H}- \ln \sdet(\Delta^{\alpha_{1}}{}_{\beta_{1}}),~
{\cal G}_{B_{0}}\right)_{\ext}
~=~(-1)^{\epsilon_{\alpha_{0}}}{\cal U}^{\alpha_{0}}_{0~\alpha_{0}B_{0}}
+{\cal O}({\cal G})~. 
\label{reducweakeq2nonminh} 
\eeq
Next one proceed as in the proof of Theorem~\ref{theorema}.
\proofbox

\noi
Equations \es{partitionfunctionreduch}{khudaverdianreduch} leads to  
\eq{1stlevelreduco} in Corollary~\ref{corollarybreduco}. So the non-minimal
approach agrees with the previous approach of Subsection~\ref{backtoreduc}.

\subsection{Higher-Stage Reducibility and Second-Class Constraints}
\label{finitestagedirac}

\begin{table} 
\caption{First-level fields in the general reducible case. Antifields are 
not shown. The last field in each row is a minimal field.} 
\label{firstleveltable}
\begin{center}
\begin{tabular}{|c||cccccccccccc|}  \hline
Grassm.&$\epsilon_{a_{5}}\!+\!1$&$\epsilon_{a_{4}}$&$\epsilon_{a_{3}}\!+\!1$&
$\epsilon_{a_{2}}$&$\epsilon_{a_{1}}\!+\!1$&$\epsilon_{a_{\pi}}$&
$\epsilon_{a_{\ext}}$&$\epsilon_{a_{1}}\!+\!1$&$\epsilon_{a_{2}}$&
$\epsilon_{a_{3}}\!+\!1$&$\epsilon_{a_{4}}$&$\epsilon_{a_{5}}\!+\!1$\\ \hline
$\pl$&$-5$&$-4$&$-3$&$-2$&$-1$&$0$&$1$&$2$&$3$&$4$&$5$&6 \\ \hline
&$\bgh_{a_{5}}$&$\bgh_{a_{4}}$&$\bgh_{a_{3}}$&$\bgh_{a_{2}}$&$\bgh_{a_{1}}$
&$\pi_{a_{\pi}}$&$\lambda^{a_{\ext}}$&$\gh^{a_{1}}$&$\gh^{a_{2}}$&
$\gh^{a_{3}}$&$\gh^{a_{4}}$&$\gh^{a_{5}}$ \\ \hline\hline
0&&&&&&&$\lambda^{\alpha_{0}}$&&&&& \\
1&&&&&$\bgh_{1,\alpha_{1}}$& $\pi_{\alpha_{1}}$&&$\gh_{1}^{\alpha_{1}}$&&&& \\
2&&&&$\bgh_{2,\alpha_{2}}$&$\bgh_{1,\alpha_{2}}$&&$\lambda^{\alpha_{2}}$&
$\gh_{1}^{\alpha_{2}}$&$\gh_{2}^{\alpha_{2}}$&&& \\
3&&&$\bgh_{3,\alpha_{3}}$&$\bgh_{2,\alpha_{3}}$&$\bgh_{1,\alpha_{3}}$&
$\pi_{\alpha_{3}}$&&$\gh_{1}^{\alpha_{3}}$&
$\gh_{2}^{\alpha_{3}}$&$\gh_{3}^{\alpha_{3}}$&& \\
4&&$\bgh_{4,\alpha_{4}}$&$\bgh_{3,\alpha_{4}}$&$\bgh_{2,\alpha_{4}}$&
$\bgh_{1,\alpha_{4}}$&&$\lambda^{\alpha_{4}}$&$\gh_{1}^{\alpha_{4}}$&
$\gh_{2}^{\alpha_{4}}$&$\gh_{3}^{\alpha_{4}}$&$\gh_{4}^{\alpha_{4}}$ &\\
5&$\bgh_{5,\alpha_{5}}$&$\bgh_{4,\alpha_{5}}$&$\bgh_{3,\alpha_{5}}$&
$\bgh_{2,\alpha_{5}}$&$\bgh_{1,\alpha_{5}}$&$\pi_{\alpha_{5}}$&&
$\gh_{1}^{\alpha_{5}}$&$\gh_{2}^{\alpha_{5}}$&$\gh_{3}^{\alpha_{5}}$&
$\gh_{4}^{\alpha_{5}}$&$\gh_{5}^{\alpha_{5}}$   \\
$\vdots$&$\vdots$&$\vdots$&$\vdots$&$\vdots$&$\vdots$&&&
$\vdots$&$\vdots$&$\vdots$&$\vdots$&$\vdots$ \\ \hline
Stage&\multicolumn{5}{c|}{Antighosts}&\multicolumn{2}{c|}{Lagr.~Mult.}&
\multicolumn{5}{c|}{Ghosts} \\ \hline
\end{tabular}
\end{center}
\end{table}

\noi
Adapting the reducible zeroth-level recipe of \Ref{BV83}
to a first-level theory of stage $s$, one introduces 
\mb{2\sum_{i=0}^{s}(2i\!+\!1)N_{i}} first-level fields and antifields as 
indicated in Table~\ref{firstleveltable}. It is useful to group together 
fields with the same Planck number assignment. Hence we write
\bea
\lambda^{a_{\ext}}&\equiv&\left\{\lambda^{\alpha_{0}};
\lambda^{\alpha_{2}};\lambda^{\alpha_{4}};\ldots\right\}~,\\
\pi_{a_{\pi}}&\equiv&\left\{\pi_{\alpha_{1}};\pi_{\alpha_{3}};
\pi_{\alpha_{5}}; \ldots\right\}~,\\
\gh^{a_{i}}&\equiv&\left\{\gh_{i}^{\alpha_{i}};\gh_{i}^{\alpha_{i+1}};
\gh_{i}^{\alpha_{i+2}};\ldots;\gh_{i}^{\alpha_{s}}\right\}
~,~~~~~~~~~~~~~~~~~~~~~~~~i=1,\ldots,s~,\\
\bgh_{a_{i}}&\equiv&\left\{\bgh_{i,\alpha_{i}};\bgh_{i,\alpha_{i+1}};
\bgh_{i,\alpha_{i+2}};\ldots;\bgh_{i,\alpha_{s}}\right\}
~,~~~~~~~~~~~~~~~~~~~i=1,\ldots,s~.
\eea
The total space is then
\bea
\Gamma^{A}_{\hl}&\equiv&\left\{\Gamma^{A_{\ext}};
\lambda^{a_{\ext}},\lambda^{*}_{a_{\ext}};
\gh^{a_{1}},\gh^{*}_{a_{1}},\bgh_{a_{1}},\bgh^{*a_{1}}; 
\gh^{a_{2}},\gh^{*}_{a_{2}},\bgh_{a_{2}},\bgh^{*a_{2}};  \ldots;
\gh^{a_{s}},\gh^{*}_{a_{s}},\bgh_{a_{s}},\bgh^{*a_{s}}\right\}~, \\
\Gamma^{A_{\ext}}&\equiv&\left\{\Gamma^{A};
\pi_{a_{\pi}},\pi^{*a_{\pi}}\right\}~.
\eea
Here the indices run
\beq
\begin{array}{rcllcrcll}
A_{\ext}&=&1,\ldots,2M_{\rm odd}&,&~~~~~~& 
a_{\pi}&=&1,\ldots,M_{\rm odd}-N&,  \\
a_{\ext}&=&1,\ldots,M_{\rm even}&,&&
a_{i}&=&1,\ldots,M_{s}-M_{i-1}&,
\end{array}
\eeq
where 
\beq 
\begin{array}{rcllrcllrcll}
M_{\rm odd}&\equiv&\sum_{\stacktwo{i=-1,\ldots,s}{i~\odd}}N_{i}&,& 
M_{\rm even}&\equiv&\sum_{\stacktwo{i=0,\ldots,s}{i~\even}}N_{i}&,& 
M_{\rm odd}&=&M_{\rm even}+N_{D}&, \\ \\
M_{i}&\equiv& \sum_{j=-1}^{i}N_{j}&,&N_{-1}&\equiv&N~. 
\end{array}
\eeq
The Planck-number operator is
\beq
\pl~=~-\left(\lambda^{a_{\ext}}\lambda^{*}_{a_{\ext}}
+\sum_{i=1}^{s}[(i\!+\!1) \gh^{*}_{a_{i}}\gh^{a_{i}}
-i \bgh_{a_{i}}\bgh^{*a_{i}}],~
\cdot~\right)_{\hl}+\hbar_{\bl} \frac{\partial }{\partial \hbar_{\bl}}~,
\eeq
or equivalently,
\beq
\begin{array}{rclcrclcrcl}
\pl(\lambda^{a_{\ext}})&=&1 &,&
\pl(\gh^{a_{i}})&=&i+1 &,&
\pl(\bgh_{a_{i}})&=&-i~, \\ \\
\pl(\lambda^{*}_{a_{\ext}})&=&-1 &,&
\pl(\gh^{*}_{a_{i}})&=&-(i+1) &,&
\pl(\bgh^{*a_{i}})&=&i~,  \\ \\
\pl(\Gamma^{A_{\ext}})&=&0 &,&
\pl(\hbar_{\bl})&=&1 &,&
\pl(\hbar)&=&0~.
\end{array}
\eeq
The standard Ansatz for the non-minimal action is
\bea
X_{D}&=&X_{D,\min}(\Gamma^{A}_{\hl\min};\hbar_{\hl})
+\sum_{\stacktwo{j=1,\ldots,s}{j~\odd}} 
\pi_{\alpha_{j}}~\bgh_{1}^{*\alpha_{j}}
+\sum_{\stacktwo{i,j=2,\ldots,s}{j-i~\even,\geq 0}}
\bgh_{i-1,\alpha_{j}}~\bgh_{i}^{*\alpha_{j}}  \cr \cr \cr
&&~~~~~~~~~~~~~~~~~~~~~~~~~~~
+\sum_{\stacktwo{j=2,\ldots,s}{j~\even}} 
\lambda^{*}_{\alpha_{j}}~\gh_{1}^{\alpha_{j}} 
+\sum_{\stacktwo{i,j=2,\ldots,s}{j-i~\odd,>0}} 
\gh^{*}_{i-1,\alpha_{j}}~\gh_{i}^{\alpha_{j}}~.
\label{reduminansatz}
\eea

\begin{table} 
\caption{The simplest choice of a rectangular gauge-fixing matrix 
\mb{\omega_{i}^{a_{i}}{}_{b_{i-1}}}, \mb{i=1,\ldots,s}.} 
\label{omegatable}
\begin{center}
\begin{tabular}{|c||cccccccc|} \hline
$a_{i}\backslash b_{i-1}$&$\beta_{i-1}$&$\beta_{i}$~&$\beta_{i+1}$
&$\beta_{i+2}$&$\beta_{i+3}$&$\beta_{i+4}$&$\cdots$&$\beta_{s}$
\\ \hline\hline
\rule[-2ex]{0ex}{5ex}$\alpha_{i}$&$\omega_{i}^{\alpha_{i}}{}_{\beta_{i-1}}$&&
$\omega_{i}^{\alpha_{i}}{}_{\beta_{i+1}}$&&&&& \\ 
\rule[-2ex]{0ex}{5ex}$\alpha_{i+1}$&&&&&&&&\\ 
\rule[-2ex]{0ex}{5ex}$\alpha_{i+2}$&&&
$\omega_{i}^{\alpha_{i+2}}{}_{\beta_{i+1}}$&&
$\omega_{i}^{\alpha_{i+2}}{}_{\beta_{i+3}}$&&&\\ 
\rule[-2ex]{0ex}{5ex}$\alpha_{i+3}$&&&&&&&&\\ 
\rule[-2ex]{0ex}{5ex}$\alpha_{i+4}$&&&&&
$\omega_{i}^{\alpha_{i+4}}{}_{\beta_{i+3}}$&&$\ddots$& \\
\rule[-2ex]{0ex}{5ex}$\vdots$&&&&&&&& \\ 
\rule[-2ex]{0ex}{5ex}$\alpha_{s}$&&&&&&&$\ddots$&
\\ \hline
\end{tabular}
\end{center}
\end{table}

\noi
A choice of the gauge fermion reads
\beq
-\Psi~=~\bgh_{a_{1}}~\omega_{1}^{a_{1}}{}_{a_{\ext}}~\lambda^{a_{\ext}}
+\sum_{i=2}^{s}\bgh_{a_{i}}~\omega_{i}^{a_{i}}{}_{a_{i-1}}~\gh^{a_{i-1}}
~,~~~~~~~~~~~~~~~~~\pl(\Psi)~=~0~,\label{reduPsi}
\eeq
where a simple choice for the matrices \mb{\omega_{i}^{a_{i}}{}_{a_{i-1}}},
\mb{i=1,\ldots,s}, is indicated in Table~\ref{omegatable}.
{}For notational reasons it is convenient to trivially extend the matrix
\mb{\omega_{1}^{a_{1}}{}_{a_{\ext}}\to\omega_{1}^{a_{1}}{}_{a_{0}}}
with zero columns such that the column index reads 
\mb{a_{0}\equiv\{ \alpha_{0};\alpha_{1};\alpha_{2};\alpha_{3};\ldots\}}
rather than 
\mb{a_{\ext}\equiv\{ \alpha_{0};\alpha_{2};\alpha_{4};\alpha_{6};\ldots\}}.
The first-level partition function is given by 
\beq
{\cal Z}^{\Psi}_{\hl D} 
~=~ \int \! d\mu_{D} \left. e^{\Ih (W_{D} + X_{D}^{\Psi})}
\right|_{\lambda^{*},\gh^{*},\bgh^{*},\pi^{*}=0}\delta(\Theta^{a})
~=~ \int \! d\mu_{D} \left. e^{\Ih (W_{D}^{\Psi} + X_{D})} \right|_{\Sigma}
\delta(\Theta^{a})~,
\label{reduWXpathintstage}
\eeq
with a measure 
\beq
d\mu_{D} ~=~ \rho_{D} [d\Gamma] [d\lambda] [d\gh] [d\bgh] [d\pi]~,
\eeq
and a gauge-fixing surface \mb{\Sigma} specified by
\bea
\lambda^{*}_{a_{\ext}}
&=&-E(\ad_{D}\Psi)\frac{\partial \Psi}{\partial\lambda^{a_{\ext}}}
~=~\bgh_{a_{1}}E(\ad_{D}\Psi)\omega_{1}^{a_{1}}{}_{a_{\ext}}~,
\label{lambda0stars} \\
\gh^{*}_{a_{i}}
&=&-E(\ad_{D}\Psi)\frac{\partial \Psi}{\partial\gh^{a_{i}}}~=~
\bgh_{a_{i+1}}E(\ad_{D}\Psi)\omega_{i+1}^{a_{i+1}}{}_{a_{i}}
~,~~~~~~~~~~~~~~~~i=1,\ldots,s-1~, \\
\gh^{*}_{a_{s}}
&=&-E(\ad_{D}\Psi)\frac{\partial \Psi}{\partial\gh^{a_{s}}}~=~0~,
\label{lambda1stars} \\
\bgh^{*a_{1}}&=&-E(\ad_{D}\Psi)\frac{\partial \Psi}{\partial \bgh_{a_{1}}}
~=~E(\ad_{D}\Psi)\omega_{1}^{a_{1}}{}_{a_{\ext}}~\lambda^{a_{\ext}} ~, \\
\bgh^{*a_{i}}&=&-E(\ad_{D}\Psi)\frac{\partial \Psi}{\partial \bgh_{a_{i}}}
~=~E(\ad_{D}\Psi)\omega_{i}^{a_{i}}{}_{a_{i-1}}~\gh^{a_{i-1}}
~,~~~~~~~~~~~~~~~~i=2,\ldots,s~, \\ 
\pi^{*a_{\pi}}
&=&-E(\ad_{D}\Psi)\frac{\partial \Psi}{\partial\pi_{a_{\pi}}}~=~0~,
\label{reduantifields}
\eea
cf.\ the prescription \e{gf1}. 

\noi
Similar to the first-stage case the generators 
\mb{\mb{Z_{i}^{\alpha_{i-1}}{}_{\beta_{i}}}} and the 
gauge-fixing matrices \mb{\omega_{i}^{a_{i}}{}_{b_{i-1}}} give rise
to Faddeev-Popov matrices \mb{\Delta_{i}^{a_{i}}{}_{b_{i}}}
as indicated in Table~\ref{deltatable}.
The pertinent rank conditions to the gauge-fixing matrices 
\mb{\omega_{i}^{a_{i}}{}_{b_{i-1}}} are such that the 
sequence of matrices \mb{\Delta^{a_{i}}{}_{b_{i}}}, \mb{i=1,\ldots,s}, 
becomes invertible. Explicit rank conditions for the 
\mb{\omega_{i}^{a_{i}}{}_{b_{i-1}}} matrices for theories of stage $1$, $2$ 
and $3$ can be found in the original paper \cite{BV83}.

\begin{table} 
\caption{A square \mb{\Delta_{i}^{a_{i}}{}_{b_{i}}} matrix
corresponding to the gauge-fixing matrices 
\mb{\omega_{i}^{a_{i}}{}_{b_{i-1}}} of Table~\ref{omegatable}.
The entry corresponding to the first row and the first column is 
\mb{\Delta_{i}^{\alpha_{i}}{}_{\beta_{i}}\equiv
\omega_{i}^{\alpha_{i}}{}_{\alpha_{i-1}}Z_{i}^{\alpha_{i-1}}{}_{\beta_{i}}}.}
\label{deltatable}
\begin{center}
\begin{tabular}{|c||ccccccc|} \hline
$a_{i}\backslash b_{i}$&$\beta_{i}$&$\beta_{i+1}$~&$\beta_{i+2}$
&$\beta_{i+3}$&$\beta_{i+4}$&$\cdots$&$\beta_{s}$ \\ \hline\hline
\rule[-2ex]{0ex}{5ex}$\alpha_{i}$&$\Delta_{i}^{\alpha_{i}}{}_{\beta_{i}}$&
$\omega_{i}^{\alpha_{i}}{}_{\beta_{i+1}}$&&&&& \\ 
\rule[-2ex]{0ex}{5ex}$\alpha_{i+1}$&
$\omega_{i+1}^{\alpha_{i+1}}{}_{\beta_{i}}$&&
$\omega_{i+1}^{\alpha_{i+1}}{}_{\beta_{i+2}}$&&&&\\ 
\rule[-2ex]{0ex}{5ex}$\alpha_{i+2}$&&
$\omega_{i}^{\alpha_{i+2}}{}_{\beta_{i+1}}$&&
$\omega_{i}^{\alpha_{i+2}}{}_{\beta_{i+3}}$&&&\\
\rule[-2ex]{0ex}{5ex}$\alpha_{i+3}$&&&
$\omega_{i+1}^{\alpha_{i+3}}{}_{\beta_{i+2}}$&&
$\omega_{i+1}^{\alpha_{i+3}}{}_{\beta_{i+4}}$&&\\ 
\rule[-2ex]{0ex}{5ex}$\alpha_{i+4}$&&&&
$\omega_{i}^{\alpha_{i+4}}{}_{\beta_{i+3}}$&&&\\
\rule[-2ex]{0ex}{5ex}$\vdots$&&&&&&&$\ddots$ \\ 
\rule[-2ex]{0ex}{5ex}$\alpha_{s}$&&&&&&$\ddots$&
\\ \hline
\end{tabular}
\end{center}
\end{table}

\begin{lemma}:~~
The path integrations over the ghost pairs 
\mb{\{\gh^{a_{1}};\bgh_{a_{1}};\ldots;\gh^{a_{s}};\bgh_{a_{s}} \}} 
can be performed explicitly. The first-level partition function thereby 
simplifies to
\beq
{\cal Z}_{D\hl}~=~\int \![d\Gamma][d\lambda][d\pi]~\rho_{D}~
e^{\Ih W_{D}+\Ih  G_{\omega,a_{\ext}}\lambda^{a_{\ext}}-H}~\delta(\Theta^{a})~
\prod_{i=1}^{s} \sdet(\Delta^{a_{i}}{}_{b_{i}})^{-(-1)^{i}}~,
\label{partitionfunctionhighreduc}  
\eeq
where \mb{G_{\omega,a_{\ext}}} are the \mb{\omega}-deformed
gauge-fixing constraints,
\bea
G_{\omega,a_{\ext}}&\equiv& \{G_{\omega,\alpha_{0}};G_{\omega,\alpha_{2}};
G_{\omega,\alpha_{4}};G_{\omega,\alpha_{6}};\ldots \}~, \\
G_{\omega,\alpha_{0}}&\equiv&G_{\alpha_{0}}
+\pi_{\alpha_{1}}~\omega_{1}^{\alpha_{1}}{}_{\alpha_{0}}~, \\
G_{\omega,\alpha_{i}}&\equiv&
\pi_{\alpha_{i+1}}~\omega_{1}^{\alpha_{i+1}}{}_{\alpha_{i}}
~,~~~~~~~~~~~~~~~~~~i=2,\ldots,s~,~~i~\even~,
\eea
and \mb{G_{\alpha_{0}}} are the original reducible constraints.
\label{lemmahighreducpathint}
\end{lemma}

\noi
{\sc Sketched Proof of Lemma~\ref{lemmahighreducpathint}}:~~
The gauge-fixed action 
\beq
S~=~(\frac{\hbar_{\bl}}{\hbar}W_{D}^{\Psi}+X_{D})|_{\Sigma}
~=~S_{0}+S_{FP}+V~,
\eeq
splits into a ``constant'' part
\beq 
S_{0}~\equiv~\frac{\hbar_{\bl}}{\hbar}W_{D}
+G_{\omega,a_{\ext}}\lambda^{a_{\ext}} 
+(i\hbar_{\bl})H
\eeq
that is independent of the ghosts and antighosts 
\mb{\{\gh^{a_{i}};\bgh_{a_{i}}\}}, a Faddeev-Popov part
\beq
S_{FP}~\equiv~\sum_{i=1}^{s}\bgh_{a_{i}}~
\Delta_{i}^{a_{i}}{}_{b_{i}}~\gh^{b_{i}}
\eeq
that is  quadratic in \mb{\{\gh^{a_{i}};\bgh_{a_{i}}\}}, and an
part $V$ that contains all interaction terms, tadpole terms and terms
quadratic in the antighosts \mb{\bgh_{a_{i}}}. The multiplicity condition 
that generalizes the rule \e{ruleofthumb} states, that for each term in $V$ 
there exists an  \mb{i=1, \ldots, s}, such that there are fewer ghosts 
\mb{\gh^{a_{i}}} than antighosts \mb{\bgh_{a_{i}}}. 
\proofbox
Next, let \mb{J_{D,\omega,\bar{\omega}}\equiv\sdet (\frac{\partial
\bar{\Gamma}^{A_{\ext}}_{\omega,\bar{\omega}}}
{\partial \Gamma^{B_{\ext}}})},
\mb{\bar{\Gamma}^{A_{\ext}}_{\omega,\bar{\omega}}\equiv
\{ F_{\bar{\omega}}^{a_{\ext}}; G_{\omega,a_{\ext}};\Theta^{a}\}} and
\mb{\Gamma^{A_{\ext}}\equiv\{
\Gamma^{A};\pi_{a_{\pi}},\pi^{*a_{\pi}} \}}. Then

\begin{theorem}:~~The Quantum Master Equation for \mb{X_{D,\min}} implies
that the quantum correction $H$ is given by the following square root formula
\beq
H ~=~- \ln\left[\left. \sqrt{\frac{J_{D,\omega,\bar{\omega}}~
\sdet( F_{\bar{\omega}}^{a_{\ext}}, G_{\omega,b_{\ext}})_{D,\ext}}
{\rho_{D}}}\right|_{\pi,\pi^{*}=0}
\prod_{i=1}^{s} \sdet(\Delta^{a_{i}}{}_{b_{i}})^{(-1)^{i}}\right]
+ {\cal O}(G_{\alpha_{0}};\Theta)~. 
\label{khudaverdianreducdirac}
\eeq
\label{theoremareducdirac}
\end{theorem}
In addition, let \mb{{\cal J}_{D}\equiv\sdet 
(\frac{\partial \bar{\Gamma}^{A_{\ext}}}
{\partial \Gamma^{B_{\ext}}})} and
\mb{\bar{\Gamma}^{A_{\ext}}\equiv
\{{\cal F}^{a_{\ext}};{\cal G}_{a_{\ext}};\Theta^{a}\}}. Then
\begin{corollary}:~~The partition function of stage \mb{s}
\beq
{\cal Z}_{D \hl}^{{\cal G}} ~=~ \int \! [d\Gamma_{{\ext}}]~e^{\Ih W_{D}}~
\delta(\pi^{*a_{\pi}})~ \delta(\Theta^{a})~
\delta({\cal G}_{a_{\ext}}) 
\sqrt{\rho_{D}~{\cal J}_{D}~\sdet( {\cal F}^{a_{\ext}}, 
{\cal G}_{b_{\ext}})_{D,\ext}}
\label{finitestagediraco}
\eeq
is independent of \mb{{\cal G}_{a_{\ext}}}'s that are in involution
\wrt the \mb{(\cdot,\cdot)_{D,\ext}}-bracket.
\label{corollarybfinitestagediraco}
\end{corollary}

\noi
Note that the alternating product of superdeterminants \cite{schwarz78}
has cancelled out of the final expression. A proof of 
Theorem~\ref{theoremareducdirac} will appear elsewhere.

\subsection{Non-Minimal Approach for Higher-Stages}
\label{secnonminapprs}

\noi
We consider the most general non-minimal solution
\bea
X_{D}&=&{\cal G}_{A_{0}}~\lambda^{A_{0}}  + (i\hbar_{\bl})~{\cal H}
+\lambda^{*}_{A_{0}}~{\cal Z}^{A_{0}}{}_{A_{1}}~\gh^{A_{1}}
+\sum_{i=1}^{s-1} \gh^{*}_{A_{i}}~Z^{A_{i}}{}_{A_{i+1}}~\gh^{A_{i+1}} \cr
&&+\Hf\lambda^{*}_{A_{0}}~{\cal U}^{A_{0}}_{0~B_{0}C_{0}}~\lambda^{C_{0}}
\lambda^{B_{0}}(-1)^{\epsilon_{B_{0}}+1}
+\sum_{i=1}^{s}\gh^{*}_{A_{i}}~{\cal U}^{A_{1}}_{i~B_{i}C_{0}}~
\lambda^{C_{0}}\gh^{B_{i}}(-1)^{\epsilon_{B_{i}}+i+1} \cr
&&+(i\hbar_{\bl})~\lambda^{*}_{A_{0}}~
{\cal V}^{A_{0}}{}_{B_{0}}~\lambda^{B_{0}}
+\ldots~. \label{reducXnonmins}
\eea
In the above \eq{reducXnonmins} we have grouped together fields of the same
Planck number,
\ie,
\beq
\begin{array}{rclcrclcrcl}
\lambda^{A_{0}}&\equiv&\twobyone{\lambda^{a_{\ext}}}{ -\bgh^{*a_{1}}}&,&
\gh^{A_{i}}&\equiv&\twobyone{\gh^{a_{i}}}{ -\bgh^{*a_{i+1}}}&,&
\gh^{A_{s}}&\equiv&\onebyone{\gh^{a_{s}}}~, \\ \\
\lambda^{*}_{A_{0}}&\equiv&\twobyone{\lambda^{*}_{a_{\ext}}}{\bgh_{a_{1}}}&,&
\gh^{*}_{A_{i}}&\equiv&\twobyone{\gh^{*}_{a_{i}}}{\bgh_{a_{i+1}}}&,&
\gh^{*}_{A_{s}}&\equiv&\onebyone{\gh^{*}_{a_{s}}}~, \\
&&&&&&i=1,\ldots,s-1&~.
\end{array}
\label{fieldgroups} 
\eeq
The first few consequences of the Quantum Master Equation read
\bea
 ({\cal G}_{A_{0}},{\cal G}_{B_{0}})_{D,\ext}
&=& {\cal G}_{C_{0}}~ {\cal U}^{C_{0}}_{0~A_{0} B_{0}}~,
\label{reducnonabelinvononmins} \\
{\cal G}_{A_{0}}{\cal Z}^{A_{0}}{}_{A_{1}}&=&0~, \label{Gredunonmins} \\
{\cal Z}^{A_{i-1}}{}_{A_{i}}
{\cal Z}^{A_{i}}{}_{A_{i+1}}&=&{\cal O}({\cal G})~, \label{Zreducs} \\
({\cal Z}^{A_{i}}{}_{B_{i+1}},{\cal G}_{B_{0}})_{D,\ext}
&=&{\cal Z}^{A_{i}}{}_{A_{i+1}}~{\cal U}^{A_{i+1}}_{i+1~B_{i+1}B_{0}} \cr
&&-(-1)^{(\epsilon_{B_{0}}+1)(\epsilon_{C_{i}}+\epsilon_{B_{i+1}})}
{\cal U}^{A_{i}}_{i~C_{i}B_{0}}~{\cal Z}^{C_{i}}{}_{B_{i+1}}
+{\cal O}({\cal G})~,\label{reducweakeq5nonmins} \\
 (\Delta_{D,\ext} {\cal G}_{B_{0}})-({\cal H},{\cal G}_{B_{0}})_{D,\ext}
&=&\sum_{i=0}^{s}(-1)^{\epsilon_{A_{i}}+i}{\cal U}^{A_{i}}_{i~A_{i}B_{0}}
 +{\cal G}_{A_{0}}~{\cal V}^{A_{0}}{}_{B_{0}}~. 
\label{reducweakeq2nonmins} 
\eea
Ignoring at first the second-class constraints, the Hessian for \mb{X} has
half rank on stationary field configurations. Putting all the antifields to
zero, the Hessian must have full rank in the field-field quadrant. It follows
that the other three quadrants of the Hessian must vanish on stationary field
configurations. In the presence of second-class constraints the same 
reasoning can be used in the physical subsector. We conclude that the extended
constraints \mb{{\cal G}_{A_{0}}\!=\!{\cal G}_{A_{0}}(\Gamma_{\ext};\hbar)}
and the extended generators 
\mb{{\cal Z}^{A_{i-1}}{}_{B_{i}}\!=\!
{\cal Z}_{i}^{A_{i-1}}{}_{B_{i}}(\Gamma_{\ext};\hbar)}
are both reducible sets of functions,
\beq
\begin{array}{rclcrclcrcl}
{\cal G}_{A_{0}}&\equiv&\onebytwo{{\cal G}_{a_{\ext}}}
{{\cal G}_{a_{1}}}&,&
{\cal Z}^{A_{i-1}}{}_{B_{i}}&=& 
\twobytwo{{\cal O}({\cal G};\Theta)}{{\cal O}({\cal G};\Theta)}
{\Delta^{a_{i}}{}_{b_{i}}}{{\cal O}({\cal G};\Theta)}&,&
{\cal Z}^{A_{s-1}}{}_{B_{s}}&=& 
\twobyone{{\cal O}({\cal G};\Theta)}{\Delta^{a_{s}}{}_{b_{s}}}~, \\
{\cal G}_{a_{1}}&=&{\cal O}({\cal G}_{a_{\ext}};\Theta)&,&&&i=1,\ldots,s-1&~.
\end{array}
\eeq
Here \mb{\Delta^{a_{i}}{}_{b_{i}}} is defined as the matrix from the
quadratic  \mb{\{\bgh_{a_{i}};\gh^{b_{i}}\}} term in the action
\e{reducXnonmins}. Hence it follows that \mb{{\cal G}_{a_{\ext}}} is 
irreducible and \mb{{\Delta}^{a_{i}}{}_{b_{i}}}, \mb{i=1,\ldots,s}, have 
maximal rank. We arrive at the following version of 
Lemma~\ref{lemmahighreducpathint}:

\begin{lemma}:~~
The path integrations over the ghost pairs 
\mb{\{\gh^{a_{1}};\bgh_{a_{1}};\ldots;\gh^{a_{s}};\bgh_{a_{s}} \}}
can be performed explicitly. The first-level partition function thereby 
simplifies to
\beq
{\cal Z}_{D\hl}~=~\int \![d\Gamma][d\lambda][d\pi]~\rho_{D}~
e^{\Ih W_{D}+\Ih  {\cal G}_{a_{\ext}}\lambda^{a_{\ext}}-{\cal H}}~
\delta(\Theta^{a})~
\prod_{i=1}^{s} \sdet(\Delta^{a_{i}}{}_{b_{i}})^{-(-1)^{i}}~.
\label{partitionfunctionhighreduch}  
\eeq
\label{lemmahighreducpathinth}
\end{lemma}

\noi
One can also prove a version of Theorem~\ref{theoremareducdirac}. For that
purpose, let \mb{{\cal J}_{D}\equiv\sdet 
(\frac{\partial \bar{\Gamma}^{A_{\ext}}}
{\partial \Gamma^{B_{\ext}}})},
\mb{\bar{\Gamma}^{A_{\ext}}\equiv
\{{\cal F}^{a_{\ext}}; {\cal G}_{a_{\ext}};\Theta^{a}\}} and
\mb{\Gamma^{A_{\ext}}\equiv\{
\Gamma^{A};\pi_{a_{\pi}},\pi^{*a_{\pi}} \}}. Then

\begin{theorem}:~~The Quantum Master Equation for the non-minimal \mb{X_{D}}
in \eq{reducXnonmins} implies that the quantum correction
\mb{{\cal H}\!=\!{\cal H}(\Gamma_{\ext};\hbar)} is given by the following
square root formula
\beq
{\cal H} ~=~- \ln\left[\sqrt{\frac{{\cal J}_{D}~
\sdet( {\cal F}^{a_{\ext}}, {\cal G}_{b_{\ext}})_{D,\ext}}{\rho_{D}}}
\prod_{i=1}^{s} \sdet(\Delta^{a_{i}}{}_{b_{i}})^{(-1)^{i}}\right]
+ {\cal O}({\cal G};\Theta)~. 
\label{khudaverdianreducdirach}
\eeq
\label{theoremareducdirach}
\end{theorem}

\noi
{\sc Proof of Theorem~\ref{theoremareducdirach}}:~~
Eq. (\ref{reducweakeq5nonmins}) can be rewritten as
\beq
( \ln \sdet(\Delta^{a_{i}}{}_{b_{i}}),~{\cal G}_{B_{0}})_{D,\ext}
~=~(-1)^{\epsilon_{a_{i}}}{\cal U}^{a_{i}}_{i~a_{i}B_{0}}
-(-1)^{\epsilon_{a_{i}}}{\cal U}^{a_{i}}_{i-1~a_{i}B_{0}}
+{\cal O}({\cal G})~,
\eeq
which with the help of \eq{reducweakeq2nonmins} becomes
\beq
(\Delta_{D,\ext} {\cal G}_{B_{0}})
-\left({\cal H}+ \sum_{i=1}^{s}(-1)^{i}\ln \sdet(\Delta^{a_{i}}{}_{b_{i}}),~
{\cal G}_{B_{0}}\right)_{D,\ext}
~=~(-1)^{\epsilon_{a_{\ext}}}{\cal U}^{a_{\ext}}_{0~a_{\ext}B_{0}}
+{\cal O}({\cal G})~. 
\label{reducweakeq2nonminhs} 
\eeq
Next one proceed as in the proof of Theorem~\ref{theorema}.
\proofbox

\noi
Equations \es{partitionfunctionhighreduch}{khudaverdianreducdirach} leads to
\eq{finitestagediraco} in Corollary~\ref{corollarybfinitestagediraco}.
So the non-minimal approach agrees with the approach of 
Subsection~\ref{finitestagedirac}.

\setcounter{equation}{0}
\section{Higher Level Formalism}
\label{sechigherlevels}

\subsection{Recursive Construction}

\noi
In the irreducible \nth-{\em level} formalism one introduces 
$N$ Lagrange multipliers \mb{\lambda_{\bn}^{\alpha}} of Grassmann parity 
\mb{\epsilon_{\alpha}^{\bn}\!\equiv\!\epsilon_{\alpha}^{\bo}\!+\!n} and $N$
antifields \mb{\lambda^{*}_{\bn\alpha}}, which we collectively call the 
\nth-{\em level fields}. The phase space variables 
\beq
\Gamma^{A}_{\hn} ~\equiv~ \left\{\Gamma^{A}_{\hnl};\lambda^{\alpha}_{\bn},
\lambda^{*}_{\bn\alpha} \right\}
\eeq
for the \nth-level formalism thus consists of fields of
levels \mb{\leq n}. The idea is roughly that the 
\nth-level Lagrange multipliers \mb{\lambda^{\alpha}}
should gauge-fix the \nlth-level first-level 
antifields \mb{\lambda^{*}_{\bnl\alpha}}, although this is just one
gauge-fixing choice out of infinitely many. 

\noi
In the \nth-level formalism, one first lifts the previous 
\nlth-level phase space \mb{\Gamma^{A}_{\hnl}} to 
a fully covariant status, \ie one allows for general coordinate 
transformations 
\mb{\Gamma^{A}_{\hnl}\to\Gamma^{\prime A}_{\hnl}
=\Gamma^{\prime A}_{\hnl}(\Gamma_{\hnl};\hbar_{\hnl})}
that preserve the Planck number symmetries of the previous levels,
so that the \nlth-level odd Laplacian becomes of the 
covariant form 
\beq
\Delta_{\hnl}~=~\frac{(-1)^{\epsilon_{A}}}{2\rho_{\hnl}}
\papal{\Gamma^{A}_{\hnl}}\rho_{\hnl} E^{AB}_{\hnl}\papal{\Gamma^{B}_{\hnl}}
\label{Deltaaugnl}
\eeq
with a symplectic metric 
\mb{E^{AB}_{\hnl}=E^{AB}_{\hnl}(\Gamma_{\hnl};\hbar_{\hnl})} and a measure
density \mb{\rho_{\hnl}=\rho_{\hnl}(\Gamma_{\hnl};\hbar_{\hnl})}. 
Secondly, one defines a \nth-level odd Laplacian 
\beq
\Delta_{\hn}~\equiv~\Delta_{\hnl} + (-1)^{\epsilon_{\alpha}^{\bn}}
\papal{\lambda^{\alpha}_{\bn}}\papal{\lambda^{*}_{\bn\alpha}}~,
\label{Deltaaugn}
\eeq 
a \nth-level Planck constant \mb{\hbar_{\bn}},
and a \nth-level Planck operator
\beq
\pl_{\bn}~=~- \left(\lambda^{\alpha}_{\bn}\lambda^{*}_{\bn\alpha},~
\cdot~\right)_{\hn}+\hbar_{\bn} \frac{\partial }{\partial \hbar_{\bn}}~.
\eeq
At even levels $n$, one has the following \nth-level 
generalization of Principle~\ref{principle3}:

\begin{principle}
The \mb{W_{\hn}}-action should satisfy three principles: 
\begin{enumerate}
\item
Planck number conservation:~~
\mb{\forall i=1, \ldots, n:~\pl_{(i)} (\frac{W_{\hn}}{\hbar_{\bn}})=0}.
\item
The Quantum Master Equation: 
\mb{\Delta_{\hn} \exp\left[\frac{i}{\hbar_{\bn}}W_{\hn}\right]=0}.
\item
The Hessian of \mb{W_{\hn}} has rank equal to half the number of
fields \mb{\Gamma^{A}_{\hn}}, \ie \mb{(n\!+\!1)N} in the irreducible case.
\end{enumerate}
\label{principle3n}
\end{principle}

\noi
The \nth-level partition function is defined as
\beq
{\cal Z}_{\hn}^{\Psi}
~=~\int\! [d\Gamma_{\hnl}][d\lambda_{\bn}]\left.\rho_{\hnl}~
e^{\Ih (W_{\hn}^{\Psi}+X_{\hnl})}\right|_{\lambda^{*}_{\bn}=0}~,
\label{nlevelz}
\eeq
with
\beq
e^{\Ih W^{\Psi}_{\hn}}
~=~e^{-[\stackrel{\rightarrow}{\Delta}_{\hn},\Psi]}e^{\Ih W_{\hn}}~.
\label{Wboltzmannexactvarn}
\eeq
It follows from standard arguments that the partition function 
\mb{{\cal Z}^{\Psi}_{\hl}} does not depend on the gauge fermion 
\mb{\Psi}. At odd levels $n$, there is a similar story where the r\^oles
of $W$ and $X$ are exchanged, up to a relative sign,
\beq
e^{\Ih X^{\Psi}_{\hn}}
~=~e^{[\stackrel{\rightarrow}{\Delta}_{\hn},\Psi]}e^{\Ih X_{\hn}}~.
\label{Xboltzmannexactvarn}
\eeq
Again the Planck number conservation limits the number of
possible \nth-level structure functions in the action
\beq
W_{\hn}~=~G_{\bnl\alpha}\lambda_{\bn}^{\alpha}+(i\hbar_{\bn})H_{\hnl}
+{\cal O}(\lambda_{\bn}^{*})~.
\label{Wplanckexpn}
\eeq
When one decomposes the \nth-level Quantum Master Equation 
in terms of the above \nlth-level structure 
functions one generates a tower of equations; the first few equations read:
\bea
 (G_{\bnl\alpha},G_{\bnl\beta})_{\hnl}
&=& G_{\bnl\gamma} U^{\gamma}_{\bnl \alpha \beta}~,
\label{nonabelinvon} \\
 (\Delta_{\hnl} G_{\bnl\beta})-(H_{\hnl},G_{\bnl\beta})_{\hnl}
&=&(-1)^{\epsilon^{\bn}_{\alpha}}U^{\alpha}_{\bnl \alpha\beta}
+  G_{\bnl\alpha}V^{\alpha}_{\bnl \beta}~, \label{weakeq2n}\\
-(\Delta_{\hnl} H_{\hnl})+\Hf (H_{\hnl},H_{\hnl})_{\hnl}
&=&V^{\alpha}_{\bnl\alpha}-G_{\bnl\alpha}\tilde{G}^{\alpha}_{\bnl}~.
\label{weakeq3n} 
\eea
There is in general no covariant on-shell expression for the higher-level
\mb{H_{\hnl}} with \mb{n \geq 2}, although partial results exist \cite{BT}.

\subsection{Recursive Reduction}

\noi
The reduction from \nth~ to \nlth-level can be demonstrated as follows:

\begin{enumerate}
\item
Go to coordinates
\beq
\Gamma^{A}_{\hn}~\to~
\left\{\Gamma^{A}_{\hnz};\lambda^{\alpha}_{\bnl},
\lambda^{*}_{\bnl\alpha};\lambda^{\alpha}_{\bn},
\lambda^{*}_{\bn\alpha}\right\}
\eeq
with Darboux coordinates at the last two levels $n$ and \mb{n\!-\!1},
and with a measure density \mb{\rho_{\hnl}\to\rho_{\hnz}}. 

\item
Choose the gauge-fixing functions 
\mb{G_{\bnl\alpha}=\frac{\hbar_{\bnl}}{\hbar}\lambda^{*}_{\bnl\alpha}},
so that \mb{U^{\gamma}_{\bnl \alpha \beta}=0}.
In this case the \eq{weakeq2n} yields that
\beq
H_{\hnl}~=~K_{\hnl}
-\lambda^{*}_{\bnl\alpha}\bar{V}^{\alpha}_{\bnl \beta}\lambda^{\beta}_{\bnl}~,
\eeq
where
\beq
\bar{V}^{\alpha}_{\bnl \beta}~\equiv~\int_{0}^{1}\! dt~V^{\alpha}_{\bnl \beta}
(\Gamma_{\hnz};t\lambda_{\bnl},\lambda^{*}_{\bnl}; \hbar_{\hnl})~,
\eeq
and where the integration ``constant'' 
\mb{K_{\hnl}=K_{\hnl}(\Gamma_{\hnz};\lambda^{*}_{\bnl}; \hbar_{\hnl})}
is independent of \mb{\lambda^{\beta}_{\bnl}}.

\item
Define the \nzth-level action as 
\beq
W_{\hnz}~=~(i\hbar_{\bnz})\left. H_{\hnl} 
\right|_{\lambda^{*}_{\bnl}=0}~,
\label{bcn}
\eeq
if $n$ is even. (If $n$ is odd, exchange \mb{X \leftrightarrow W}.) 
The action \mb{W_{\hnz}} defined this way does not depend on 
\mb{\hbar_{\bnl}}, because of Planck number conservation 
\mb{\pl_{\bnl}(W_{\hnz})=0}, and moreover \mb{W_{\hnz}} 
satisfies the \mb{(n\!-\!2)^{\prime}}{\em th}-level Quantum Master Equation
\beq
\Delta_{\hnz} \exp\left[\frac{i}{\hbar_{\bnz}} W_{\hnz} \right]~=~0~,
\eeq
as a result of (one of the consequences of) the 
\nth-level Quantum Master Equation \eq{weakeq3n}. 
Note that other consequences 
of the \nth-level Quantum Master Equation 
would in general impose other conditions on \mb{W_{\hnz}}.
(If the \nzth-level action \mb{W_{\hnz}} is already 
known independently, the \eq{bcn} should be interpreted as a boundary 
condition.)

\item
Choose the gauge fermion \mb{\Psi} independent of the 
\nth-level Lagrange multiplier antifields 
\mb{\lambda^{*}_{\bn\alpha}}. Then \mb{\Psi} is independent of the
Lagrange multipliers \mb{\lambda_{\bn}^{\alpha}} as well, because of Planck
number conservation \mb{\pl_{\bn}(\Psi)\!=\!0}. (Here we have implicitly 
assumed that there are only non-negative powers of 
\nth-level objects \mb{\{\Gamma_{\bn}^{A};\hbar_{\bn}\}} 
present inside \mb{\Psi}.) 
Using the symmetry \e{deltasym} of the \mb{\Delta_{\hn}}-operator, one derives
\bea
{\cal Z}_{\hn}^{\Psi}
&=&\int\! [d\Gamma_{\hn}]~\rho_{\hnz}~
e^{\Ih W_{\hn}} ~e^{[\stackrel{\rightarrow}{\Delta}_{\hn},\Psi]}
\delta(\lambda^{*}_{\bn})e^{\Ih X_{\hnl}} \cr
&=&\int\! [d\Gamma_{\hn}]~\rho_{\hnz}~
e^{\Ih( \lambda^{*}_{\bnl\alpha}\lambda_{\bn}^{\alpha}+ W_{\hnz})}~
\delta(\lambda^{*}_{\bn})~
e^{[\stackrel{\rightarrow}{\Delta}_{\hn},\Psi]}e^{\Ih X_{\hnl}} \cr
&=&\int\! [d\Gamma_{\hnz}][d\lambda_{\bnl}]\left. \rho_{\hnz}~ 
e^{\Ih (W_{\hnz}+X_{\hnl}^{\Psi})}\right|_{\lambda^{*}_{\bnl}=0}
~=~{\cal Z}_{\hnl}^{\Psi}~,
\label{nm1levelz}
\eea
which completes the reduction step.
\end{enumerate}

\noi
The \nth-level formalism can also be set up in the case of reducible
gauge-fixing constraints and second-class constraints. In the reducible case
one introduces the relevant stages of minimal and non-minimal fields, in
accordance with the general field-antifield prescription, as we saw in 
Section~\ref{secreduc}.

\noi
Hence the multi-level formalism consists in recursively building master 
actions \mb{W_{\ho}\!\equiv\! W}, \mb{X_{\hl}\!\equiv\! X}, \mb{W_{\hz}}, 
\mb{X_{\htr}}, \mb{\ldots}. By zig-zagging through the $W$- and $X$-parts it 
becomes a simple matter to re-use the constructions of the previous levels 
and to create a manifestly gauge-independent formalism.

\section{Conclusion}
\label{secconc}

\noi
Driven by the wish to develop the Lagrangian quantization program into its
most general and axiomatized formulation, we have in this paper 
focused on three aspects. First, we have given a more fully geometric
description of the multi-level formalism, in all generality. Second,
we have explored the new symmetric formulation which puts the ``action"
$W$ on the same footing as the ``gauge-fixing" $X$ to yield the full
quantum action that enters in the Boltzmann factor of the functional
integral. One particular aspect of this symmetry concerns the algebras
behind the Master Actions $W$ and $X$.
On one hand there is the gauge-generating algebra behind $W$, which is
associated with gauge symmetries of the classical action, and is
known to accommodate both open and reducible gauge-algebras. 
On the other hand there is the gauge-fixing algebra behind $X$,
which is a quantum mechanical feature, with no analog at the level of the
classical theory, that usually carries just an irreducible algebra.
We have in detail demonstrated how to permit a reducible gauge-fixing
algebra, and calculated the associated measure or gauge volume from first 
principles, namely by solving the Quantum Master Equation. 
Several consistency checks on the formalism were performed along the way
by reduction methods. 
Third, we included an extensive discussion of antisymplectic second-class
constraints in the field-antifield formalism, and demonstrated manifest
invariance under reparametrization of the second-class constraints. 
Second-class constraints hold surprisingly many features in common with
the gauge-fixing constraints \mb{G_{\alpha}}, and often do they appear
side-by-side in the formulas. In this way, the second-class constraints
merge effortlessly with the multi-level formalism, even in the case of
reducible gauge-fixing algebras.

\noi
While there are still many more aspects of the new and more axiomatic
formulation of the Lagrangian quantization prescription that need to
be explored, it is already now apparent that there is a rich and beautiful
algebraic structure behind. This algebra has its root in the one single
object from which all is derived: the Grassmann-odd nilpotent
$\Delta$-operator, and its associated Quantum Master Equation.

\noi
{\sc Note added:}~After the paper appeared on the archive we became aware of 
\Ref{reshetnyak} where a reducible gauge-fixing $X$ arose in a superfield 
context.

\noi
{\sc Acknowledgement:}~I.A.B.\ and K.B.\ would like to thank the Niels Bohr 
Institute for the warm hospitality extended to them there. The work of I.A.B.\
is supported partially by grants RFBR 05-01-00996 and RFBR 05-02-17217, and 
the work of K.B.\ is supported partially by the Ministry of Education of the
Czech Republic under the project MSM 0021622409. 

\noi
{\sc Correction added after publication in Nuclear Physics B:}~{}From three
lines above eq.\ (4.33) down to three lines below eq.\ (4.34):~The three 
consecutive sentences, which start with ``For every system of unitarizing
coordinates \ldots'', should be discarded. This is because the statement that
eq.\ (4.34) can be solved recursively to all orders in \mb{\Theta}, is, in
general, incorrect. Nevertheless, transversal coordinate systems do exist
locally, which is all that is needed for the ensuing discussion.


\end{document}